\documentclass[a4paper,11pt]{article}
\usepackage{jheppub} 

\usepackage[T1]{fontenc} 
\usepackage{tikz-feynman}

\title{\boldmath Spin Dependent Gravitational Tail Memory in D=4}

\author[a]{Debodirna Ghosh}
\author{and}
\author[b]{Biswajit Sahoo}

\affiliation[a]{Chennai Mathematical Institute,\\ Siruseri, Chennai, India}
\affiliation[b]{Fields and Strings Laboratory, Institute of Physics,\\ Ecole Polytechnique Federale de Lausanne (EPFL),\\ CH-1015 Lausanne, Switzerland}

\emailAdd{debodirna@cmi.ac.in}
\emailAdd{biswajit.sahoo@epfl.ch}

\abstract{We derive the leading spin-dependent gravitational tail memory, which appears at the second post-Minkowskian (2 PM) order and behaves as $u^{-2}$ for large retarded time $u$. This result follows from classical soft graviton theorem at order $\omega\ln\omega$ as a low-frequency expansion of gravitational waveform with frequency $\omega$. First, we conjecture the result from the classical limit of quantum soft graviton theorem up to sub-subleading order in soft expansion and then we derive it for a classical scattering process without any reference to the soft graviton theorem. The final result of the gravitational waveform in the direct derivation completely agrees with the conjectured result. 
}

\begin{document} 
\maketitle
\flushbottom

\section{Introduction}

The observation of the permanent displacement between the mirrors of the gravitational wave detector relative to their initial distance after the passage of full gravitational radiation produced from an astrophysical scattering event is known as gravitational memory. Non-oscillatory sources, e.g. scattering of two un-bounded compact objects in hyperbolic orbit contributes to "linear memory", which can be read off from the matter energy-momentum tensor determined in terms of the trajectories of the scattered objects \cite{mem1,mem2,mem3}. In companion with this, there will also be "non-linear memory" due to gravitational radiation from the gravitational waves produced during the scattering process, which can be read off studying gravitational energy-momentum tensor \cite{Payne,mem4,christodoulou,thorne,bondi}. The nonlinear memory effect is always present for any gravitational scattering process whether it is bounded or not. In the recent future there is a hope of direct detection of gravitational memory in the upcoming gravitational wave detectors \cite{1410.3323,1003.3486}.

\begin{center}
\begin{figure}[h!]\label{scattering_setup}
	\includegraphics[scale=0.5]{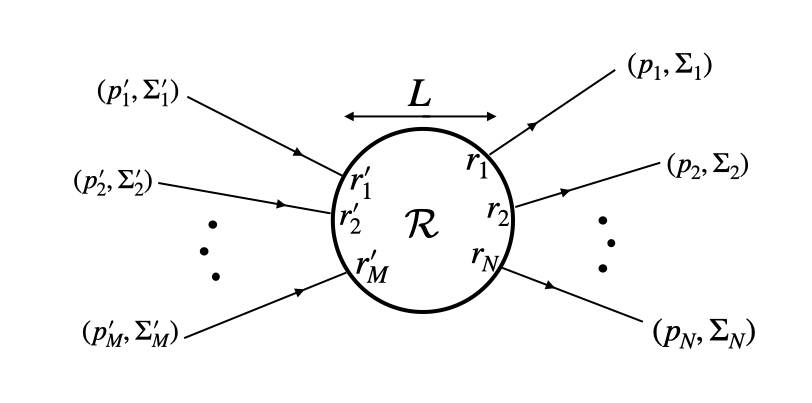}
	\caption{The setup of the gravitational scattering event.}
\end{figure}
\end{center}

\begin{center}
\begin{figure}[h!]
	\includegraphics[scale=0.25]{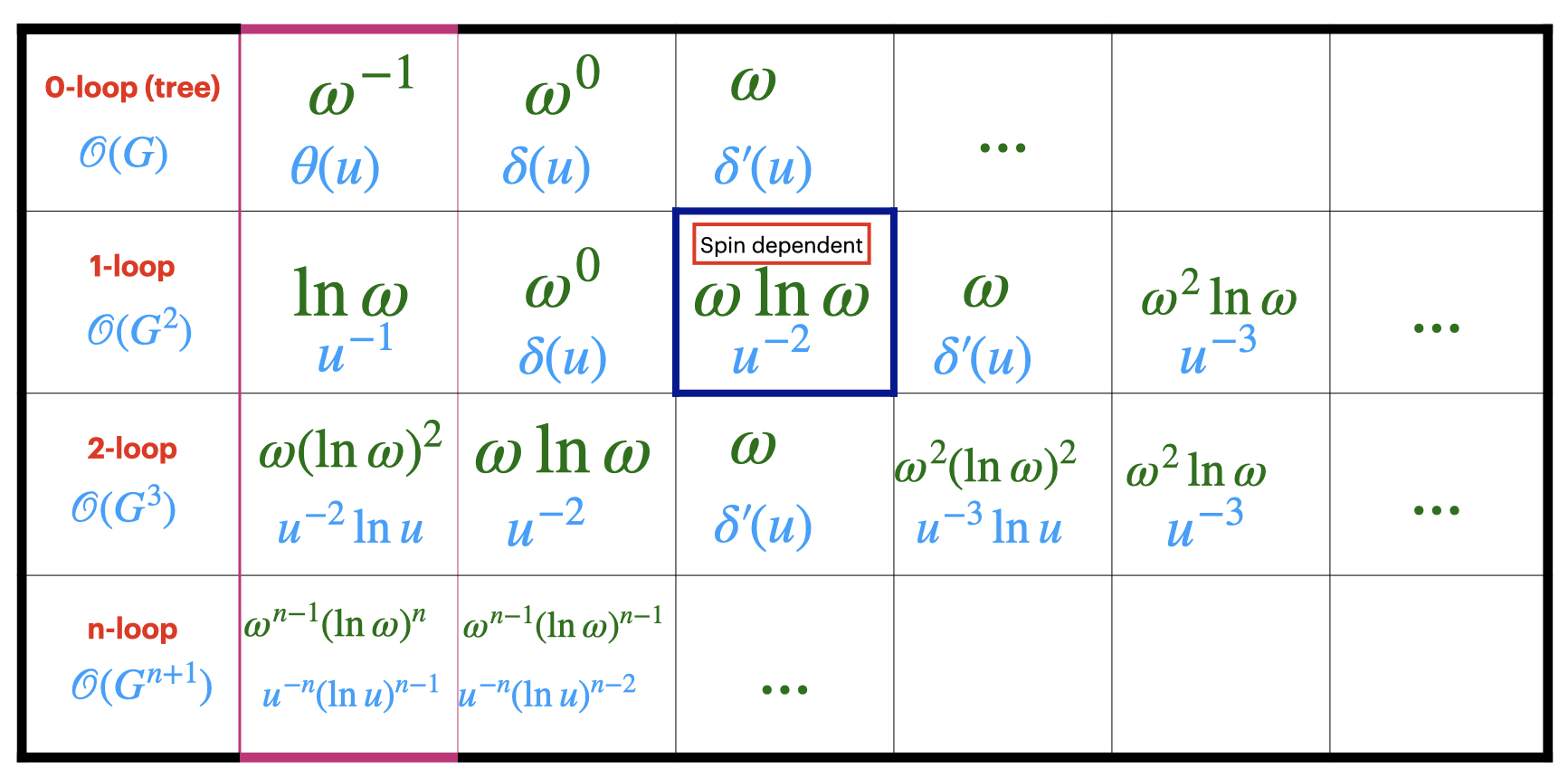}
	\caption{This table summarizes different orders of low frequency gravitational waveform $\widetilde{e}^{\mu\nu}(\omega,R,\textbf{n})$ in $\omega\rightarrow 0$ limit [written in green] and their relations to post-Minkowskian (PM) expansion. It also describe how the late and early time gravitational waveform $e^{\mu\nu}(u,R\hat{n})$  behaves at large retarded time $u$ [written in blue]. The spin dependent order $\omega\ln\omega$ gravitational waveform indicated inside the blue cell is the primary interest of this paper. }\label{table}
\end{figure}
\end{center}

In recent years there has been a proposal of another kind of gravitational memory known as "tail memory", which describes how the mirrors of gravitational wave detector behaves at large retarded time before reaching their permanent displaced position\cite{1806.01872,1808.03288,1912.06413,2008.04376,2105.08739}. The late and early time gravitational waveforms responsible for gravitational tail memories are first conjectured from the classical limit of soft graviton theorem with some infrared regulator prescription\cite{1801.07719,1804.09193,1808.03288} and then derived explicitly in the name of classical soft graviton theorem\cite{1912.06413,2008.04376}. It has also been shown that these late and early time behaviours of gravitational waveform are related to the radiative mode of low frequency gravitational waveform via Fourier transformation in frequency variable. In \cite{1912.06413,2008.04376} the setup considered for the scattering event is schematically described in Fig.\ref{scattering_setup}. Consider $M$ number of objects are coming in from asymptotic past with masses $\lbrace m'_a\rbrace$, momenta $\lbrace p_a^{\prime}\rbrace$, spins $\lbrace \Sigma'_a\rbrace$ going through some unspecified interaction within region $\mathcal{R}$ of spacetime size $L$ and disperse to $N$ number of objects with masses $\lbrace m_a\rbrace$, momenta $\lbrace p_a\rbrace$, spins $\lbrace \Sigma_a\rbrace$. These incoming and outgoing objects can also be massless radiations. We choose the origin of coordinate system (i.e. scattering centre) well inside this region $\mathcal{R}$ and we place the gravitational wave detector at a distance $R$ from this origin, along the direction $\hat{n}$. 
Now we define the deviation of metric from Minkowski background as:
\be
h_{\mu\nu}(x)\ \equiv\ \f{1}{2}\big(g_{\mu\nu}(x)-\eta_{\mu\nu}\big)\hspace{1cm},\hspace{1cm}e_{\mu\nu}(x)\equiv h_{\mu\nu}(x)-\f{1}{2}\eta_{\mu\nu}\eta^{\alpha\beta}h_{\alpha\beta}(x)
\ee
The time Fourier transform of the trace reversed metric is defined as:
\be
\widetilde{e}^{\mu\nu}(\omega, \vec{x})&=&\ \int_{-\infty}^\infty dt\ e^{i\omega t}\ e^{\mu\nu}(t,\vec{x})
\ee
With this scattering setup the goal is to determine the radiative mode of $\widetilde{e}^{\mu\nu}(\omega, R\hat{n})$ in the frequency range $R^{-1}<< \omega<<L^{-1} $. In \cite{1912.06413,2008.04376} the first three non-analytic contribution in $\omega\rightarrow 0$ limit has been evaluated which behave like $\omega^{-1}, \ln\omega$ and $\omega(\ln\omega)^2$ and the coefficients depend only on the incoming and outgoing momenta and $\hat{n}$. These low-frequency gravitational waveforms contribute to DC gravitation memory, $u^{-1}$ tail memory, and $u^{-2}\ln u$ tail memory respectively. To derive these results the authors developed an iterative prescription of solving the geodesic equation of scattered objects and Einstein equation treating gravitational constant $G$ as an iterative parameter. Since $GM\omega$ is a dimensionless quantity in the unit where speed of light is unity, expansion in the power of $\omega$ is equivalent to expansion in the power of $G$, where $M$ represents the mass or momentum of scattered objects. Generalizing this prescription the structure of leading non-analytic contribution in each iterative order has been conjectured in \cite{2008.04376}, which has been summarized in the second column of the table given in Fig.\ref{table}. 
If the scattered objects carry spin $\Sigma$, along with $GM\omega$ there will be another dimensionless quantity $G\Sigma\omega^{2}$. This simple dimensional analysis tells us that the spin dependence of gravitational waveform at any iterative order in $G$ will carry an extra factor of $\omega$ relative to the spin-independent leading non-analytic contribution at that order. With this observation, from the table in Fig.\ref{table} we see that at the order $G$ of the gravitational waveform, spin dependence appears at order $\omega^{0}$. Since the order $\omega^{0}$ term is analytic in $\omega\rightarrow 0$ limit it does not contribute to displacement kind of memory\cite{1502.06120}. Now in the next iterative order i.e. in the order $G^2$ of the gravitational waveform the spin dependence comes at order $\omega\ln\omega$, which is non-analytic in $\omega\rightarrow 0$ limit and contributes to order $u^{-2}$ tail memory\footnote{Here we want to emphasize that the order $u^{-2}$ tail memory is there even for the scattering of non-spinning objects as $G\mathbf{L}\omega^{2}$ for $\mathbf{L}$ being the orbital angular momentum of the scattered object, is also a dimensionless quantity like $G\Sigma\omega^{2}$. Hence in our final result of order $u^{-2}$ tail memory, setting the spins of the scattered objects to zero, one can read off the result of the gravitational waveform for non-spinning object's scattering. }. The existence of spin-dependent $u^{-2}$ tail memory was first pointed out in the subsection-(5.3) of \cite{2008.04376}, using some naive analysis of matter and gravitational energy-momentum tensor, indicated within the blue cell in the table of Fig.\ref{table}. From the table, it is also clear that the spin-dependent order $\omega\ln\omega$ waveform is not exact, but receives corrections from order $G^3$ as well. The order $G^{3}$ correction to $\omega\ln\omega$ gravitational waveform is expected to be spin-independent.

In this paper, our main goal is to derive the $\omega\ln\omega$ gravitational waveform at order $G^{2}$ and the gravitational tail memory it predicts. The main result of the paper is summarised in eq.\eqref{future_memory} and eq.\eqref{past_memory}. In \S\ref{S:prediction} we make a conjecture on the $\omega\ln\omega$ gravitational waveform from the classical limit of sub-subleading soft graviton theorem and discuss various applications of the result in different limits. Then in \S\ref{S:derivation} we derive the $\omega\ln\omega$ gravitational waveform directly for a classical scattering process without any reference to soft graviton theorem. Various appendices discuss intermediate steps required in the analysis of \S\ref{S:derivation}. Finally, in \S\ref{S:conclusion} we conclude the paper by discussing the novel features of our result, its theoretical and observational importance, and possibility of re-derivation of our result with other available prescriptions in the literature.
\section{Prediction of spin dependent gravitational waveform from classical limit of soft graviton theorem}\label{S:prediction}
Emboldened by the success of predicting long wavelength gravitational waveform and gravitational tail memory from classical limit of quantum soft graviton theorem at subleading and sub-subleading orders \cite{1808.03288,1912.06413}, here we proceed to conjecture the leading spin dependent gravitational tail memory in $D=4$ spacetime dimensions.

In spacetime dimensions $D>4$, the universal piece of quantum soft graviton operator up to sub-subleading order takes the following form\cite{1706.00759}: 
\be
\mathbf{S}^{gr}_{uni}&=&\ \sum_{a=1}^{M+N} \Bigg[\f{\varepsilon_{\mu\nu}p_a^\mu p_a^\nu}{p_a.k}+\f{\varepsilon_{\mu\nu}p_a^\mu k_\rho \widehat{\mathbf{J}}_a^{\rho\nu}}{p_a.k}+\f{1}{2}\f{\varepsilon_{\mu\nu} k_\rho k_\sigma \widehat{\mathbf{J}}_a^{\rho\mu}\widehat{\mathbf{J}}_a^{\sigma\nu}}{p_a.k}\Bigg]
\ee
Note that we have not included the non-universal piece in the sub-subleading soft factor which is determined in terms of non-minimal coupling of hard particle field to soft graviton field via Riemann tensor along with the three point 1PI vertex involving two hard particles and a soft graviton\cite{1706.00759}. In the above expression of soft graviton operator $\varepsilon$ represents the polarization tensor of the soft graviton with momentum $k=\omega \mathbf{n}$, where $\omega$ is the energy of the outgoing soft graviton. The scattering process involves $M$ number of ingoing hard particles and $N$ number of outgoing hard particles. Momentum of hard particle-a is denoted by $p_a$ and angular momentum operator is given by:
\be
\widehat{\mathbf{J}}_{a}^{\mu\nu}\equiv p_a^\nu \f{\p}{\p p_{a\mu}}-p_a^\mu \f{\p}{\p p_{a\nu}}+\ \widehat{\mathbf{\Sigma}}_a^{\mu\nu}
\ee
where $\widehat{\mathbf{\Sigma}}_a^{\mu\nu}$ represents the quantum spin generator of Lorentz group for hard particle-a. We are following the convention that the outgoing particles carry positive energy and the incoming particles carry negative energy.

From the classical limit of quantum soft graviton theorem up to sub-subleading order in D=4, we  get the following radiative mode of gravitational waveform with frequency $\omega$ at distance $R=|\vec{x}|$ from the scattering centre   \cite{1801.07719, 1804.09193, 1808.03288}:
\begingroup
\allowdisplaybreaks
\be
\widetilde{e}^{\mu\nu}(\omega,\vec{x})
&=&\ (-i)\ \f{2G}{R}\ e^{i\omega R}\ exp\Big[-2iG \ln\lbrace(\omega+i\epsilon)R\rbrace\sum_{b=1}^{N}p_{b}.k \Big]\non\\
&&\times \sum_{a=1}^{M+N} \Bigg[\f{p_a^\mu p_a^\nu}{p_a.k}-i\ \f{p_a^{(\mu} k_\rho \mathbf{J}_a^{\rho\nu)}}{p_a.k}-\f{1}{2}\f{\varepsilon_{\mu\nu} k_\rho k_\sigma \mathbf{J}_a^{\rho\mu}\mathbf{J}_a^{\sigma\nu}}{p_a.k}\Bigg]\label{e_trial}
\ee
\endgroup
where $\mathbf{J}_a^{\mu\nu}$ is the classical angular momentum of particle-a expressed in terms of trajectory $X_a$ and classical spin $\Sigma_a$ of particle-a by the following relation:
\be
\mathbf{J}_a^{\mu\nu}= X_a^{\mu}p_a^\nu -X_a^\nu p_a^\mu +\Sigma_a^{\mu\nu}
\ee
In eq.\eqref{e_trial}, the symmetrization convention we are following is: $A^{(\alpha}B^{\beta)}\equiv \f{1}{2}(A^\alpha B^\beta +A^\beta B^\alpha)$. It is well known that in four spacetime dimensions the asymptotic trajectories of the scattered particles receive logarithmic correction due to long range gravitational force \cite{1804.09193,1808.03288} (e.g. if the asymptotic trajectory of particle-a is represented by $X_a^{\mu}(\sigma)=r_a^{\mu}+v_a^{\mu}\sigma +Y_a^{\mu}(\sigma)$ for proper time $\sigma$, then $Y_a(\sigma)$  behaves like $\ln|\sigma|$ for $\sigma\rightarrow \pm\infty$ ). So following the prescription of \cite{1804.09193,1808.03288}, if we replace $\ln|\sigma|$ by $-\ln(\omega +i\epsilon \eta_a)$ in the classical angular momentum of eq.\eqref{e_trial}, it predicts the correct gravitational waveform, which has been independently verified in \cite{1912.06413, 2008.04376} up to sub-subleading order non-spinning particle scattering. Here once again using the same prescription, from eq.\eqref{e_trial} we get,
\begingroup
\allowdisplaybreaks
\be
&&\widetilde{e}^{\mu\nu}(\omega,\vec{x})\non\\
&=&\ (-i)\ \f{2G}{R}\ e^{i\omega R}\ \exp\Big[-2iG \ln\lbrace(\omega+i\epsilon)R\rbrace\sum_{b=1}^{N}p_{b}.k \Big]\non\\
&&\times \Bigg[\sum_{a=1}^{M+N}\f{p_{a}^{\mu}p_{a}^{\nu}}{p_{a}.k}+\sum_{a=1}^{M+N}\f{p_{a}^{(\mu}k_{\rho}}{p_{a}.k}\Bigg{\lbrace}\Bigg(p_{a}^{\nu)}\f{\p}{\p p_{a\rho}}-p_{a}^{\rho}\f{\p}{\p p_{a\nu)}}\Bigg)K_{gr}^{cl}-i\Big(r_{a}^{\rho}p_{a}^{\nu)}-r_{a}^{\nu)}p_{a}^{\rho}+\Sigma_{a}^{\rho\nu)}\Big)\Bigg{\rbrace}\non\\
&&\ +\ \f{1}{2}\sum_{a=1}^{M+N}\f{k_{\rho}k_{\sigma}}{p_{a}.k}\Bigg{\lbrace}\Bigg(p_{a}^{\mu}\f{\p}{\p p_{a\rho}}-p_{a}^{\rho}\f{\p}{\p p_{a\mu}}\Bigg)K_{gr}^{cl}-i\Big(r_{a}^{\rho}p_{a}^{\mu}-r_{a}^{\mu}p_{a}^{\rho}+\Sigma_{a}^{\rho\mu}\Big)\Bigg{\rbrace}\non\\
&&\times \Bigg{\lbrace}\Bigg(p_{a}^{\nu}\f{\p}{\p p_{a\sigma}}-p_{a}^{\sigma}\f{\p}{\p p_{a\nu}}\Bigg)K_{gr}^{cl}-i\Big(r_{a}^{\sigma}p_{a}^{\nu}-r_{a}^{\nu}p_{a}^{\sigma}+\Sigma_{a}^{\sigma\nu}\Big)\Bigg{\rbrace}\ \Bigg]
\ee
\endgroup
where the expression of $K_{gr}^{cl}$ is given by\cite{2008.04376}:
\begingroup
\allowdisplaybreaks
\be
K_{gr}^{cl}\ &=&\ -\f{i}{2}\ (8\pi G)\sum_{\substack{b,c\\b\neq c}}\int_{\omega}^{L^{-1}}\f{d^4\ell}{(2\pi)^4}\ G_{r}(\ell)\f{1}{p_{b}.\ell+i\epsilon}\f{1}{p_{c}.\ell-i\epsilon}\ \Big{\lbrace}(p_{b}.p_{c})^2-\f{1}{2}p_{b}^2p_{c}^2\Big{\rbrace}\non\\
&=&\ -\f{i}{2}\ (2G)\ \sum_{\substack{b,c\\ b\neq c\\ \eta_{b}\eta_{c}=1}}\ \ln\Big{\lbrace}L(\omega+i\epsilon\eta_{b})\Big{\rbrace}\ \f{(p_{b}.p_{c})^{2}-\f{1}{2}p_{b}^{2}p_{c}^{2}}{\sqrt{(p_{b}.p_{c})^{2}-p_{b}^2 p_{c}^2}}
\ee
\endgroup
In the above expression $\eta_b=+1$ if the particle-b is outgoing and $\eta_b=-1$ if the particle-b is ingoing in the scattering event. In the expression of $K_{gr}^{cl}$, $L^{-1}$ denotes the UV energy scale and for the value of $\omega$ below this $UV$ scale our result for the gravitational waveform can be trusted. Roughly, the length scale $L\simeq |r_a -r_b|$ is of the order of the impact parameters for a scattering event involving un-bounded hyperbolic orbits of compact objects. 

 Now expanding the exponential in small $\omega$ limit we find the following order $\mathcal{O}(\omega\ln\omega)$ contribution of the gravitational waveform at order $G^2$:
\begingroup
\allowdisplaybreaks
\be
\Delta_{(G^2)}^{(\omega\ln\omega)}\ \widetilde{e}^{\mu\nu}(\omega,\vec{x})
&=&\ (-i)\ \f{2G}{R}\ \exp\Big\lbrace{i\omega R-2iG \ln R\sum\limits_{b=1}^{N}p_{b}.k}\Big\rbrace \Bigg[-2G \ln\lbrace\omega+i\epsilon\rbrace\sum_{b=1}^{N}p_{b}.k\non\\
&&\times \sum_{a=1}^{M+N}\f{p_{a}^{(\mu}k_{\rho}}{p_{a}.k}\Big(r_{a}^{\rho}p_{a}^{\nu)}-r_{a}^{\nu)}p_{a}^{\rho}+\Sigma_{a}^{\rho\nu)}\Big)\non\\
&&-\f{i}{2}\sum_{a=1}^{M+N}\f{k_{\rho}k_{\sigma}}{p_{a}.k}\Bigg{\lbrace}\Bigg(p_{a}^{\mu}\f{\p}{\p p_{a\rho}}-p_{a}^{\rho}\f{\p}{\p p_{a\mu}}\Bigg)K_{gr}^{cl}\times \Big(r_{a}^{\sigma}p_{a}^{\nu}-r_{a}^{\nu}p_{a}^{\sigma}+\Sigma_{a}^{\sigma\nu}\Big)\non\\
&&\ +\ \Bigg(p_{a}^{\nu}\f{\p}{\p p_{a\sigma}}-p_{a}^{\sigma}\f{\p}{\p p_{a\nu}}\Bigg)K_{gr}^{cl}\times \Big(r_{a}^{\rho}p_{a}^{\mu}-r_{a}^{\mu}p_{a}^{\rho}+\Sigma_{a}^{\rho\mu}\Big)\Bigg{\rbrace}\Bigg]\ \label{waveform_expectation}
\ee
\endgroup
Explicitly the above equation can be written in terms of incoming scattering data $\lbrace p_a^{\prime} ,\Sigma_a^{\prime}, r_a^{\prime}\rbrace $ and outgoing scattering data $\lbrace p_a ,\Sigma_a , r_a\rbrace $ in the following way:
\begingroup
\allowdisplaybreaks
\be
&&\Delta_{(G^2)}^{(\omega\ln\omega)}\ \widetilde{e}^{\mu\nu}(\omega,R, \mathbf{n})\non\\
&=&\ \f{G^2}{R} \exp\Big\lbrace{i\omega R-2iG \omega \ln R\sum\limits_{b=1}^{N}p_{b}.\mathbf{n}}\Big\rbrace\non\\
&&\times \Bigg[ 4i \omega \ln (\omega +i\epsilon)\sum_{b=1}^{N}p_b.\mathbf{n}\Bigg\lbrace \sum_{a=1}^{N}\f{p_{a}^{(\mu}\mathbf{n}_{\rho}}{p_{a}.\mathbf{n}}\Big(r_{a}^{\rho}p_{a}^{\nu)}-r_{a}^{\nu)}p_{a}^{\rho}+\Sigma_{a}^{\rho\nu)}\Big)\non\\
&& \ -\sum_{a=1}^{M}\f{p^{'(\mu}_a \mathbf{n}_{\rho}}{p'_{a}.\mathbf{n}}\Big(r_{a}^{'\rho}p_{a}^{'\nu)}-r_{a}^{'\nu)}p_{a}^{'\rho}+\Sigma_{a}^{'\rho\nu)}\Big)\Bigg\rbrace\non\\
&& +i\omega \ln(\omega +i\epsilon)\sum_{a=1}^{N}\sum_{\substack{b=1\\ b\neq a}}^{N}\f{p_a.p_b}{[(p_a.p_b)^2 -p_a^2 p_b^2 ]^{3/2}}\lbrace 2(p_a.p_b)^2 -3p_a^2 p_b^2 \rbrace \f{\mathbf{n}_\rho \mathbf{n}_\sigma}{p_a.\mathbf{n}}\non\\
&& \Big\lbrace (p_a^\mu p_b^\rho -p_a^\rho p_b^\mu)(r_{a}^{\sigma}p_{a}^{\nu}-r_{a}^{\nu}p_{a}^{\sigma}+\Sigma_{a}^{\sigma\nu}) \ +\ (p_a^\nu p_b^\sigma -p_a^\sigma p_b^\nu)(r_{a}^{\rho}p_{a}^{\mu}-r_{a}^{\mu}p_{a}^{\rho}+\Sigma_{a}^{\rho\mu})\Big\rbrace\non\\
&& +i\omega \ln(\omega -i\epsilon)\sum_{a=1}^{M}\sum_{\substack{b=1\\ b\neq a}}^M \f{p'_a.p'_b}{[(p'_a.p'_b)^2 -p_a^{'2} p_b^{'2} ]^{3/2}}\lbrace 2(p'_a.p'_b)^2 -3p_a^{'2} p_b^{'2} \rbrace \f{\mathbf{n}_\rho \mathbf{n}_\sigma}{p'_a.\mathbf{n}}\non\\
&& \Big\lbrace (p_a^{'\mu} p_b^{'\rho} -p_a^{'\rho} p_b^{'\mu})(r_{a}^{'\sigma}p_{a}^{'\nu}-r_{a}^{'\nu}p_{a}^{'\sigma}+\Sigma_{a}^{'\sigma\nu}) \ +\ (p_a^{'\nu} p_b^{'\sigma} -p_a^{'\sigma} p_b^{'\nu})(r_{a}^{'\rho}p_{a}^{'\mu}-r_{a}^{'\mu}p_{a}^{'\rho}+\Sigma_{a}^{'\rho\mu})\Big\rbrace\Bigg]\non\\ \label{expectation_e}
\ee
\endgroup
where $k^{\mu}=\omega \mathbf{n}^\mu=\omega (1,\hat{n})$ with $\hat{n}$ being the unit vector denoting the direction of gravitational radiation. For the scattering of non-spinning objects, we can read off the order $\omega\ln\omega$ gravitational waveform after setting $\Sigma_a=0$ and $\Sigma'_a=0$ in the above relation.
\subsection{Spin dependent tail memory}\label{S:memory_final_I}
 Performing Fourier transform in $\omega$ variable of expression in eq.\eqref{expectation_e}, we find the following expressions for late and early time gravitational waveforms:
\begingroup
\allowdisplaybreaks
\be
&&\Delta_{(G^2)}^{(1/u^2)}\ e^{\mu\nu}(u,\vec{x}=R \hat{n})\non\\
&=&\ -\f{G^2}{R} \ \f{1}{u^2}\Bigg[4\sum_{b=1}^{N}p_b.\mathbf{n}\Bigg\lbrace \sum_{a=1}^{N}\f{p_{a}^{(\mu}\mathbf{n}_{\rho}}{p_{a}.\mathbf{n}}\Big(r_{a}^{\rho}p_{a}^{\nu)}-r_{a}^{\nu)}p_{a}^{\rho}+\Sigma_{a}^{\rho\nu)}\Big)\non\\
&& \ -\sum_{a=1}^{M}\f{p^{\prime (\mu}_a \mathbf{n}_{\rho}}{p'_{a}.\mathbf{n}}\Big(r_{a}^{\prime\rho}p_{a}^{\prime\nu)}-r_{a}^{\prime\nu)}p_{a}^{\prime\rho}+\Sigma_{a}^{\prime\rho\nu)}\Big)\Bigg\rbrace\non\\
&& + \sum_{a=1}^{N}\sum_{\substack{b=1\\ b\neq a}}^{N}\f{p_a.p_b}{[(p_a.p_b)^2 -p_a^2 p_b^2 ]^{3/2}}\lbrace 2(p_a.p_b)^2 -3p_a^2 p_b^2 \rbrace \f{\mathbf{n}_\rho \mathbf{n}_\sigma}{p_a.\mathbf{n}} \Big\lbrace (p_a^\mu p_b^\rho -p_a^\rho p_b^\mu)(r_{a}^{\sigma}p_{a}^{\nu}-r_{a}^{\nu}p_{a}^{\sigma}+\Sigma_{a}^{\sigma\nu}) \non\\
&&\ +\ (p_a^\nu p_b^\sigma -p_a^\sigma p_b^\nu)(r_{a}^{\rho}p_{a}^{\mu}-r_{a}^{\mu}p_{a}^{\rho}+\Sigma_{a}^{\rho\mu})\Big\rbrace \Bigg] \ ,\hspace{1cm}\hbox{for $u\rightarrow +\infty$}\label{future_memory}
\ee
\be
&&\Delta_{(G^2)}^{(1/u^2)}\ e^{\mu\nu}(u,\vec{x}=R \hat{n})\non\\
&=&  \f{G^2}{R} \ \f{1}{u^2}\sum_{a=1}^{M}\sum_{\substack{b=1\\ b\neq a}}^{M}\f{p'_a.p'_b}{[(p'_a.p'_b)^2 -p_a^{\prime 2} p_b^{\prime 2} ]^{3/2}}\lbrace 2(p'_a.p'_b)^2 -3p_a^{\prime 2} p_b^{\prime 2} \rbrace \f{\mathbf{n}_\rho \mathbf{n}_\sigma}{p'_a.\mathbf{n}}\non\\
&& \Big\lbrace (p_a^{\prime\mu} p_b^{\prime\rho} -p_a^{\prime\rho} p_b^{\prime\mu})(r_{a}^{\prime\sigma}p_{a}^{\prime\nu}-r_{a}^{\prime\nu}p_{a}^{\prime\sigma}+\Sigma_{a}^{\prime\sigma\nu}) \ +\ (p_a^{\prime\nu} p_b^{\prime\sigma} -p_a^{\prime\sigma} p_b^{\prime\nu})(r_{a}^{\prime\rho}p_{a}^{\prime\mu}-r_{a}^{\prime\mu}p_{a}^{\prime\rho}+\Sigma_{a}^{\prime\rho\mu})\Big\rbrace \ ,\non\\
&&\hspace{8cm} \hbox{for $u\rightarrow -\infty$}\label{past_memory}
\ee
\endgroup
where retarded time $u$ is given by $u=t-R+2G \ln R\ \sum\limits_{b=1}^{N}p_b\cdot\mathbf{n}$ .
The above expressions predict the order $u^{-2}$ tail memory along with the known order $u^{-1}$ and $u^{-2}\ln u$ tail memories. For non-spinning object scattering the order $\mathcal{O}(u^{-2})$ tail memory is non-vanishing and can be read off by setting $\Sigma_a=0$ and $\Sigma'_a=0$ in the above expressions. At this stage the order $\mathcal{O}(u^{-2})$ gravitational tail memory seems to be non-universal even at order $G^2$, as the result depends not only on the asymptotic data i.e. incoming and outgoing momenta and spins of scattered objects but also on the choice of the region $\mathcal{R}$ through $r_a$. Let us fix two particular time slices before and after the scattering event such that before and after this time, the value of the kinetic energy of all the particles exceeds the value of their potential energy due to interaction among themselves. This way we can fix the boundary of the region $\mathcal{R}$. In our notation this boundary corresponds to $\sigma=0$ for the particle's trajectory i.e. $X_a(\sigma=0)=r_a$. Now if we want to change our definition of region $\mathcal{R}$ to a different time slice say $\sigma=\sigma_0$, it would not affect our result above as under this choice $r_a$ will be shifted by $r_a+\f{p_a}{m_a}\sigma_0$ and that does not affect $(r_a^\alpha p_a^\beta - r_a^\beta p_a^\alpha)$ combination. Hence this property suggests that in principle, it is possible to write the result only in terms of the asymptotic data. As a piece of evidence, consider a special case of $2\rightarrow 2$ scattering event where object-1 is very heavy and object-2 is a probe of small mass and suppose they scatter with a large impact parameter. Now for this process, if we choose the scattering centre at the origin of object-1, then $r_1=0 $ and $r_2$ is the impact parameter which is determinable in terms of asymptotic scattering data.
\subsection{Spin dependent tail memory rewritten}\label{S:rewritten}
Consider a classical scattering process where out of $N$ outgoing objects $\widetilde{N}$ number of objects are massive and rest are massless radiation including gravitational wave. For this process in \cite{2105.08739}, it has been shown that the late time gravitational memory at order $u^{-1}$ and $u^{-2}\ln u$ can be fully expressed in terms of the momenta of incoming massive objects, incoming massless radiation and outgoing massive objects only. So the information about outgoing massless radiation is not needed to compute the late time gravitational memory at order $u^{-1}$ and $u^{-2}\ln u$. In this subsection, we show that the late time gravitational memory at order $u^{-2}$, given in eq.\eqref{future_memory} can also be rewritten in such a way that it would not carry any information of outgoing massless radiation.

We denote the momenta of outgoing massive objects by $\widetilde{p}_a$, the spins of outgoing massive objects by $\widetilde{\Sigma}_a$ and the coordinates of the intersection points with the boundary of region $\mathcal{R}$ of outgoing massive objects by $\widetilde{r}_a$ for $a=1,\cdots , \widetilde{N}$. Then after some manipulation using conservation of asymptotic momenta and asymptotic angular momenta the expression in eq.\eqref{future_memory} can be rewritten as:
\begingroup
\allowdisplaybreaks
\be
&&\Delta_{(G^2)}^{(1/u^2)}\ e^{\mu\nu}(u,\vec{x}=R \hat{n})\non\\
&=&\ -\f{G^2}{R} \ \f{1}{u^2}\Bigg[4\sum_{b=1}^{\widetilde{N}}\widetilde{p}_b.\mathbf{n}\Bigg\lbrace \sum_{a=1}^{\widetilde{N}}\f{\widetilde{p}_{a}^{(\mu}\mathbf{n}_{\rho}}{\widetilde{p}_{a}.\mathbf{n}}\Big(\widetilde{r}_{a}^{\rho}\widetilde{p}_{a}^{\nu)}-\widetilde{r}_{a}^{\nu)}\widetilde{p}_{a}^{\rho}+\widetilde{\Sigma}_{a}^{\rho\nu)}\Big)\Bigg\rbrace\non\\
&& \ -4(P'\cdot \mathbf{n})\Bigg\lbrace\sum_{a=1}^{M}\f{p^{\prime (\mu}_a \mathbf{n}_{\rho}}{p'_{a}.\mathbf{n}}\Big(r_{a}^{\prime\rho}p_{a}^{\prime\nu)}-r_{a}^{\prime\nu)}p_{a}^{\prime\rho}+\Sigma_{a}^{\prime\rho\nu)}\Big)\Bigg\rbrace\non\\
&& + \sum_{a=1}^{\widetilde{N}}\sum_{\substack{b=1\\ b\neq a}}^{\widetilde{N}}\f{\widetilde{p}_a.\widetilde{p}_b}{[(\widetilde{p}_a.\widetilde{p}_b)^2 -\widetilde{p}_a^2 \widetilde{p}_b^2 ]^{3/2}}\lbrace 2(\widetilde{p}_a.\widetilde{p}_b)^2 -3\widetilde{p}_a^2 \widetilde{p}_b^2 \rbrace \f{\mathbf{n}_\rho \mathbf{n}_\sigma}{\widetilde{p}_a.\mathbf{n}} \Big\lbrace (\widetilde{p}_a^\mu \widetilde{p}_b^\rho -\widetilde{p}_a^\rho \widetilde{p}_b^\mu)(\widetilde{r}_{a}^{\sigma}\widetilde{p}_{a}^{\nu}-\widetilde{r}_{a}^{\nu}\widetilde{p}_{a}^{\sigma}+\widetilde{\Sigma}_{a}^{\sigma\nu}) \non\\
&&\ +\ (\widetilde{p}_a^\nu \widetilde{p}_b^\sigma -\widetilde{p}_a^\sigma \widetilde{p}_b^\nu)(\widetilde{r}_{a}^{\rho}\widetilde{p}_{a}^{\mu}-\widetilde{r}_{a}^{\mu}\widetilde{p}_{a}^{\rho}+\widetilde{\Sigma}_{a}^{\rho\mu})\Big\rbrace \non\\
&& -4\widetilde{P}^{(\mu}\ \mathbf{n}_\rho \sum_{a=1}^{\widetilde{N}}\Big(\widetilde{r}_{a}^{\rho}\widetilde{p}_{a}^{\nu)}-\widetilde{r}_{a}^{\nu)}\widetilde{p}_{a}^{\rho}+\widetilde{\Sigma}_{a}^{\rho\nu)}\Big)\non\\
&& +4P^{\prime (\mu}\ \mathbf{n}_\rho \sum_{a=1}^{M} \Big(r_{a}^{\prime\rho}p_{a}^{\prime\nu)}-r_{a}^{\prime\nu)}p_{a}^{\prime\rho}+\Sigma_{a}^{\prime\rho\nu)}\Big)
\Bigg] \ ,\hspace{1cm}\hbox{for $u\rightarrow +\infty$}\label{rewritten}
\ee
where $P^{\prime\mu}=\sum\limits_{a=1}^{M} p_a^{\prime\mu}$ and $\widetilde{P}^{\mu}=\sum\limits_{a=1}^{\widetilde{N}}\widetilde{p}_a^{\mu}$. The above expression does not carry any information about massless outgoing radiation. 

Now consider binary blackhole merger process, where there is only one incoming object (i.e. $M=1$) which is the bound state of two blackholes and in the final state there are one massive blackhole (i.e. $\widetilde{N}=1$) and lots of gravitational radiation. Hence for this process the third and fourth lines after the equality vanishes in eq.\eqref{rewritten}. Also the first term within the square bracket cancels the second last term as well as the second term within the square bracket cancels the last term in \eqref{rewritten}. Hence the late time gravitational memory at order $G^2 u^{-2}$ given in \eqref{rewritten} vanishes for blackhole merger process. But as discussed in the introduction, the $u^{-2}$ tail memory receives correction at order $G^3$, which has not been derived or conjectured yet. So we do not know whether order $G^3 u^{-2}$ memory also vanishes or not for blackhole merger process.

Now let us consider a gravitational scattering process where all the incoming and outgoing objects are massless particles/radiation. For this event after using conservation of asymptotic momenta and asymptotic angular momenta, the expression in eq.\eqref{future_memory} and eq.\eqref{past_memory} takes the following simple form:
\be
&&\Delta_{(G^2)}^{(1/u^2)}\ e^{\mu\nu}(u\rightarrow +\infty,\vec{x}=R \hat{n})\ = \ - \ \Delta_{(G^2)}^{(1/u^2)}\ e^{\mu\nu}(u\rightarrow -\infty,\vec{x}=R \hat{n})\non\\
&=& -\f{4G^2}{R} \f{1}{u^2}\Bigg[P^{\prime(\mu}\mathbf{n}_\rho \Big(r_{a}^{\prime\rho}p_{a}^{\prime\nu)}-r_{a}^{\prime\nu)}p_{a}^{\prime\rho}+\Sigma_{a}^{\prime\rho\nu)}\Big) -(P^\prime \cdot \mathbf{n})\sum_{a=1}^{M}\f{p^{\prime (\mu}_a \mathbf{n}_{\rho}}{p'_{a}.\mathbf{n}}\Big(r_{a}^{\prime\rho}p_{a}^{\prime\nu)}-r_{a}^{\prime\nu)}p_{a}^{\prime\rho}+\Sigma_{a}^{\prime\rho\nu)}\Big)\Bigg]\non\\
\ee
From the above expression it is clear that for massless particle scattering the order $\mathcal{O}(G^2 u^{-2})$ gravitational tail memory is completely determined in terms of the scattering data of ingoing particles only. So the result is independent of scattering angles even if the ingoing massless particles/radiation form a blackhole along with some massless gravitational radiation in the final state. In \cite{2105.08739} this feature has been established for $u^{-1}$ and $u^{-2}\ln u$ tail memories as well.

\section{Derivation of spin dependent gravitational waveform}\label{S:derivation}
We consider a classical scattering process where $M$ number of spinning macroscopic objects are coming in from asymptotic past, go through some complicated process involving fusion, splitting etc. within a finite region of spacetime $\mathcal{R}$, and finally disperse to $N$ number of objects including finite energy radiation flux as shown in Fig.\ref{scattering_setup}. For this kind of scattering event, we are interested in determining the late and early time gravitational waveform, which is also related to the radiative mode of low-frequency gravitational waveform via Fourier transformation in the time variable. We choose the region $\mathcal{R}$ to be sufficiently large so that outside this region only long-range gravitational interaction is present. Let the spacetime size of the region $\mathcal{R}$ be $L$. Now we choose the scattering centre well inside the region $\mathcal{R}$ and place the gravitational wave detector at a distance $R$ from the scattering centre along the direction $\hat{n}$. With this setup we want to determine the $\f{1}{R}$ component of the gravitational waveform with frequency $\omega$ within the range $R^{-1}<<\omega <<L^{-1}$. 

We proceed by defining the deviation of the metric from flat background and it's trace reversed component in the following way,
\be
h_{\mu\nu}(x)\ \equiv\ \f{1}{2}\big(g_{\mu\nu}(x)-\eta_{\mu\nu}\big)\hspace{1cm},\hspace{1cm}e_{\mu\nu}(x)\equiv h_{\mu\nu}(x)-\f{1}{2}\eta_{\mu\nu}\eta^{\alpha\beta}h_{\alpha\beta}(x)
\ee
In the above expression $\eta $ represents the Minkowski metric with mostly positive signature. In several literatures including \cite{1611.03493, 1801.07719, 1804.09193, 1906.08288, 1912.06413} the following relation between radiative mode of gravitational waveform and the Fourier transform of the total energy-momentum tensor has been derived in four spacetime dimensions:
\be
\widetilde{e}^{\mu\nu}(\omega, R, \hat{n})\ \simeq \f{2G}{R}\ e^{i\omega R }\ \widehat{T}^{\mu\nu}(k)\label{eT}
\ee
where under $\simeq$ sign we are neglecting the terms with higher powers of $R^{-1}$. In the relation above $\vec{x}=R\hat{n}$, $k=\omega(1,\hat{n})\equiv \omega\mathbf{n}$ and the gravitational radiation is considered to be outgoing. The expressions of $\widetilde{e}^{\mu\nu}$ and $\widehat{T}^{\mu\nu}$ are given by,
\be
\widetilde{e}^{\mu\nu}(\omega, \vec{x})&=&\ \int_{-\infty}^\infty dt\ e^{i\omega t}\ e^{\mu\nu}(t,\vec{x})\\
\widehat{T}^{\mu\nu}(k)&=&\int d^4x\ e^{-ik.x}\ T^{\mu\nu}(x)\ +\ \hbox{boundary terms at $\infty$}\label{That}
\ee
where $T^{\mu\nu}(x)$ is the total (matter+gravitational) energy-momentum tensor which appears in the RHS of linearised Einstein's equation. In the above relation the Fourier transform of energy-momentum tensor is defined inside the region $|\vec{x}|<<R$ or equivalently we may need to add appropriate boundary terms at $\infty$ to make the integral well defined\cite{1801.07719, 1804.09193, 1906.08288, 1912.06413}. 

\subsection{General setup and strategy}
Consider the incoming particles have masses $\lbrace m'_a\rbrace$, velocities $\lbrace v'_a\rbrace$, momenta $\lbrace p'_a =m'_a v'_a\rbrace$ and spins $\lbrace \Sigma'_a\rbrace$ at asymptotic past for $a=1,2,\cdots,M$ and the outgoing particles have masses $\lbrace m_a\rbrace$, velocities $\lbrace v_a\rbrace$, momenta $\lbrace p_a =m_a v_a\rbrace$ and spins $\lbrace \Sigma_a\rbrace$ at asymptotic future for $a=1,2,\cdots,N$. Let $X'_{a}(\sigma)$ denotes the trajectory of the incoming particles in the affine parameter range $-\infty<\sigma\leq 0$ for $a=1,2,\cdots,M$ and $X_{a}(\sigma)$ denotes the trajectory of the outgoing particles in the affine parameter range $0\leq \sigma<\infty$ for $a=1,2,\cdots,N$. Now to treat incoming and outgoing particles uniformly, we treat the incoming particles as some extra outgoing particles under the following identifications:
\be
m_{N+a}&=&m'_a\ ,\ v^{\mu}_{N+a}=-v'^{\mu}_a\ ,\ p^{\mu}_{N+a}=-p'^{\mu}_a\ , \ \Sigma_{N+a}^{\mu\nu}=-\Sigma^{'\mu\nu}_a\ ,\ X_{N+a}^{\mu}(\sigma)=X^{'\mu}_{a}(-\sigma)\non\\
&&\hbox{for\ $a=1,2,\cdots,M$ and  $0\leq \sigma <\infty$}
\ee
The trajectories and spins of the scattered objects satisfy the following boundary conditions,
\be
&&X_{a}^{\mu}(\sigma=0)\ =\ r_{a}^{\mu}\ ,\ \frac{dX_{a}^{\mu}(\sigma)}{d\sigma}\Bigg{|}_{\sigma\rightarrow\infty} =\ v_a^{\mu}\ ,\label{Bdycon}\\
 &&\text{and}\hspace{1cm} \Sigma_a^{\mu\nu}(\sigma)\Big{|}_{\sigma\rightarrow\infty}=\Sigma_a^{\mu\nu}\hspace{1cm} \hbox{for $a=1,2,\cdots,M+N$.}\label{spin_Bdycon}
\ee
with $r_{a}$ being the coordinate on the boundary of region $\mathcal{R}$, where the trajectory of a'th particle intersects. Now outside the region $\mathcal{R}$, the movement of the scattered objects can be well captured by the following matter energy-momentum tensor\cite{1906.08288,1712.09250,1803.02405,1709.06016,1812.06895,Tulczyjew,Dixon,0511061, 0511133, 0604099, 0804.0260, 1504.04290}\footnote{We are using the following definition of energy-momentum tensor for the world-line action $S_X$ : 
\be
T^{X\alpha\beta}(x)=2\f{\delta S_X}{\delta g_{\alpha\beta}(x)}\ ,
\ee
which differs with the canonical definition of energy-momentum tensor by a multiplicative factor of $\sqrt{-det\ g}$.},
\be
 T^{X\alpha\beta}(x)\ &=&\ \sum_{a=1}^{M+N}\int_{0}^{\infty} d\sigma\ \Bigg[ \ m_{a}\f{d X_{a}^{\alpha}(\sigma)}{d \sigma}\f{d X_{a}^{\beta}(\sigma)}{d \sigma}\ \delta^{(4)}\big(x-X_{a}(\sigma)\big)\Bigg]\non\\
 &&\ +\ \sum_{a=1}^{M+N}\sqrt{-det\ g(x)}\ \nabla_\gamma \Bigg[ \int_{0}^{\infty} d\sigma\ \f{d X_{a}^{(\alpha}(\sigma)}{d \sigma}\ \Sigma_{a}^{\beta)\gamma}(\sigma)\ \f{\delta^{(4)}\big(x-X_{a}(\sigma)\big)}{\sqrt{-det\ g(x)}}\Bigg]\non\\
 &&\ +\ \cdots
 \ee
where unspecified $``\cdots"$ terms contain two or more covariant derivatives derivatives operating on the $\sigma$ integral containing delta function and carry the information of the internal structures of the macroscopic objects in terms of gravitational multipole moments and tidal responses, which we will not affect our result to the order we are working with. The symmetrization convention used above is defined as: $A^{(\alpha}B^{\beta)}=\f{1}{2}\big(A^{\alpha}B^{\beta}+A^{\beta}B^{\alpha}\big)$. Now using the definition of covariant derivative on a rank-3 tensor and the property $\Gamma^{\gamma}_{\gamma\delta}=\f{1}{\sqrt{-det\ g}}\p_{\delta}\big(\sqrt{-det\ g}\ \big)$, the above matter energy-momentum tensor simplifies to:
\be
 T^{X\alpha\beta}(x)\ &=&\ \sum_{a=1}^{M+N}\int_{0}^{\infty} d\sigma\ \Bigg[ \ m_{a}\f{d X_{a}^{\alpha}(\sigma)}{d \sigma}\f{d X_{a}^{\beta}(\sigma)}{d \sigma}\ \delta^{(4)}\big(x-X_{a}(\sigma)\big)\non\\
 &&\ +\f{d X_{a}^{(\alpha}(\sigma)}{d \sigma}\ \Sigma_{a}^{\beta)\gamma}(\sigma)\ \p_{\gamma}\delta^{(4)}\big(x-X_{a}(\sigma)\big)\non\\
&&\ +\ \Gamma^{(\alpha}_{\gamma\delta}(X_a(\sigma)) \f{dX_a^{\delta}(\sigma)}{d\sigma}\Sigma_a^{\beta )\gamma}(\sigma)\delta^{(4)}\big(x-X_{a}(\sigma)\big)\Bigg]\label{TXspin}
 \ee
 With the above matter energy-momentum tensor, we have to solve Einstein equation to derive the background metric,
\be
&&\sqrt{-\det g}\ \Bigg(R^{\alpha\beta}-\f{1}{2}R\ g^{\alpha\beta}\Bigg)\ =\ 8\pi G\ T^{X\alpha\beta}\label{EinEq}
\ee
We also have to solve geodesic equation and the equation for the time evolution of spin as given below\cite{Papapetrou,Mathisson,Waldspin,0511133,0511061,0804.0260,2105.12454}:
\be
&&\f{d^{2}X_{a}^{\alpha}(\sigma)}{d\sigma^{2}} + \Gamma^{\alpha}_{\beta\gamma}(X_{a}(\sigma)) \f{dX_{a}^{\beta}(\sigma)}{d\sigma} \f{dX_{a}^{\gamma}(\sigma)}{d\sigma} = -\f{1}{2m_a}R^{\alpha}\ _{\nu\rho\sigma}(X_a(\sigma))\Sigma_a^{\rho\sigma}(\sigma) \f{dX_{a}^{\nu}(\sigma)}{d\sigma}
\label{geodesicEq}\\
&&\f{d\Sigma_a^{\mu\nu}(\sigma)}{d\sigma}+\Gamma^\mu_{\alpha\beta}(X_a(\sigma))\Sigma^{\alpha\nu}(\sigma)\f{dX_a^\beta(\sigma)}{d\sigma} +\Gamma^{\nu}_{\alpha\beta}(X_a(\sigma))\Sigma^{\mu\alpha}(\sigma)\f{dX_a^\beta(\sigma)}{d\sigma} = 0\label{spin_eq}
\ee
The above equations follow from Mathisson-Papapetrou equations with some correction terms, which are explicitly derived in appendix-\ref{app:Spin_Geo}. Also in appendix-\ref{app:Spin_Geo} the above equations has beed derived demanding the covariant conservation of the canonical version of the matter energy-momentum tensor in eq.\eqref{TXspin}.

Now in terms of trace reversed metric fluctuation the Einstein's equation \eqref{EinEq} takes the following form:
\be
\eta^{\rho\sigma}\p_{\rho}\p_{\sigma}e^{\alpha\beta}(x)\ &=&\ -8\pi G\ \Big(T^{X\alpha\beta}(x)\ +\ T^{h\alpha\beta}(x)\Big)\ \equiv \ -8\pi G\ T^{\alpha\beta}(x)\label{LinearEEq}
\ee
where $T^{h\alpha\beta}(x)$ is the gravitational energy-momentum tensor defined as,
\be
T^{h\alpha\beta}\ &\equiv &\ -\f{1}{8\pi G}\Bigg[\sqrt{-\det g}\ \Bigg(R^{\alpha\beta}-\f{1}{2}R\ g^{\alpha\beta}\Bigg)\  +\ \eta^{\rho\sigma}\p_{\rho}\p_{\sigma}e^{\alpha\beta}\Bigg]\label{gravEMtensor}
\ee
Now we briefly sketch the strategy of our computation:
\begin{itemize}
	\item We solve the equations \eqref{LinearEEq}, \eqref{geodesicEq} and \eqref{spin_eq} iteratively considering gravitational constant $G$ as an iterative parameter, in post-Minkowskian(PM) sense.
	\item At zeroth iterative order we set the initial value of the metric fluctuation $e^{\mu\nu}(x)=0$ and consider the scattered objects travel in asymptotic linearised trajectory $X_{a}^{\mu}(\sigma)=r_{a}^{\mu}+v_{a}^{\mu}\sigma$ with constant value of spin $\Sigma_a^{\mu\nu}(\sigma)=\Sigma_a^{\mu\nu}$.
	\item Now we want to compute the Fourier transform of matter energy-momentum tensor given in eq.\eqref{TXspin} and keep terms up to subleading order in $\omega$ expansion. We divide the integration region in eq.\eqref{That} into two parts: one inside the region $\mathcal{R}$ denoted by $|\vec{x}|\leq L$ where we only need to use the conservation of energy-momentum tensor, and another outside the region $\mathcal{R}$ denoted by $|\vec{x}|\geq L$ where we only need to use the linearized trajectory approximation of scattered objects\cite{1906.08288}.
	\item Now using eq.\eqref{eT} we can read off order $\mathcal{O}(G)$ gravitational waveform up to subleading order in small $\omega$. Here the leading contribution in small $\omega$ expansion contributes to gravitational DC memory after Fourier transformation in time variable. But the subleading terms in $\omega$ expansion which depends on spin, does not contribute to displacement kind of memory as it is analytic in $\omega\rightarrow 0$ limit.
	\item In the first iterative order we use the $\mathcal{O}(G)$ metric fluctuation as background metric and solve geodesic equation \eqref{geodesicEq} with boundary condition \eqref{Bdycon} to find the correction to linearised trajectory. At this order we also need to solve the spin evolution equation \eqref{spin_eq} with boundary condition \eqref{spin_Bdycon}.
	\item Now using this corrected trajectory, spin correction and order $\mathcal{O}(G)$ background metric we compute the first iterative correction to the matter and gravitational energy momentum tensor outside region $\mathcal{R}$\footnote{Since at this order we are only interested to extract the non-analytic terms in $\omega$ for $\omega^{-1}>>L$, Fourier transform of energy-momentum tensor inside the region $\mathcal{R}$ does not contribute to non-analytic terms in $\omega$.}. This first iterative correction to  energy-momentum tensor contributes to non-analytic terms as $\omega\rightarrow 0$ at order $\mathcal{O}(\omega^{n}\ln\omega)$ for $n\geq 0$ along with analytic terms.
	\item Using eq.\eqref{eT} with the corrected energy-momentum tensor we get the order $\mathcal{O}(G^2)$ gravitational waveform. At this order the leading non-analytic contribution in small $\omega$ expansion behaves like $\ln\omega$ and this has been evaluated in \cite{1912.06413}, which contributes to $u^{-1}$ tail memory. So in this article our main goal is to systematically evaluate the order $\omega\ln\omega$ contribution to the gravitational waveform which turns out to be spin dependent and after Fourier transform contributes to $u^{-2}$ tail memory as first time pointed out in \cite{2008.04376}.
	\item Now if we do the analysis in the next iterative order i.e. at order $\mathcal{O}(G^3)$ of gravitational waveform, we see that it corrects the order $\omega\ln\omega$ gravitational waveform\cite{2008.04376}. Possibly there might be corrections also from order $\mathcal{O}(G^n)$  for $n\geq 4$ to order $\omega\ln\omega$ gravitational waveform, which seems to be absent by naive dimensional analysis(e.g. see table in Fig.\ref{table}). Hence in our analysis we are only able to derive the order $\omega\ln\omega$ contribution of gravitational waveform at order $\mathcal{O}(G^2)$. So our result should be thought of as 2PM contribution of order $\omega\ln\omega$ gravitational waveform, which is also equivalent to 2PM contribution of order $u^{-2}$ gravitational tail memory.
	\end{itemize}
\subsection{Order $\mathcal{O}(G)$ gravitational waveform}
In this subsection we review the analysis of \cite{1906.08288} in four spacetime dimensions\footnote{ There are some sign differences relative to \cite{1906.08288} due to the fact that we are considering the velocities of outgoing particles to be positive, where as there the velocities of outgoing particles were considered to be negative.}, which is necessary for our analysis of next subsection.  Following the strategy, let $\Delta_{(0)}\widehat{T}_{<}^{X}$ denotes the contribution of Fourier transform of eq.\eqref{TXspin} inside region $\mathcal{R}$ and $\Delta_{(0)}\widehat{T}_{>}^{X}$ denotes the contribution of Fourier transform of eq.\eqref{TXspin} outside region $\mathcal{R}$ at order $\mathcal{O}(G^0)$.

After using integration by parts the Fourier transform of eq.\eqref{TXspin} outside region $\mathcal{R}$ becomes,
\be
\widehat{T}_{>}^{X\alpha\beta}(k)
&=&\ \sum_{a=1}^{M+N}\int_{0}^{\infty}d\sigma\ e^{-ik.X_{a}(\sigma)} \Bigg[m_a \f{dX_{a}^{\alpha}(\sigma)}{d\sigma}\f{dX_{a}^{\beta}(\sigma)}{d\sigma}+i\f{dX_{a}^{(\alpha}(\sigma)}{d\sigma}\Sigma_{a}^{\beta)\gamma}k_{\gamma}\non\\
&&\hspace{2cm} +\ \Gamma^{(\alpha}_{\gamma\delta}(X_a(\sigma)) \f{dX_a^{\delta}(\sigma)}{d\sigma}\Sigma_a^{\beta )\gamma}(\sigma)\Bigg]\non\\
&& -\sum_{a=1}^{M+N}\int_{0}^{\infty}d\sigma\int d^4x \ \delta(|\vec{x}|-L)\delta^{(4)}(x-X_{a}(\sigma))\ e^{-ik.X_{a}(\sigma)}\non\\
&&\hspace{2cm}\times \ \f{dX_{a}^{(\alpha}(\sigma)}{d\sigma}\Sigma_{a}^{\beta)\gamma}\tilde{n}_{\gamma}\label{TXoutside}
\ee
where $\tilde{n}^{\alpha}\simeq \big(0,\f{\vec{x}}{|\vec{x}|}\big)$ is the outward unit vector on the boundary $\p\mathcal{R}$. In the last two lines we get the boundary contribution using $\int_{|x|\geq L} d^4x\ \p_{\gamma}A^{\alpha\beta\gamma}(x)=-\int d^4x\ \delta(|\vec{x}|-L)\tilde{n}_{\gamma}A^{\alpha\beta\gamma}(x)$. Now to evaluate the boundary contribution(last two lines) of the above expression with trajectory $X_{a}(\sigma)=r_a+v_a\sigma$,  we first do the integration over $x$ using the delta function $\delta^{(4)}(x-X_{a}(\sigma))$, which substitutes $x=r_a+v_a\sigma$ everywhere inside the integrand. Now we do the $\sigma$ integral using the delta function $\delta(|\vec{r}_a+\vec{v}_a\sigma|-L)$, which contributes to $1/(v_a.\tilde{n}_a)$ where $\tilde{n}_{a}^{\alpha}=\big(0,\f{\vec{r}_a}{|\vec{r}_a|}\big)$. There is no modulus sign in the contribution from delta function as for outgoing particles $v_a.\tilde{n}_a$ is positive. Hence from eq.\eqref{TXoutside} we get\footnote{Here we ignore the term containing Christoffel connection inside the square bracket of eq.\eqref{TXoutside}, as it will not contribute at order $\mathcal{O}(G^0)$. },
\begingroup
\allowdisplaybreaks
\be
\Delta_{(0)}\widehat{T}_{>}^{X\alpha\beta}(k)&=&\ \sum_{a=1}^{M+N} e^{-ik.r_a}\Bigg[\f{m_av_a^{\alpha}v_a^{\beta}}{i(v_a.k-i\epsilon)}+\f{v_a^{(\alpha}\Sigma_a^{\beta)\gamma}k_\gamma}{v_a.k-i\epsilon}\Bigg]\non\\
&&\ -\ \sum_{a=1}^{M+N} e^{-ik.r_a}\ \f{1}{\tilde{n}_a.v_a}\ v_a^{(\alpha}\Sigma_a^{\beta)\gamma}\tilde{n}_{a\gamma}\non\\
&=&\ (-i)\sum_{a=1}^{M+N}\Bigg[\f{p_{a}^{\alpha}p_{a}^{\beta}}{p_{a}.k-i\epsilon}+i\f{p_{a}^{(\alpha}\Sigma_{a}^{\beta)\gamma}k_{\gamma}}{p_{a}.k-i\epsilon}-ik.r_{a}\ \f{p_{a}^{\alpha}p_{a}^{\beta}}{p_{a}.k-i\epsilon}\Bigg]\non\\
&&\ -\sum_{a=1}^{M+N} \f{1}{\tilde{n}_a.p_a}\ p_a^{(\alpha}\Sigma_a^{\beta)\gamma}\tilde{n}_{a\gamma} \ +\ \mathcal{O}(\omega)\label{TXout}
\ee
\endgroup
where in the last two lines we only kept the terms up to subleading order in $\omega$ expansion and used $p_{a}=m_a v_a$. The $i\epsilon$ prescription is fixed demanding the finiteness of $\widehat{T}^{X}$ from the $\infty$ range of $\sigma$ integration.

Now to derive the Fourier transform of matter energy-momentum tensor inside the region $\mathcal{R}$, let us start with:
\be
-ik_{\alpha}\widehat{T}_{<}^{X\alpha\beta}(k)\ &=&\ \int_{|x|\leq L}d^4x\ \f{\p}{\p x^{\alpha}}\Big[e^{-ik.x}\Big]\ T_{in}^{X\alpha\beta}(x)\non\\
&=&\ \int d^4x\ \delta(|\vec{x}|-L)\ e^{-ik.x}\ \tilde{n}_{\alpha}T_{in}^{X\alpha\beta}(x)
\ee
where $T_{in}^{X\alpha\beta}(x)$ represent the matter energy-momentum tensor inside region $\mathcal{R}$ which is not same as eq.\eqref{TXspin} in general. To get the second line from the first line above, we use integration by parts and consider the conservation law of matter energy-momentum tensor $\p_{\alpha}T_{in}^{X\alpha\beta}=0$ at linearized order. Since the full contribution is written as just a boundary term on $\p\mathcal{R}$, $T_{in}^{X}$ will match with the expression in eq.\eqref{TXspin} at the boundary. So in the last line above we substitute the expression \eqref{TXspin} in place of $T_{in}^{X}$ and then do integration by parts to remove the derivative over the delta function. Finally after performing the $x$ integration using the delta function, we get\footnote{Again here we are ignoring the term containing Christoffel connection as it will not contribute at order $\mathcal{O}(G^0)$ in the analysis of $\widehat{T}^{X\alpha\beta}(k)$.}
\be
-ik_{\alpha}\widehat{T}_{<}^{X\alpha\beta}(k)\ &=&\ \sum_{a=1}^{M+N}\int_{0}^{\infty}d\sigma\ e^{-ik.X_{a}(\sigma)}\ \tilde{n}_\alpha\Bigg[\delta\big(|\vec{X}_{a}(\sigma)|-L\big)m_a\f{dX_a^\alpha}{d\sigma}\f{dX_a^{\beta}}{d\sigma}\non\\
&&\ -\tilde{n}_{\gamma}\delta'\big(|\vec{X}_a(\sigma)|-L\big)\f{dX_{a}^{(\alpha}}{d\sigma}\Sigma_{a}^{\beta)\gamma}+ik_{\gamma}\delta\big(|\vec{X}_a(\sigma)|-L\big)\f{dX_{a}^{(\alpha}}{d\sigma}\Sigma_{a}^{\beta)\gamma}\Bigg]\non\\
\ee
Now we want to evaluate the above expression using asymptotic linearized trajectory $X_{a}(\sigma)=r_a+v_a\sigma$. To do that we have to use the following two identities of delta function:
\be
\delta\big(|\vec{r}_a+\vec{v}_a\sigma|-L\big)=\f{1}{\tilde{n}_a.v_a}\delta(\sigma)\ ,\ \delta'\big(|\vec{r}_a+\vec{v}_a\sigma|-L\big)=\f{1}{(\tilde{n}_a.v_a)^2}\delta'(\sigma)
\ee
Hence using these properties and performing the $\sigma$ integration we get,
\be
&&-ik_{\alpha}\Delta_{(0)}\widehat{T}_{<}^{X\alpha\beta}(k)\non\\
 &=&\ \sum_{a=1}^{M+N}e^{-ik.r_a}\Bigg[p_{a}^{\beta}-\f{i}{2}\f{k.v_a}{\tilde{n}_a.v_a}\Sigma_{a}^{\beta\gamma}\tilde{n}_{a\gamma}+\f{i}{2}\Sigma_{a}^{\beta\gamma}k_{\gamma}\ +\f{i}{2}\f{v_{a}^{\beta}}{\tilde{n}_{a}.v_a}\ \tilde{n}_{a\alpha}\Sigma_{a}^{\alpha\gamma}k_{\gamma}\Bigg]\non\\
&=&\ \sum_{a=1}^{M+N}\Bigg[-ik.r_a\ p_{a}^{\beta}-\f{i}{2}\f{k.p_a}{\tilde{n}_a.p_a}\Sigma_{a}^{\beta\gamma}\tilde{n}_{a\gamma}+\f{i}{2}\Sigma_{a}^{\beta\gamma}k_{\gamma}\ +\f{i}{2}\f{p_{a}^{\beta}}{\tilde{n}_{a}.p_a}\ \tilde{n}_{a\alpha}\Sigma_{a}^{\alpha\gamma}k_{\gamma}\Bigg]\ +\mathcal{O}(\omega^2)
\ee
Now from the above expression after stripping out the $k_{\alpha}$ and using total angular momentum conservation relation $\sum\limits_{a}\Big(\Sigma_{a}^{\alpha\beta}+r_a^{\alpha}p_a^{\beta}-r_{a}^{\beta}p_a^{\alpha}\Big)=0$ we get\footnote{The stripping out of $k_{\alpha}$ is unique up to this order as illustrated in \cite{1906.08288}. },
\be
\Delta_{(0)}\widehat{T}_{<}^{X\alpha\beta}(k)&=&\sum_{a=1}^{M+N}\Bigg[r_a^{(\alpha}p_{a}^{\beta)}+\f{p_a^{(\alpha}}{\tilde{n}_a.p_a}\Sigma_a^{\beta)\gamma}\tilde{n}_{a\gamma}\Bigg]\ +\ \mathcal{O}(\omega)\label{TXin}
\ee
Summing over the contributions of eq.\eqref{TXout} and eq.\eqref{TXin} we find the following order $\mathcal{O}(G^{0})$ Fourier transformed energy-momentum tensor:
 \be
 \Delta_{(0)}\widehat{T}^{X\alpha\beta}(k)\ &=&\ (-i)\sum_{a=1}^{M+N}\f{1}{(p_{a}.k-i\epsilon)}\ \Big[ p_{a}^{\alpha}p_{a}^{\beta}\ +\ ip_{a}^{(\alpha}J_{a}^{\beta)\gamma}k_{\gamma}\Big]\ +\ \mathcal{O}(\omega)\label{zerothT}
\ee 
where $J_{a}^{\beta\gamma}=r_{a}^{\beta}p_{a}^{\gamma}-r_{a}^{\gamma}p_{a}^{\beta}+\Sigma_{a}^{\beta\gamma}$ is the total classical angular momentum tensor of particle-a. Here we can see that neither $r_{a}^{\mu}$ nor $\Sigma_{a}^{\mu\nu}$ are unambiguously defined as there exists transformation $r_{a}^{\mu}\rightarrow r_{a}^{\mu}+c_{a}^{
\mu}$ and $\Sigma_{a}^{\mu\nu}\rightarrow \Sigma_{a}^{\mu\nu}-c_{a}^{\mu}p_{a}^{\nu}+c_{a}^{\nu}p_{a}^{\mu}$, under which total angular momentum $J_{a}^{\mu\nu}$ remains invariant. So to give an unambiguous covariant definition of spin angular momentum we are using $p_{a\mu}\Sigma_{a}^{\mu\nu}=0$ from the beginning, which is known as Tulczyjew-Dixon spin supplementary condition\cite{Fokker, Tulczyjew}.

Now using the relation \eqref{eT} we get the radiative mode of gravitational waveform at order $\mathcal{O}(G)$ up to subleading order in $\omega$ expansion,
\be
\Delta_{(0)}\widetilde{e}^{\mu\nu}(\omega ,R, \hat{n})\ \simeq -i\f{2G}{R}\ e^{i\omega R}\sum_{a=1}^{M+N}\f{1}{(p_{a}.k-i\epsilon)}\ \Big[ p_{a}^{\mu}p_{a}^{\nu}\ +\ ip_{a}^{(\mu}J_{a}^{\nu)\rho}k_{\rho}\Big]\ 
\ee
Now performing Fourier transform in time variable we get,
\be
\Delta_{(0)}e^{\mu\nu}(u,R,\hat{n})\ &\simeq &\ -\f{2G}{R}\sum_{a=1}^{M+N}\Bigg[\f{p_a^{\mu}p_a^{\nu}}{p_a\cdot\mathbf{n}}\ H(u)\ -\ \f{p_{a}^{(\mu}J_{a}^{\nu)\rho}\mathbf{n}_{\rho}}{p_a\cdot\mathbf{n}}\ \delta(u)\Bigg]
\ee
where $u=t-R$ is the retarded time and $H(u)$ is the Heaviside theta function. Above the first term within the square bracket contributes to gravitational DC memory\cite{mem1,mem2,mem3,mem4,christodoulou,thorne,bondi,1411.5745,1712.01204,1502.06120,1806.01872,1912.06413}. On the other hand the second term which depends on spin angular momenta does not contribute to displacement gravitational memory as it is localized near $u=0$\footnote{But this angular momenta dependent term can contribute to a different kind of `spin memory' induced by radiative angular momentum flux\cite{1502.06120,1702.03300,2007.11562}. This order $\mathcal{O}(\omega^{0})$ term in the gravitational waveform receives corrections from order $\mathcal{O}(G^{n})$ for $n\geq 2$ which turns out to be theory dependent and depends on the structures of scattered objects. So the full (all PM order) contribution of order $\mathcal{O}(\omega^{0})$ gravitational waveform is non-universal\cite{1912.06413,2008.04376}.}. Hence to derive the leading spin dependent memory we need to go to next iterative order and hope to get a spin dependent non-analytic contribution in small $\omega$ from the Fourier transform of energy-momentum tensor.

\subsection{Order $\mathcal{O}(G^2)$ gravitational waveform}\label{S:order2waveform}
Due to the asymptotic linearized trajectory of scattered object-b, the metric fluctuation can be read off from the solution of linearized Einstein equation \eqref{EinEq} with the momentum space energy-momentum tensor \eqref{zerothT},
\be
 h^{(b)}_{\alpha\beta}(x)
&=& -(8\pi G)\ \int \f{d^{4}\ell}{(2\pi)^{4}}\ e^{i\ell\cdot x}\ G_{r}(\ell)\ \f{1}{i(p_{b}.\ell-i\epsilon)}\ \Big[ p_{b\alpha}p_{b\beta}-\f{1}{2}p_{b}^{2}\eta_{\alpha\beta}\non\\
&& +\ ip_{b(\alpha}J_{b,\beta)\gamma}\ell^{\gamma}-\f{i}{2}p_{b}^{\delta}J_{b,\delta\gamma}\ell^{\gamma}\eta_{\alpha\beta}\Big]
\label{metric}
\ee
where $G_r(\ell)$ is the momentum space retarded Greens function given as $G_r(\ell)=\lbrace (\ell^{0}+i\epsilon)^2 -\vec{\ell}^{2}\rbrace^{-1}$. We should denote the above metric fluctuation as $\Delta_{(0)}h^{(b)}_{\alpha\beta}$ following our convention, but just to reduce notational complexity we remove the $\Delta_{(0)}$ piece.
The Christoffel connection for the metric above is given by,
\begingroup
\allowdisplaybreaks
\be
&&\Gamma^{(b)\mu}_{\nu\rho}(x)\non\\ 
&=&\ -\ (8\pi G)\int \f{d^{4}\ell}{(2\pi)^{4}}\ e^{i\ell.x}\ G_{r}(\ell)\ \f{1}{p_{b}.\ell-i\epsilon}\ \Big[ \ell_{\nu}\Big(p_{b\rho}p_{b}^{\mu}-\f{1}{2}\delta^{\mu}_{\rho}p_{b}^{2}\Big)\ +\ \ell_{\rho}\Big(p_{b\nu}p_{b}^{\mu}-\f{1}{2}\delta^{\mu}_{\nu}p_{b}^{2}\Big)\non\\
&&\ -\ \ell^{\mu}\Big(p_{b\nu}p_{b\rho}-\f{1}{2}p_{b}^{2}\eta_{\nu\rho}\Big)\Big]\non\\
&& -i(8\pi G)\int \f{d^{4}\ell}{(2\pi)^{4}}\ e^{i\ell.x}\ G_{r}(\ell)\ \f{1}{p_{b}.\ell-i\epsilon}\ \Big[\ell_{\nu}p_{b(\rho}J_{b}^{\mu)\alpha}\ell_{\alpha}+\ell_{\rho}p_{b(\nu}J_{b}^{\mu)\alpha}\ell_{\alpha}-\ell^{\mu}p_{b(\nu}J_{b,\rho)\alpha}\ell^{\alpha}\non\\
&&\ -\f{1}{2}\Big(\ell_{\nu}\delta_{\rho}^{\mu}+\ell_{\rho}\delta_{\nu}^{\mu}-\ell^{\mu}\eta_{\nu\rho}\Big)p_{b}^{\delta}J_{b,\delta\gamma}\ell^{\gamma}\Big]\label{Christoffel}
\ee
\endgroup
Now due to the non-flat background metric, consider the leading correction to the asymptotic straight-line trajectory of particle-a outside the region $\mathcal{R}$ to be,
\be
X_{a}^{\mu}(\sigma)\ &=&\ r_{a}^{\mu}+v_{a}^{\mu}\sigma\ +Y_{a}^{\mu}(\sigma)\label{Trajectory}
\ee
with boundary conditions:
\be 
Y_{a}^{\mu}(\sigma)\Bigg{|}_{\sigma=0}=0\hspace{1cm}   \ ,\hspace{1cm}\hbox{and} \hspace{1cm}\f{d Y_{a}^{\mu}(\sigma)}{d \sigma}\Bigg{|}_{\sigma\rightarrow\infty}=0\label{BoundaryC}
\ee
Similarly due to the non-flat background metric, there will also be a correction of spin outside the region $\mathcal{R}$, which is given by,
\be
\Sigma_a^{\mu\nu}(\sigma)=\Sigma_a^{\mu\nu}\ +\ S_a^{\mu\nu}(\sigma)
\ee
with boundary condition:
\be
S_a^{\mu\nu}(\sigma)=0\ ,\ \hbox{for}\ |\sigma|\rightarrow \infty\label{spinBC1}
\ee
 Now the order $\mathcal{O}(G)$ contribution of $Y_{a}^{\mu}(\sigma)$ and $S_a^{\mu\nu}(\sigma)$ satisfies the equations below, which follows from eq.\eqref{geodesicEq} and eq.\eqref{spin_eq}.
\be
\f{d^{2}Y_{a}^{\mu}(\sigma)}{d \sigma^{2}}\ &=&\ -\Gamma^{\mu}_{\nu\rho}(r_{a}+v_{a}\sigma)v_{a}^{\nu}v_{a}^{\rho}-\f{1}{2m_a}\Big(\p_\rho \Gamma^{\mu}_{\nu\sigma}(r_a+v_a\sigma)-\p_\sigma \Gamma^{\mu}_{\nu\rho}(r_a+v_a\sigma)\Big)\Sigma_a^{\rho\sigma}v_a^\nu \non\\
\f{dS_a^{\mu\nu}(\sigma)}{d\sigma}&=& -\Gamma^{\mu}_{\sigma\rho}(r_a+v_a\sigma)\Sigma_a^{\rho\nu}v_a^\sigma +\Gamma^{\nu}_{\sigma\rho}(r_a+v_a\sigma)\Sigma_a^{\rho\mu}v_a^\sigma \label{geodesic_step}
\ee
where $\Gamma^{\mu}_{\nu\rho}(r_{a}+v_{a}\sigma)=\sum\limits_{\substack{b=1\\ b\neq a}}^{M+N}\Gamma^{(b)\mu}_{\nu\rho}(r_{a}+v_{a}\sigma)$ .
After doing integration with the specified boundary conditions of eq.\eqref{BoundaryC} and eq.\eqref{spinBC1}, we get
\be
\f{d Y_{a}^{\mu}(\sigma)}{d \sigma}\ &=&\ \int_{\sigma}^{\infty}d\sigma'\ \Big[\Gamma^{\mu}_{\nu\rho}(r_{a}+v_{a}\sigma')v_{a}^{\nu}v_{a}^{\rho}+\f{1}{2m_a}\Big(\p_\rho \Gamma^{\mu}_{\nu\sigma}(r_a+v_a\sigma')-\p_\sigma \Gamma^{\mu}_{\nu\rho}(r_a+v_a\sigma')\Big)\Sigma_a^{\rho\sigma}v_a^\nu\Big]\non\\ \label{dY}\\
S_a^{\mu\nu}(\sigma)&=&\ \int_{\sigma}^{\infty}d\sigma'\Big[\Gamma^{\mu}_{\sigma\rho}(r_a+v_a\sigma')\Sigma_a^{\rho\nu}v_a^\sigma -\Gamma^{\nu}_{\sigma\rho}(r_a+v_a\sigma')\Sigma_a^{\rho\mu}v_a^\sigma\Big]\label{Sa}
\ee
\subsubsection{Analysis of matter energy-momentum tensor}
In this section we compute the order $\mathcal{O}(G)$ correction of the Fourier transformation of matter energy-momentum tensor due to the corrected trajectory and spin of the particle. Since only the non-analytic terms in $\omega$ contributes to gravitational memory, it is sufficient to analyze the Fourier transformation of matter energy-momentum tensor outside region $\mathcal{R}$ without the boundary contribution as given in eq.\eqref{TXoutside}. Contributions from the boundary terms and inside the region $\mathcal{R}$ always reproduce terms, which are analytic  in $\omega$. Hence neglecting the last two lines of eq.\eqref{TXoutside} and substituting the corrected trajectory and spin we get,
\be
\widehat{T}_{>}^{X\mu\nu}(k)&\simeq & \sum_{a=1}^{M+N}\int _{0}^{\infty}d\sigma e^{-ik.(v_{a}\sigma+r_{a})}\ \Big{\lbrace}1-ik.Y_{a}(\sigma)+\cdots\Big{\rbrace}\Bigg[m_{a}\Big(v_{a}^{\mu}v_{a}^{\nu}+v_{a}^{\mu}\f{dY_{a}^{\nu}(\sigma)}{d\sigma}\non\\
&&\ +v_{a}^{\nu}\f{dY_{a}^{\mu}(\sigma)}{d\sigma}\Big)+i\Big(v_{a}^{(\mu}\Sigma_{a}^{\nu)\alpha}+v_{a}^{(\mu}S_{a}^{\nu)\alpha}(\sigma)+\f{dY_{a}^{(\mu}(\sigma)}{d\sigma}\Sigma_{a}^{\nu)\alpha}\Big)k_{\alpha}\non\\
&&\ +\ \Gamma^{(\mu}_{\alpha\beta}(r_a+v_a\sigma)\Sigma_a^{\nu)\alpha}v_a^\beta\ \Bigg]
\ee
The order $\mathcal{O}(G)$ correction to the matter energy momentum tensor,
\begingroup
\allowdisplaybreaks
\be
&&\Delta_{(1)}\widehat{T}^{X\mu\nu}(k)\non\\
 &=&\ \sum_{a=1}^{M+N}\int _{0}^{\infty}d\sigma e^{-ik.(v_{a}\sigma+ r_{a})}\ \Bigg[-ik.Y_{a}(\sigma)\ m_{a}v_{a}^{\mu}v_{a}^{\nu}\ +m_{a}v_{a}^{\mu}\f{dY_{a}^{\nu}(\sigma)}{d\sigma}\non\\
&&\ +m_{a}v_{a}^{\nu}\f{dY_{a}^{\mu}(\sigma)}{d\sigma}\ +i\f{dY_{a}^{(\mu}(\sigma)}{d\sigma}\Sigma_{a}^{\nu)\alpha}k_{\alpha} + k.Y_{a}(\sigma)\ v_{a}^{(\mu}\Sigma_{a}^{\nu)\alpha}k_{\alpha}\Bigg]\non\\
&&+\sum_{a=1}^{M+N}\int_0^{\infty}d\sigma e^{-ik.(r_a +v_a\sigma)} \Big[iv_a^{(\mu}S_a^{\nu)\alpha}(\sigma)k_\alpha +\Gamma^{(\mu}_{\alpha\beta}(r_a+v_a\sigma)\Sigma_a^{\nu)\alpha}v_a^{\beta}\Big]\label{TX1_inter}
\ee
\endgroup
Now performing integration by parts and using the boundary conditions of eq.\eqref{BoundaryC} we get,
\be
\Delta_{(1)}\widehat{T}^{X\mu\nu}(k)\ &=& \sum_{a=1}^{M+N} e^{-ik.r_{a}}\Bigg[-\f{p_{a}^{\mu}p_{a}^{\nu}}{p_{a}.k}k_{\beta}+p_{a}^{\mu}\delta_{\beta}^{\nu}+p_{a}^{\nu}\delta_{\beta}^{\mu}+i\delta_{\beta}^{(\mu}\Sigma_{a}^{\nu)\alpha}k_{\alpha}-i\f{k_{\beta}}{k.p_{a}}p_{a}^{(\mu}\Sigma_{a}^{\nu)\alpha}k_{\alpha}\Bigg]\non\\
&&\ \times \ \int_{0}^{\infty}d\sigma\ e^{-ik.v_{a}\sigma}\ \f{dY_{a}^{\beta}(\sigma)}{d\sigma}\non\\
&&+\sum_{a=1}^{M+N}\int_0^{\infty}d\sigma e^{-ik.(r_a +v_a\sigma)} \Big[iv_a^{(\mu}S_a^{\nu)\alpha}(\sigma)k_\alpha +\Gamma^{(\mu}_{\alpha\beta}(r_a+v_a\sigma)\Sigma_a^{\nu)\alpha}v_a^{\beta}\Big]\label{DeltaTX}
\ee

We evaluate the above integrations using the results of eq.\eqref{dY}, eq.\eqref{Sa} and eq.\eqref{Christoffel} and get,
\begingroup
\allowdisplaybreaks
\be
I^{\beta}&\equiv & \int_{0}^{\infty}d\sigma\ e^{-ik.v_{a}\sigma}\ \f{dY_{a}^{\beta}(\sigma)}{d\sigma}\non\\
&=& (8\pi G)\sum_{\substack{b=1\\b\neq a}}^{M+N}\int \f{d^{4}\ell}{(2\pi)^{4}}\ e^{i\ell.r_{a}}\ G_{r}(\ell)\f{1}{p_{b}.\ell -i\epsilon}\ \f{1}{p_{a}.\ell+i\epsilon}\ \f{1}{p_{a}.(\ell-k)+i\epsilon}\non\\
&&\times \Big[2p_{a}.\ell(p_{a}.p_{b}p_{b}^{\beta}-\f{1}{2}p_{b}^{2}p_{a}^{\beta})-\ell^{\beta}\lbrace (p_{a}.p_{b})^{2}-\f{1}{2}p_{a}^{2}p_{b}^{2}\rbrace\Big]\non\\
&&\ +i(8\pi G)\ \sum_{\substack{b=1\\b\neq a}}^{M+N}\int \f{d^{4}\ell}{(2\pi)^{4}}\ e^{i\ell.r_{a}}\ G_{r}(\ell)\f{1}{p_{b}.\ell -i\epsilon}\ \f{1}{p_{a}.\ell+i\epsilon}\ \f{1}{p_{a}.(\ell-k)+i\epsilon}\non\\
&&\times \Big[p_{a}.\ell p_{a}.p_{b}J_{b}^{\beta\alpha}\ell_{\alpha}+p_{a}.\ell p_{b}^{\beta}p_{a\rho}J_{b}^{\rho\alpha}\ell_{\alpha}-\ell^{\beta}p_{a}.p_{b}p_{a\rho}J_{b}^{\rho\alpha}\ell_{\alpha}-\f{1}{2}\big{\lbrace}2p_{a}.\ell p_{a}^{\beta}-\ell^{\beta}p_{a}^2\big{\rbrace}\non\\
&&\ \times p_{b}^{\delta}J_{b,\delta\gamma}\ell^{\gamma}\Big]\non\\
&&+i(8\pi G)\sum_{\substack{b=1\\ b\neq a}}^{M+N}\int \f{d^4\ell}{(2\pi)^4} e^{i\ell.r_a}G_{r}(\ell)\ \f{1}{p_b.\ell -i\epsilon}\ \f{1}{p_a.\ell +i\epsilon}\ \f{1}{p_a.(\ell -k)+i\epsilon}\non\\
&&\ \Big[ p_a.\ell\ \ell_\rho \Sigma_a^{\rho\sigma}p_{b\sigma}p_{b}^{\beta}-\f{1}{2}p_b^2 p_a.\ell \ell_{\rho}\Sigma_a^{\rho\beta}-\ell^{\beta}p_a.p_b \ell_{\rho}\Sigma_a^{\rho\sigma}p_{b\sigma}\Big]\label{loopint}
\ee
\endgroup
The leading non-analytic contribution in small $\omega$ expansion from eq.\eqref{DeltaTX} is of the order of $\ln\omega$ which is determined in \cite{1912.06413}. Our goal in this sub-section is to extract the $\mathcal{O}(\omega\ln\omega)$ contribution from eq.\eqref{DeltaTX} in the integration region $\omega<<|\ell^{\mu}|<<L^{-1}$. In turn this demands that, we have to extract the order $\mathcal{O}(\ln\omega)$ and order $\mathcal{O}(\omega\ln\omega)$ contributions from the integral expression in eq.\eqref{loopint} which we denote by $I_{1}^{\beta}$ and $I_{2}^{\beta}$ respectively.
\be
I_{1}^{\beta}\ &=&\ (8\pi G)\sum_{\substack{b=1\\b\neq a}}^{M+N}\int_{\omega}^{L^{-1}} \f{d^{4}\ell}{(2\pi)^{4}}\ G_{r}(\ell)\f{1}{p_{b}.\ell -i\epsilon}\ \f{1}{(p_{a}.\ell+i\epsilon)^{2}}\non\\
&&\times \Big[2p_{a}.\ell(p_{a}.p_{b}p_{b}^{\beta}-\f{1}{2}p_{b}^{2}p_{a}^{\beta})-\ell^{\beta}\lbrace (p_{a}.p_{b})^{2}-\f{1}{2}p_{a}^{2}p_{b}^{2}\rbrace\Big]\non\\
&=&\ i\ \f{\p}{\p p_{a\beta}}K_{gr}^{cl}
\ee
\begingroup
\allowdisplaybreaks
\be
I_{2}^{\beta}\ &=&\ i(8\pi G)\ (p_{a}.k)\sum_{\substack{b=1\\b\neq a}}^{M+N}\int_{\omega}^{L^{-1}} \f{d^{4}\ell}{(2\pi)^{4}}\ \ell.r_{a}\ G_{r}(\ell)\f{1}{p_{b}.\ell -i\epsilon}\ \f{1}{(p_{a}.\ell+i\epsilon)^{3}}\non\\
&&\times \Big[2p_{a}.\ell(p_{a}.p_{b}p_{b}^{\beta}-\f{1}{2}p_{b}^{2}p_{a}^{\beta})-\ell^{\beta}\lbrace (p_{a}.p_{b})^{2}-\f{1}{2}p_{a}^{2}p_{b}^{2}\rbrace\Big]\non\\
&&\ + i(8\pi G)\ (p_{a}.k)\sum_{\substack{b=1\\b\neq a}}^{M+N}\int_{\omega}^{L^{-1}} \f{d^{4}\ell}{(2\pi)^{4}}\ G_{r}(\ell)\f{1}{p_{b}.\ell -i\epsilon}\ \f{1}{(p_{a}.\ell+i\epsilon)^{3}}\non\\
&&\times \Big[p_{a}.\ell p_{a}.p_{b}J_{b}^{\beta\alpha}\ell_{\alpha}+p_{a}.\ell p_{b}^{\beta}p_{a\rho}J_{b}^{\rho\alpha}\ell_{\alpha}-\ell^{\beta}p_{a}.p_{b}p_{a\rho}J_{b}^{\rho\alpha}\ell_{\alpha}-\f{1}{2}\big{\lbrace}2p_{a}.\ell p_{a}^{\beta}-\ell^{\beta}p_{a}^2\big{\rbrace}\ p_{b}^{\delta}J_{b,\delta\gamma}\ell^{\gamma}\Big]\non\\
&&+i(8\pi G)\sum_{\substack{b=1\\ b\neq a}}^{M+N}\int_{\omega}^{L^{-1}} \f{d^4\ell}{(2\pi)^4} G_{r}(\ell)\ \f{1}{p_b.\ell -i\epsilon}\ \f{1}{(p_a.\ell +i\epsilon)^3}\ \ p_a.k\non\\
&&\times \Big[ p_a.\ell\ \ell_\rho \Sigma_a^{\rho\sigma}p_{b\sigma}p_{b}^{\beta}-\f{1}{2}p_b^2 p_a.\ell \ell_{\rho}\Sigma_a^{\rho\beta}-\ell^{\beta}p_a.p_b \ell_{\rho}\Sigma_a^{\rho\sigma}p_{b\sigma}\Big]\non\\
&=&\ iG\  (p_{a}.k)\ \ln\lbrace(\omega+i\epsilon\eta_{a})L\rbrace\sum_{\substack{b=1\\b\neq a\\ \eta_{a}\eta_{b}=1}}^{M+N}\f{1}{[(p_{a}.p_{b})^{2}-p_{a}^{2}p_{b}^{2}]^{3/2}}\ \Big[ 5(p_a.p_b)^2 p_b.r_a p_b^\beta -4p_a.p_b p_b^2 p_a.r_a p_b^\beta\non\\
&& \ -2p_b^2 p_a.p_b p_b.r_a p_a^\beta +2(p_b^2)^2 p_a.r_a p_a^\beta -(p_a.p_b)^2 p_b^2 r_a^\beta -\f{1}{2}p_a^2 p_b^2 p_b.r_a p_b^\beta +\f{1}{2}p_a^2 (p_b^2)^2 r_a^\beta \Big]\non\\
&&\ -3iG\  (p_{a}.k)\ \ln\lbrace(\omega+i\epsilon\eta_{a})L\rbrace\sum_{\substack{b=1\\b\neq a\\ \eta_{a}\eta_{b}=1}}^{M+N}\f{1}{[(p_{a}.p_{b})^{2}-p_{a}^{2}p_{b}^{2}]^{5/2}}\big(p_a.p_b p_b.r_a -p_b^2 p_a.r_a\big)\non\\
&&\times \Big[(p_a.p_b)^3 p_b^\beta -(p_a.p_b)^2 p_b^2 p_a^\beta -\f{1}{2}p_a^2 p_b^2 p_a.p_b p_b^\beta +\f{1}{2}p_a^2 (p_b^2)^2 p_a^\beta\Big]\non\\
&& + iG\  (p_{a}.k)\ \ln\lbrace(\omega+i\epsilon\eta_{a})L\rbrace\sum_{\substack{b=1\\b\neq a\\ \eta_{a}\eta_{b}=1}}^{M+N}\f{1}{[(p_{a}.p_{b})^{2}-p_{a}^{2}p_{b}^{2}]^{3/2}}\Big[ (p_a.p_b)^2 p_b^2 r_b^{\beta}-5(p_a.p_b)^2 p_b.r_b p_b^\beta \non\\
&& +4p_a.p_b p_b^2 p_a.r_bp_b^\beta +2p_a.p_b p_b.r_b p_b^2 p_a^\beta -2(p_b^2)^2 p_a.r_b p_a^\beta +\f{1}{2}p_a^2 p_b^2 p_b.r_b p_b^\beta -\f{1}{2}p_a^2 (p_b^2)^2 r_b^\beta\non\\
&& -p_a.p_b p_b^2 \Sigma_b^{\beta\alpha}p_{a\alpha}\Big]\non\\
&&  -3iG\  (p_{a}.k)\ln\lbrace(\omega+i\epsilon\eta_{a})L\rbrace\sum_{\substack{b=1\\b\neq a\\ \eta_{a}\eta_{b}=1}}^{M+N}\f{1}{[(p_{a}.p_{b})^{2}-p_{a}^{2}p_{b}^{2}]^{5/2}}\Big[(p_a.p_b)^3 p_b^2 p_a.r_b p_b^\beta +(p_a.p_b)^3 p_b^2 p_b.r_b p_a^\beta\non\\
&& -\f{1}{2}p_a^2 (p_b^2)^2 p_a.p_b p_a.r_b p_b^\beta -\f{1}{2}p_a^2 (p_b^2)^2 p_a.p_b p_b.r_b p_a^\beta -(p_a.p_b)^4 p_b.r_b p_b^\beta +\f{1}{2}(p_a.p_b)^2 p_a^2 p_b^2 p_b.r_b p_b^\beta \non\\
&& -(p_a.p_b)^2 (p_b^2)^2 p_a.r_b p_a^\beta +\f{1}{2}p_a^2 (p_b^2)^3 p_a.r_b p_a^\beta\Big]
\ee
\endgroup
where\cite{2008.04376},
\be
K_{gr}^{cl}\ &=&\ -\f{i}{2}\ (8\pi G)\sum_{\substack{b,c\\b\neq c}}\int_{\omega}^{L^{-1}}\f{d^4\ell}{(2\pi)^4}\ G_{r}(\ell)\f{1}{p_{b}.\ell+i\epsilon}\f{1}{p_{c}.\ell-i\epsilon}\ \Big{\lbrace}(p_{b}.p_{c})^2-\f{1}{2}p_{b}^2p_{c}^2\Big{\rbrace}\non\\
&=&\ -\f{i}{2}\ (2G)\ \sum_{\substack{b,c\\ b\neq c\\ \eta_{b}\eta_{c}=1}}\ \ln\Big{\lbrace}L(\omega+i\epsilon\eta_{b})\Big{\rbrace}\ \f{(p_{b}.p_{c})^{2}-\f{1}{2}p_{b}^{2}p_{c}^{2}}{\sqrt{(p_{b}.p_{c})^{2}-p_{b}^2 p_{c}^2}}\label{Kgrcl}
\ee
Above $\eta_b=+1$ if particle-b is outgoing and $\eta_b=-1$ if particle-b is ingoing. We also need to evaluate the last line of eq.\eqref{DeltaTX}, which contributes at order $\mathcal{O}(\omega\ln\omega)$ in the integration region $\omega<<|\ell^{\mu}|<<L^{-1}$,
\begingroup
\allowdisplaybreaks
\be
&&\sum_{a=1}^{M+N}\int_0^{\infty}d\sigma e^{-ik.(r_a +v_a\sigma)} \Big[iv_a^{(\mu}S_a^{\nu)\alpha}(\sigma)k_\alpha +\Gamma^{(\mu}_{\alpha\beta}(r_a+v_a\sigma)\Sigma_a^{\nu)\alpha}v_a^{\beta}\Big]\non\\
&\simeq &\ +iG \sum_{a=1}^{M+N}\sum_{\substack{b=1\\ b\neq a\\ \eta_a\eta_b =1}}^{M+N}\ln \big\lbrace (\omega +i\epsilon \eta_a)L\big\rbrace  \f{1}{[(p_a.p_b)^2 -p_a^2 p_b^2]^{3/2}}\non\\
&&\Big[ \lbrace (p_a.p_b)^2 -p_a^2 p_b^2\rbrace p_{b\rho}\Sigma_a^{\rho\alpha}k_\alpha (p_a^\mu p_b^\nu +p_a^\nu p_b^\mu )-p_b^2 \lbrace (p_a.p_b)^2 -p_a^2 p_b^2\rbrace (p_a^\mu \Sigma_a^{\nu\alpha}k_\alpha +p_a^\nu \Sigma_a^{\mu\alpha}k_\alpha)\non\\
&& -\lbrace (p_a.p_b)^2 -p_a^2 p_b^2\rbrace p_b.k (p_a^\mu p_{b\rho}\Sigma_a^{\rho\nu}+p_a^\nu p_{b\rho}\Sigma_a^{\rho\mu})+p_a.p_b p_b^2 p_{b\rho}\Sigma_a^{\rho\alpha}k_\alpha p_a^\mu p_a^\nu\non\\
&& +(p_a.k)\lbrace (p_a.p_b)^2 -p_a^2 p_b^2\rbrace \big(p_b^{\mu}p_{b\alpha}\Sigma_a^{\alpha\nu}+p_b^{\nu}p_{b\alpha}\Sigma_a^{\alpha\mu}\big)\Big]\label{spin_part}
\ee
\endgroup
Now using these results of integrations, the order $\mathcal{O}(\omega\ln\omega)$ contribution to matter energy-momentum tensor from eq.\eqref{DeltaTX} becomes,
\begingroup
\allowdisplaybreaks
\be
&&\Delta_{(1)}^{(\omega\ln\omega)}\widehat{T}^{X\mu\nu}(k)\non\\
&=&\ -\sum_{a=1}^{M+N}(k.r_{a})\Bigg[\f{p_{a}^{\mu}p_{a}^{\nu}}{p_{a}.k}k^{\alpha}\f{\p}{\p p_{a}^{\alpha}}K_{gr}^{cl}-\f{1}{2}p_{a}^{\mu}\f{\p}{\p p_{a\nu}}K_{gr}^{cl}-\f{1}{2}p_{a}^{\nu}\f{\p}{\p p_{a\mu}}K_{gr}^{cl}\Bigg]\non\\
&&\ +\f{1}{2}\sum_{a=1}^{M+N}\Bigg[\f{1}{p_{a}.k}\Big{\lbrace}p_{a}^{\mu}\Sigma_{a}^{\nu\alpha}k_{\alpha}+p_{a}^{\nu}\Sigma_{a}^{\mu\alpha}k_{\alpha}\Big{\rbrace}k^{\beta}\f{\p}{\p p_{a}^{\beta}}K_{gr}^{cl}-\Sigma_{a}^{\nu\alpha}k_{\alpha}\f{\p}{\p p_{a\mu}}K_{gr}^{cl}-\Sigma_{a}^{\mu\alpha}k_{\alpha}\f{\p}{\p p_{a\nu}}K_{gr}^{cl}\Bigg]\non\\
&&\ -\sum_{a=1}^{M+N}\Bigg[\f{p_{a}^{\mu}p_{a}^{\nu}}{p_{a}.k}k_{\beta}I_{2}^{\beta}-p_{a}^{\mu}I_{2}^{\nu}-p_{a}^{\nu}I_{2}^{\mu}\Bigg]\ +\sum_{a=1}^{M+N}(k.r_{a})\Bigg[\f{1}{2}p_{a}^{\mu}\f{\p}{\p p_{a\nu}}K_{gr}^{cl}+\f{1}{2}p_{a}^{\nu}\f{\p}{\p p_{a\mu}}K_{gr}^{cl}\Bigg]\non\\
&&\ +iG \sum_{a=1}^{M+N}\sum_{\substack{b=1\\ b\neq a\\ \eta_a\eta_b =1}}^{M+N}\ln \big\lbrace (\omega +i\epsilon \eta_a)L\big\rbrace  \f{1}{[(p_a.p_b)^2 -p_a^2 p_b^2]^{3/2}}\non\\
&&\Big[ \lbrace (p_a.p_b)^2 -p_a^2 p_b^2\rbrace p_{b\rho}\Sigma_a^{\rho\alpha}k_\alpha (p_a^\mu p_b^\nu +p_a^\nu p_b^\mu )-p_b^2 \lbrace (p_a.p_b)^2 -p_a^2 p_b^2\rbrace (p_a^\mu \Sigma_a^{\nu\alpha}k_\alpha +p_a^\nu \Sigma_a^{\mu\alpha}k_\alpha)\non\\
&& -\lbrace (p_a.p_b)^2 -p_a^2 p_b^2\rbrace p_b.k (p_a^\mu p_{b\rho}\Sigma_a^{\rho\nu}+p_a^\nu p_{b\rho}\Sigma_a^{\rho\mu})+p_a.p_b p_b^2 p_{b\rho}\Sigma_a^{\rho\alpha}k_\alpha p_a^\mu p_a^\nu\non\\
&& +(p_a.k)\lbrace (p_a.p_b)^2 -p_a^2 p_b^2\rbrace \big(p_b^{\mu}p_{b\alpha}\Sigma_a^{\alpha\nu}+p_b^{\nu}p_{b\alpha}\Sigma_a^{\alpha\mu}\big)\Big] \label{TXfinal_exp}
\ee
\endgroup

\subsubsection{Analysis of gravitational energy-momentum tensor}
Here we will compute the spin dependent gravitational energy-momentum tensor for the metric fluctuation given in eq.\eqref{metric}. Fourier transform of the order $G$ gravitational energy-momentum tensor takes the following form:
\be
\Delta_{(1)}\widehat{T}^{h\mu\nu}(k)\ &=&\ -(8\pi G)\sum_{a,b=1}^{M+N}\int \f{d^{4}\ell}{(2\pi)^{4}}\ G_{r}(k-\ell)G_{r}(\ell)\ \f{1}{p_{b}.\ell-i\epsilon}\ \f{1}{p_{a}.(k-\ell)-i\epsilon}\non\\
&&\times \Big{\lbrace} p_{b\alpha}p_{b\beta}-\f{1}{2}p_{b}^{2}\eta_{\alpha\beta}+ip_{b(\alpha}J_{b,\beta)\gamma}\ell^{\gamma}-\f{i}{2}\eta_{\alpha\beta}p_{b}^{\delta}J_{b,\delta\gamma}\ell^{\gamma}\Big{\rbrace}\ \mathcal{F}^{\mu\nu, \alpha\beta , \rho\sigma}(k,\ell)\non\\
&&\times \Big{\lbrace}p_{a\rho}p_{a\sigma}-\f{1}{2}p_{a}^{2}\eta_{\rho\sigma}+ip_{a(\rho}J_{a,\sigma)\delta}(k-\ell)^{\delta}-\f{i}{2}\eta_{\rho\sigma}p_{a}^{\kappa}J_{a,\kappa\tau}(k-\ell)^{\tau}\Big{\rbrace}\label{Thathspin}
\ee
where\cite{1912.06413,2008.04376},
\begingroup
\allowdisplaybreaks
\be \label{F-full}
&&\mathcal{F}^{\mu\nu , \alpha\beta , \rho\sigma }(k,\ell)\non\\
 &=&\ 2\Big[\f{1}{2}\ell^{\mu}(k-\ell)^{\nu}\eta^{\alpha\rho}\eta^{\beta\sigma}+(k-\ell)^{\mu}(k-\ell)^{\nu}\eta^{\alpha\rho}\eta^{\beta\sigma}-(k-\ell)^{\nu}(k-\ell)^{\beta}\eta^{\alpha\rho}\eta^{\mu\sigma}
\non\\
&& 
 -(k-\ell)^{\mu}(k-\ell)^{\beta}\eta^{\alpha\rho}\eta^{\nu\sigma}+(k-\ell)^{\alpha}(k-\ell)^{\beta}\eta^{\mu\rho}\eta^{\nu\sigma}+(k-\ell).\ell\, \eta^{\nu\beta}\eta^{\alpha\rho}\eta^{\mu\sigma}
 \non\\
&&-\ell^{\rho}(k-\ell)^{\alpha}\eta^{\nu\beta}\eta^{\mu\sigma}-\f{1}{2}(k-\ell)^{2}\eta^{\mu\alpha}\eta^{\nu\beta}\eta^{\rho\sigma}\ +\eta^{\mu\alpha}\eta^{\beta\rho}\eta^{\nu\sigma}(k-\ell)^{2}\ +\ \eta^{\nu\alpha}\eta^{\beta\rho}\eta^{\mu\sigma}(k-\ell)^{2}
\Big]\non\\
&&\ -\eta^{\mu\nu}\Big[\f{3}{2}(k-\ell).\ell\, 
\eta^{\alpha\rho}\eta^{\beta\sigma}+2(k-\ell)^{2}\eta^{\alpha\rho}\eta^{\beta\sigma}-\ell^{\sigma}(k-\ell)^{\alpha}\eta^{\beta\rho}\Big]\non\\
&&\ -\eta^{\alpha\beta}(k-\ell)^{2}\ \eta^{\mu\rho}\eta^{\nu\sigma}+\f{1}{2}\eta^{\alpha\beta}(k-\ell)^{2}\ \eta^{\rho\sigma}\eta^{\mu\nu}
\ee
\endgroup
follows from the quadratic part of gravitational energy-momentum tensor defined in eq.\eqref{gravEMtensor}. In the integration region $R^{-1}<<|\ell^{\mu}|<<\omega$, at leading order $\mathcal{F}(k,\ell)$ approximates in the following form up to gauge equivalence,
\be
\mathcal{F}^{\mu\nu, \alpha\beta , \rho\sigma}(k,\ell)\simeq -2k^{\beta}k^{\sigma}\eta^{\alpha\rho}\eta^{\mu\nu}+2k^{\alpha}k^{\beta}\eta^{\mu\rho}\eta^{\nu\sigma}
\ee
Hence in the integration region $R^{-1}<<|\ell^{\mu}|<<\omega$ the order $\mathcal{O}(\omega\ln\omega)$ contribution from eq.\eqref{Thathspin} turns out to be,
\begingroup
\allowdisplaybreaks
\be
&&\Delta_{(1)}^{(\omega\ln\omega)}\widehat{T}_{1}^{h\mu\nu}(k)\non\\
 &=&\  -(8\pi G)\ \sum_{a,b=1}^{M+N}\f{1}{p_{a}.k}\Big{\lbrace} p_{b\alpha}p_{b\beta}-\f{1}{2}p_{b}^{2}\eta_{\alpha\beta}\Big{\rbrace}\big[-2k^{\beta}k^{\sigma}\eta^{\alpha\rho}\eta^{\mu\nu}+2k^{\alpha}k^{\beta}\eta^{\mu\rho}\eta^{\nu\sigma} \big]\non\\
&&\times \Big{\lbrace}ip_{a(\rho}J_{a,\sigma)\delta}k^{\delta}-\f{i}{2}\eta_{\rho\sigma}p_{a}^{\kappa}J_{a,\kappa\tau}k^{\tau}\Big{\rbrace}\int_{R^{-1}}^{\omega}\f{d^{4}\ell}{(2\pi)^{4}}\ G_{r}(\ell)\f{1}{p_{b}.\ell -i\epsilon}\f{1}{2k.\ell+i\epsilon}\label{interm}
\ee
Now after using the result of the integration\cite{1912.06413} 
\be
&&\int_{R^{-1}}^{\omega}\f{d^{4}\ell}{(2\pi)^{4}}\ G_{r}(\ell)\f{1}{p_{b}.\ell -i\epsilon}\f{1}{k.\ell+i\epsilon}\label{interm}
=\f{1}{4\pi}\ \delta_{\eta_{b},1}\ \f{1}{p_{b}.k}\ \ln\big{\lbrace}(\omega+i\epsilon)R\big{\rbrace}
\ee
we get,
\be
&&\Delta_{(1)}^{(\omega\ln\omega)}\widehat{T}_{1}^{h\mu\nu}(k)\non\\
&=& -iG\ \ln\big{\lbrace}(\omega+i\epsilon)R\big{\rbrace}\sum_{b=1}^{N} \sum_{a=1}^{M+N}\Bigg[-\eta^{\mu\nu}p_{b}^{\rho}J_{a,\rho\sigma}k^{\sigma}+\f{p_b.k}{p_a.k}\Big(p_a^\mu J_{a}^{\nu\rho}k_\rho +p_a^{\nu}J_a^{\mu\rho}k_{\rho}\Big)\Bigg]
\ee
After using the total angular momentum conservation relation $\sum\limits_{a=1}^{M+N}J_{a}^{\rho\sigma}=0$ we get,
\be
\Delta_{(1)}^{(\omega\ln\omega)}\widehat{T}_{1}^{h\mu\nu}(k)\ &=&\ -2iG\ \ln\big{\lbrace}(\omega+i\epsilon)R\big{\rbrace}\sum_{b=1}^{N}p_{b}.k\ \sum_{a=1}^{M+N}\f{1}{p_{a}.k}\ p_{a}^{(\mu}J_{a}^{\nu)\alpha}k_{\alpha}\label{Thphase}
\ee
Next we analyze the expression in eq.\eqref{Thathspin} in the integration region $\omega<<|\ell^{\mu}|<<L^{-1}$ and extract the order $\mathcal{O}(\omega\ln\omega)$ contribution. First let us substitute the following identity in eq.\eqref{Thathspin},
\be
G_{r}(k-\ell)G_{r}(\ell)\ &=&\ G_{r}(k-\ell)G_{r}(-\ell)-2\pi i\delta(\ell^2)\big[H(\ell^0)-H(-\ell^0)\big]G_{r}(k-\ell)
\ee
and focus on the contribution coming from $G_{r}(k-\ell)G_{r}(-\ell)$  part. As described in appendix-\ref{radiation}, the gravitational energy-momentum tensor with $\delta(\ell^2)\big[H(\ell^0)-H(-\ell^0)\big]G_{r}(k-\ell)$ part does not contribute to order $\mathcal{O}(\omega\ln\omega)$ in the integration region $\omega<<|\ell^{\mu}|<<L^{-1}$. Following the discussion of appendix-B in \cite{1912.06413}, a part of this contribution can be identified with the soft radiation from the finite energy real gravitational radiation, which is already taken care of in the hard particle sums of the earlier expressions. So in the integration region $\omega<<|\ell^{\mu}|<<L^{-1}$, the order $\mathcal{O}(\omega\ln\omega)$ contribution from gravitational energy-momentum tensor with $G_{r}(k-\ell)G_{r}(-\ell)$ part turns out to be,
\begingroup
\allowdisplaybreaks
\be
&&\Delta_{(1)}^{(\omega\ln\omega)}\widehat{T}_{2}^{h\mu\nu}(k)\non\\
&=&\ (8\pi G)\ \sum_{a,b=1}^{M+N}\int_{\omega}^{L^{-1}}\f{d^{4}\ell}{(2\pi)^{4}}\  \Big{\lbrace}G_{r}(-\ell)\Big{\rbrace}^{3}\ (-2k.\ell)\ \f{1}{p_{b}.\ell -i\epsilon}\f{1}{p_{a}.\ell +i\epsilon}\non\\
&&\times\Bigg[ \Big{\lbrace} ip_{b}^{(\alpha}J_{b}^{\beta)\gamma}\ell_{\gamma}-\f{i}{2}\eta^{\alpha\beta}p_{b\kappa}J_{b}^{\kappa\gamma}\ell_{\gamma}\Big{\rbrace}\Delta_{(\ell\ell)}\mathcal{F}^{\mu\nu}\ _{\alpha\beta,\rho\sigma}(k,\ell)\Big{\lbrace}p_{a}^{\rho}p_{a}^{\sigma}-\f{1}{2}p_{a}^{2}\eta^{\rho\sigma}\Big{\rbrace}\non\\
&&\ + \Big{\lbrace} p_{b}^{\alpha}p_{b}^{\beta}-\f{1}{2}p_{b}^{2}\eta^{\alpha\beta}\Big{\rbrace}\ \Delta_{(\ell\ell)}\mathcal{F}^{\mu\nu}\ _{\alpha\beta,\rho\sigma}(k,\ell)\Big{\lbrace}-ip_{a}^{(\rho}J_{a}^{\sigma)\delta}\ell_{\delta}+\f{i}{2}\eta^{\rho\sigma}p_{a\kappa}J_a^{\kappa\delta}\ell_{\delta}\Big{\rbrace}\Bigg]\non\\
&&\ +(8\pi G)\ \sum_{a,b=1}^{M+N}\int_{\omega}^{L^{-1}}\f{d^{4}\ell}{(2\pi)^{4}}\  \Big{\lbrace}G_{r}(-\ell)\Big{\rbrace}^{2}\  \f{1}{p_{b}.\ell -i\epsilon}\f{1}{(p_{a}.\ell +i\epsilon)^{2}}\ (p_{a}.k)\non\\
&&\times\Bigg[ \Big{\lbrace} ip_{b}^{(\alpha}J_{b}^{\beta)\gamma}\ell_{\gamma}-\f{i}{2}\eta^{\alpha\beta}p_{b\kappa}J_{b}^{\kappa\gamma}\ell_{\gamma}\Big{\rbrace}\Delta_{(\ell\ell)}\mathcal{F}^{\mu\nu}\ _{\alpha\beta,\rho\sigma}(k,\ell)\Big{\lbrace}p_{a}^{\rho}p_{a}^{\sigma}-\f{1}{2}p_{a}^{2}\eta^{\rho\sigma}\Big{\rbrace}\non\\
&&\ + \Big{\lbrace} p_{b}^{\alpha}p_{b}^{\beta}-\f{1}{2}p_{b}^{2}\eta^{\alpha\beta}\Big{\rbrace}\ \Delta_{(\ell\ell)}\mathcal{F}^{\mu\nu}\ _{\alpha\beta,\rho\sigma}(k,\ell)\Big{\lbrace}-ip_{a}^{(\rho}J_{a}^{\sigma)\delta}\ell_{\delta}+\f{i}{2}\eta^{\rho\sigma}p_{a\kappa}J_a^{\kappa\delta}\ell_{\delta}\Big{\rbrace}\Bigg]\non\\
&&\ +(8\pi G)\ \sum_{a,b=1}^{M+N}\int_{\omega}^{L^{-1}}\f{d^{4}\ell}{(2\pi)^{4}}\  \Big{\lbrace}G_{r}(-\ell)\Big{\rbrace}^{2}\  \f{1}{p_{b}.\ell -i\epsilon}\f{1}{p_{a}.\ell +i\epsilon}\non\\
&&\times\Bigg[ \Big{\lbrace} ip_{b}^{(\alpha}J_{b}^{\beta)\gamma}\ell_{\gamma}-\f{i}{2}\eta^{\alpha\beta}p_{b\kappa}J_{b}^{\kappa\gamma}\ell_{\gamma}\Big{\rbrace}\Delta_{(k\ell)}\mathcal{F}^{\mu\nu}\ _{\alpha\beta,\rho\sigma}(k,\ell)\Big{\lbrace}p_{a}^{\rho}p_{a}^{\sigma}-\f{1}{2}p_{a}^{2}\eta^{\rho\sigma}\Big{\rbrace}\non\\
&&\ + \Big{\lbrace} p_{b}^{\alpha}p_{b}^{\beta}-\f{1}{2}p_{b}^{2}\eta^{\alpha\beta}\Big{\rbrace}\ \Delta_{(k\ell)}\mathcal{F}^{\mu\nu}\ _{\alpha\beta,\rho\sigma}(k,\ell)\Big{\lbrace}-ip_{a}^{(\rho}J_{a}^{\sigma)\delta}\ell_{\delta}+\f{i}{2}\eta^{\rho\sigma}p_{a\kappa}J_a^{\kappa\delta}\ell_{\delta}\Big{\rbrace}\Bigg]\non\\
&&\ + (8\pi G)\ \sum_{a,b=1}^{M+N}\int_{\omega}^{L^{-1}}\f{d^{4}\ell}{(2\pi)^{4}}\  \Big{\lbrace}G_{r}(-\ell)\Big{\rbrace}^{2}\ \f{1}{p_{b}.\ell -i\epsilon}\f{1}{p_{a}.\ell +i\epsilon}\non\\
&&\times \Big{\lbrace} p_{b}^{\alpha}p_{b}^{\beta}-\f{1}{2}p_{b}^{2}\eta^{\alpha\beta}\Big{\rbrace}\Delta_{(\ell\ell)}\mathcal{F}^{\mu\nu}\ _{\alpha\beta,\rho\sigma}(k,\ell)\Big{\lbrace}ip_{a}^{(\rho}J_{a}^{\sigma)\delta}k_{\delta}-\f{i}{2}\eta^{\rho\sigma}p_{a\kappa}J_a^{\kappa\delta}k_{\delta}\Big{\rbrace}\label{Thomegalnomega}
\ee
\endgroup
In the above expression, $\Delta_{(\ell\ell)}\mathcal{F}(k,\ell)$ and $\Delta_{(k\ell)}\mathcal{F}(k,\ell)$ corresponds to the order $\mathcal{O}(\ell\ell)$ and order $\mathcal{O}(k\ell)$ contributions of  $\mathcal{F}(k,\ell)$ respectively, which can be easily extractable from eq.\eqref{F-full}. The expression in eq.\eqref{Thomegalnomega} is explicitly evaluated in Appendix-\ref{A} and the final result after summing over the contributions of eq.\eqref{L1L3final}, \eqref{L2final} and \eqref{L4final} becomes:
\be
&&\Delta_{(1)}^{(\omega\ln\omega)}\widehat{T}_{2}^{h\mu\nu}(k)\non\\
&=& \f{iG}{2}\sum_{\substack{a,b=1 \\ b\neq a\\ \eta_a \eta_b =1}}^{M+N}\f{1}{\big[(p_a.p_b)^2 -p_a^2 p_b^2\big]^{5/2}}\ln\big\lbrace L(\omega +i\epsilon \eta_a)\big\rbrace \non\\
&&\Big[ p_b.k \Big\lbrace 6(p_a.p_b)^3 p_b^\mu p_a^\nu p_b.r_a p_a^2 -6(p_a.p_b)^4 p_b^\mu p_a^\nu p_a.r_a -3(p_a.p_b)^2 p_a^2 p_b^2 p_a^\mu p_a^\nu p_b.r_a \non\\
&& +3(p_a.p_b)^3 p_b^2 p_a^\mu p_a^\nu p_a.r_a -6(p_a.p_b)^2 (p_a^2)^2 p_b^\mu p_b^\nu p_b.r_a +6(p_a.p_b)^3 p_a^2 p_b^\mu p_b^\nu p_a.r_a\non\\
&& +3p_a.p_b (p_a^2)^2 p_b^2 p_a^\mu p_b^\nu p_b.r_a -3(p_a.p_b)^2 p_a^2 p_b^2 p_a^\mu p_b^\nu p_a.r_a -3p_a.p_b (p_a^2)^2 p_b^2 p_b^\mu p_a^\nu p_b.r_a\non\\
&& +3(p_a.p_b)^2 p_a^2 p_b^2 p_b^\mu p_a^\nu p_a.r_a +3p_b^2 (p_a^2)^3 p_b^\mu p_b^\nu p_b.r_a -3p_b^2 (p_a^2)^2 p_a.p_b p_b^\mu p_b^\nu p_a.r_a\Big\rbrace\non\\
&& -p_a.k \Big\lbrace 3p_b^2 (p_a.p_b)^3 p_a^\mu p_b^\nu p_a.r_a -3p_b^2 (p_a.p_b)^2 p_a^\mu p_b^\nu p_a^2 p_b.r_a +3p_a^2 p_b^2 (p_a.p_b)^2 p_b^\mu p_b^\nu p_a.r_a \non\\
&& -3(p_a^2)^2 p_b^2 p_a.p_b p_b^\mu p_b^\nu p_b.r_a -3(p_b^2)^2 (p_a.p_b)^2 p_a^\mu p_a^\nu p_a.r_a +3(p_b^2)^2 p_a^2 p_a.p_b p_a^\mu p_a^\nu p_b.r_a \non\\
&& -3(p_b^2)^2 p_a^2 p_a.p_b p_b^\mu p_a^\nu p_a.r_a +3(p_a^2)^2 (p_b^2)^2 p_b^\mu p_a^\nu p_b.r_a +6(p_a.p_b)^3 p_b^\mu p_b^\nu p_a^2 p_b.r_a \non\\
&& -6(p_a.p_b)^4 p_b^\mu p_b^\nu p_a.r_a -6(p_a.p_b)^2 p_a^2 p_b^2 p_b^\mu p_a^\nu p_b.r_a +6(p_a.p_b)^3 p_b^2 p_b^\mu p_a^\nu p_a.r_a\Big\rbrace\Big]\non\\
&& +\f{iG}{2}\sum_{\substack{a,b=1 \\ b\neq a\\ \eta_a \eta_b =1}}^{M+N}\f{1}{\big[(p_a.p_b)^2 -p_a^2 p_b^2\big]^{3/2}}\ln\big\lbrace L(\omega +i\epsilon \eta_a)\big\rbrace \non\\
&& \Big[ -2p_a.p_b p_b.k p_b.r_a p_a^2 p_a^\mu p_b^\nu +5p_b^2 p_a.k p_a.p_b p_a.r_a p_a^\mu p_b^\nu -4(p_a.p_b)^2 p_b.r_a p_a.k p_a^\mu p_b^\nu \non\\
&& +2(p_a.p_b)^3 r_a.k p_a^\mu p_b^\nu -3p_a^2 p_b^2 p_a.p_b r_a.k p_a^\mu p_b^\nu \non\\
&& -p_a^2 p_b^2 p_b.k p_a.r_a p_a^\mu p_b^\nu -8p_a.p_b p_b.k p_a^2 p_b.r_a p_a^\nu p_b^\mu +10 (p_a.p_b)^2 p_b.k p_a.r_a p_a^\nu p_b^\mu \non\\
&& -2p_a^2 p_b^2 p_b.r_a p_a.k p_a^\nu p_b^\mu -4(p_a.p_b)^2 p_b.r_a p_a.k p_a^\nu p_b^\mu -3p_a^2 p_b^2 p_a.p_b r_a.k p_a^\nu p_b^\mu \non\\
&& +8p_a.p_b p_b^2 p_a.r_a p_a.k p_a^\nu p_b^\mu +2(p_a.p_b)^3 r_a.k p_a^\nu p_b^\mu -10(p_a.p_b)^2 p_a.r_a p_a.k p_b^\mu p_b^\nu \non\\
&& +8p_a.p_b p_a^2 p_b.r_a p_a.k p_b^\mu p_b^\nu -4 p_b.k p_a^2 p_a.r_a p_a.p_b p_b^\mu p_b^\nu +4 p_b.k (p_a^2)^2 p_b.r_a p_b^\mu p_b^\nu \non\\
&& +2(p_a.p_b)^3 p_a.k p_b^\mu r_a^\nu -3p_a^2 p_b^2 p_a.p_b p_a.k p_b^\mu r_a^\nu +2p_a.k (p_a.p_b)^3 p_b^\nu r_a^\mu \non\\
&& -3p_a.k p_a^2 p_b^2 p_a.p_b p_b^\nu r_a^\mu -2p_b.k(p_a.p_b)^2 p_a^2 p_b^\nu r_a^\mu +(p_a^2)^2 p_b^2 p_b^\nu r_a^\mu p_b.k \non\\
&& +4p_b.k (p_a.p_b)^2 p_b.r_a p_a^\mu p_a^\nu -5p_b.k p_a.p_b p_b^2 p_a.r_a p_a^\mu p_a^\nu -2p_b^2 p_a.p_b p_a.k p_b.r_a p_a^\mu p_a^\nu \non\\
&& -(p_b^2)^2 p_a.k p_a.r_a p_a^\mu p_a^\nu -2p_b.k (p_a.p_b)^3 p_a^\nu r_a^\mu +3p_b.k p_a^2 p_b^2 p_a.p_b p_a^\nu r_a^\mu \non\\
&& +2p_b^2 (p_a.p_b)^2 p_a.k p_a^\nu r_a^\mu -p_a^2 (p_b^2)^2 p_a.k p_a^\nu r_a^\mu -2p_b.k (p_a.p_b)^3 p_a^\mu r_a^\nu \non\\
&& +3p_b.k p_a^2  p_b^2 p_a.p_b p_a^\mu r_a^\nu +2p_b^2 (p_a.p_b)^2 p_a.k p_a^\mu r_a^\nu -p_a^2 (p_b^2)^2 p_a.k p_a^\mu r_a^\nu \non\\
&& +2(p_a.p_b)^2 p_a^2 r_a.k p_b^\mu p_b^\nu -2(p_a.p_b)^2 p_b^2 r_a.k p_a^\mu p_a^\nu +p_a^2 p_b^2 p_a.r_a p_a.k p_b^\mu p_b^\nu \non\\
&& -(p_a^2)^2 p_b^2 r_a.k p_b^\mu p_b^\nu +3p_a^2 (p_b^2)^2 r_a.k p_a^\mu p_a^\nu -p_b.k p_a^2 p_b^2 p_b.r_a p_a^\mu p_a^\nu \non\\
&& -2p_b.k (p_a.p_b)^2 p_a^2 p_b^\mu r_a^\nu -p_b.k p_a^2 p_b^2 p_a.r_a p_b^\mu p_a^\nu +p_b.k (p_a^2)^2 p_b^2 p_b^\mu r_a^\nu \non\\
&& +4p_b.k (p_a.p_b)^2 p_a.r_a p_a^\mu p_b^\nu +p_a.k p_a^2 p_b^2 p_b.r_a p_a^\mu  p_b^\nu \non\\
&& -2\lbrace (p_a.p_b)^2 -p_a^2 p_b^2 \rbrace p_a^\mu p_b^\nu p_{b\alpha}\Sigma_a^{\alpha\beta}k_\beta -2\lbrace (p_a.p_b)^2 -p_a^2 p_b^2 \rbrace p_a^\nu p_b^\mu p_{b\alpha}\Sigma_a^{\alpha\beta}k_\beta \non\\
&& -2\lbrace (p_a.p_b)^2 -p_a^2 p_b^2 \rbrace p_a.k p_b^\mu p_{b\alpha}\Sigma_a^{\alpha\nu} -2 \lbrace (p_a.p_b)^2 -p_a^2 p_b^2 \rbrace p_a.k p_b^\nu p_{b\alpha}\Sigma_a^{\alpha\mu} \non\\
&& +2 \lbrace (p_a.p_b)^2 -p_a^2 p_b^2 \rbrace p_b.k p_a^\nu p_{b\alpha}\Sigma_a^{\alpha\mu} +2 \lbrace (p_a.p_b)^2 -p_a^2 p_b^2 \rbrace p_b.k p_a^\mu p_{b\alpha} \Sigma_a^{\alpha\nu}\non\\
&& +2 \lbrace (p_a.p_b)^2 -p_a^2 p_b^2 \rbrace p_b^2 p_a^\nu \Sigma_a^{\mu\alpha}k_\alpha +2 \lbrace (p_a.p_b)^2 -p_a^2 p_b^2 \rbrace p_b^2 p_a^\mu \Sigma_a^{\nu\alpha}k_\alpha \non\\
&& +2 p_a.p_b p_a^2 p_b^\mu p_b^\nu p_{b\alpha}\Sigma_a^{\alpha\beta}k_\beta -2p_a.p_b p_a^2 p_b.k p_b^\mu p_{b\alpha}\Sigma_a^{\alpha\nu }-2 p_b^2 p_a.p_b p_a^\mu p_a^\nu p_{b\alpha}\Sigma_a^{\alpha\beta}k_\beta \non\\
&& -2p_b.k p_a.p_b p_a^2 p_b^\nu p_{b\alpha}\Sigma_a^{\alpha\mu}\Big]\label{Thfinal}
\ee
\subsubsection{Total energy-momentum tensor and gravitational waveform at order $\mathcal{O}(\omega\ln\omega)$}

Summing over the contributions of eq.\eqref{TXfinal_exp}, eq.\eqref{Thfinal} and eq.\eqref{Thphase}, we get the Fourier transform the total energy-momentum tensor at order $\mathcal{O}(G\ \omega\ln\omega)$:
\be
\Delta_{(1)}^{(\omega\ln\omega)}\widehat{T}^{\mu\nu}(k)
&=& -2iG\ \ln\big{\lbrace}(\omega+i\epsilon)R\big{\rbrace}\sum_{b=1}^{N}p_{b}.k\ \sum_{a=1}^{M+N}\f{1}{p_{a}.k}\ p_{a}^{(\mu}J_{a}^{\nu)\alpha}k_{\alpha}\non\\
&&-\f{1}{2}\sum_{a=1}^{M+N}\f{k_{\rho}k_{\sigma}}{p_{a}.k}\Bigg{\lbrace}\Bigg(p_{a}^{\mu}\f{\p}{\p p_{a\rho}}-p_{a}^{\rho}\f{\p}{\p p_{a\mu}}\Bigg)K_{gr}^{cl}\times \Big(r_{a}^{\sigma}p_{a}^{\nu}-r_{a}^{\nu}p_{a}^{\sigma}+\Sigma_{a}^{\sigma\nu}\Big)\non\\
&&\ +\ \Bigg(p_{a}^{\nu}\f{\p}{\p p_{a\sigma}}-p_{a}^{\sigma}\f{\p}{\p p_{a\nu}}\Bigg)K_{gr}^{cl}\times \Big(r_{a}^{\rho}p_{a}^{\mu}-r_{a}^{\mu}p_{a}^{\rho}+\Sigma_{a}^{\rho\mu}\Big)\Bigg{\rbrace}
\label{FinalT}
\ee
Using the relation in eq.\eqref{eT}, the radiative mode of gravitational waveform at order $\mathcal{O}(G^2 \omega\ln\omega)$ for the above derived energy-momentum tensor takes the following form:
\begingroup
\allowdisplaybreaks
\be
&&\Delta_{(G^2)}^{(\omega\ln\omega)}\ \widetilde{e}^{\mu\nu}(\omega,\vec{x})\non\\
&=&\  \f{2G}{R}\ \exp\lbrace i\omega R\rbrace \Bigg[2iG \ln\lbrace(\omega+i\epsilon)R\rbrace\sum_{b=1}^{N}p_{b}.k\non\\
&&\times \sum_{a=1}^{M+N}\f{p_{a}^{(\mu}k_{\rho}}{p_{a}.k}\Big(r_{a}^{\rho}p_{a}^{\nu)}-r_{a}^{\nu)}p_{a}^{\rho}+\Sigma_{a}^{\rho\nu)}\Big)\non\\
&&-\f{1}{2}\sum_{a=1}^{M+N}\f{k_{\rho}k_{\sigma}}{p_{a}.k}\Bigg{\lbrace}\Bigg(p_{a}^{\mu}\f{\p}{\p p_{a\rho}}-p_{a}^{\rho}\f{\p}{\p p_{a\mu}}\Bigg)K_{gr}^{cl}\times \Big(r_{a}^{\sigma}p_{a}^{\nu}-r_{a}^{\nu}p_{a}^{\sigma}+\Sigma_{a}^{\sigma\nu}\Big)\non\\
&&\ +\ \Bigg(p_{a}^{\nu}\f{\p}{\p p_{a\sigma}}-p_{a}^{\sigma}\f{\p}{\p p_{a\nu}}\Bigg)K_{gr}^{cl}\times \Big(r_{a}^{\rho}p_{a}^{\mu}-r_{a}^{\mu}p_{a}^{\rho}+\Sigma_{a}^{\rho\mu}\Big)\Bigg{\rbrace}\Bigg]\label{emunu_final}
\ee
\endgroup
Now if we compare the above expression with the conjectured result given in eq.\eqref{waveform_expectation}, we observe that our result in the direct derivation completely agrees with the conjectured waveform. 
\section{Conclusion}\label{S:conclusion}
In this paper, we derive the leading spin-dependent gravitational tail memory which behaves like $u^{-2}$ for retarded time $u\rightarrow\pm \infty$. First, we predict the result from the classical limit of the soft graviton theorem, then derive it for a general gravitational scattering process involving spinning objects. The final result of leading spin-dependent gravitational tail memory has been summarized in \S\ref{S:memory_final_I} and \S\ref{S:rewritten}. Here we are pointing out the novel features of our result, its theoretical and observational importance and possible way of re-deriving our result with other available prescriptions:
\begin{enumerate}
\item Even when the scattered objects do not carry any intrinsic spins still a large part of the order $\mathcal{O}(G^2 \omega\ln\omega)$ gravitational waveform is non-vanishing and can be read off from eq.\eqref{expectation_e} by setting $\Sigma_a=\Sigma'_a=0$. This result is fully determined in terms of the asymptotic momenta, asymptotic orbital angular momenta, and the direction cosine and frequency of gravitational wave emission.
\item Our result of order $u^{-2}$ gravitational memory as given in eq.\eqref{future_memory} and eq.\eqref{past_memory} has several theoretical and observational importance in the current era of gravitational wave physics. First of all from the observation of this tail memory one can read off the spins of scattered objects, which was not possible for earlier known DC memory or even $u^{-1}$ and $u^{-2}\ln u $ tail memories. Secondly for a black hole binary merger process even if the blackholes carry spin, still in \S\ref{S:rewritten} we have shown that gravitational tail memory at this order vanishes, which is another non-trivial prediction from general relativity. In \S\ref{S:rewritten} we also have shown that if the scattering event carries some massless particles or radiation along with high-frequency gravitational wave in the final state, still the waveform can be rewritten in such a way that it does not carry any information about the outgoing massless particles or radiation.
\item In the recent past there has been a lot of progress in deriving various classical observables including gravitational radiation for $2\rightarrow 2$ scattering of spinning bodies under weak gravitational interaction\cite{1712.09250,1803.02405,1906.09260,2104.03957}. There are two differences between our analysis and the analysis done in the papers cited above. The first one is about setting the boundary conditions: to derive the low-frequency gravitational waveform, we give both initial and final data for the scattered objects. This has the advantage that we don't have to specify the kind of interaction and strength within the region $\mathcal{R}$. On the other hand in the references \cite{1712.09250,1803.02405,1906.09260,2104.03957}, one specifies only the initial data for the scattered objects and derive various classical observables assuming weak gravitational interaction responsible for the scattering event. Another difference is that in the references \cite{1712.09250,1803.02405,1906.09260,2104.03957} the gravitational radiation has been expressed as an integral form without the evaluation of the integrals in low-frequency limit (soft region), as done here. But recently in \cite{ 2007.02077} an attempt has been taken to relate these two approaches for non-spinning $2\rightarrow 2$ particle scattering event under weak gravitational interaction. It is possible to generalize this idea for spinning particle scattering and re-derive our result.
\end{enumerate}

\acknowledgments
We are deeply thankful to Alok Laddha, Arnab Priya Saha, Ashoke Sen, and Amitabh Virmani for many enlightening discussions, encouragements, and useful comments on the draft. The work of D.G. is supported by a grant to CMI from the Infosys foundation. The work of B.S. is supported by the Simons Foundation grant 488649 (Simons
Collaboration on the Nonperturbative Bootstrap) and by the Swiss National Science Foundation through the National Centre of Competence in Research SwissMAP. B.S. also thanks to the organizers
and participants of the workshop {\it{ Gravitational scattering, inspiral, and radiation }} at the Galileo Galilei Institute for Theoretical Physics in Florence, for hospitality
and fruitful exchanges where a part of this work has been presented. 

\appendix
\section{Review of geodesic equation and spin evolution of spinning object}\label{app:Spin_Geo}
In this appendix we review the geodesic equation for spinning object as well as  the time evolution of the spin. These equations are known as Mathisson-Papapetrou equations\cite{Mathisson, Papapetrou} in general relativity and the covariant form of those are discovered by Tulczyjew and Dixon\cite{Tulczyjew, Dixon}. The Mathisson-Papapetrou equations take the following form:
\be
\f{DP^\mu}{D\sigma}&=&-\f{1}{2}R^{\mu}\ _{\nu\rho\sigma}u^\nu \Sigma^{\rho\sigma}\label{trajectory}\\
\f{D\Sigma^{\mu\nu}}{D\sigma}&=& P^\mu u^\nu -P^\nu u^\mu\label{spin_evolution}
\ee
where $P^\mu$ is the kinematical momentum and $\Sigma^{\mu\nu}$ is the spin angular momentum of the object measured along the world line as a function of affine parameter $\sigma$. The four velocity is denoted by $u^\mu$ and defined as $u^{\mu}=\f{dx^\mu}{d\sigma}$, for $x^\mu(\sigma)$ representing the world line and $u^\mu u_\mu =-1$ if $\sigma$ is the proper time. $\f{D}{D\sigma}$ represents the covariant derivative along the world line. Let us define the kinematic mass $m\equiv -P\cdot u$, which is in general not a constant of motion\cite{km}. Now contracting with $u_\nu$ from eq.\eqref{spin_evolution} we get,
\be
P^\mu = mu^\mu -\f{D\Sigma^{\mu\nu}}{D\sigma}u_\nu\label{P}
\ee
To understand how the kinematical mass evolves along the trajectory, let us take derivative over $m$ w.r.t. $\sigma$,
\be
\f{dm}{d\sigma}=\f{Dm}{D\sigma}=-\f{DP^\mu}{D\sigma}u_\mu -P^\mu \f{Du_\mu}{D\sigma}= \f{Du_\mu}{D\sigma} \f{D\Sigma^{\mu\nu}}{D\sigma}u_\nu \label{dm}
\ee
where to get the last expression after equality, we substituted the results of eq.\eqref{trajectory} and eq.\eqref{spin_evolution}. Now we substitute the expression of $P^\mu$ in eq.\eqref{trajectory} and use the expressions in eq.\eqref{dm} and \eqref{spin_evolution}. After this substitutions we simplify eq.\eqref{trajectory} using $u_{\alpha}\f{Du^\alpha}{D\sigma}=0$ and get the following simplified trajectory equation,
\be
m\f{D u^\mu}{D\sigma}-\f{D^2 \Sigma^{\mu\nu}}{D\sigma^2}u_\nu = -\f{1}{2}R^{\mu}\ _{\nu\rho\sigma}u^\nu \Sigma^{\rho\sigma}\label{trajectory_exact}
\ee
Note that to derive the above trajectory equation we do not have to use any spin supplementarity condition(SSC).

Now let us focus on the simplification of the spin evolution equation given in eq.\eqref{spin_evolution}, using the trajectory equation \eqref{trajectory_exact} and spin supplementarity condition $u_\mu \Sigma^{\mu\nu}=0$. We start by substituting the expression of $P^\mu$ in eq.\eqref{spin_evolution} and get,
\be
\f{D\Sigma^{\mu\nu}}{D\sigma}&=& -\f{D\Sigma^{\mu\alpha}}{D\sigma}u_\alpha u^\nu +\f{D\Sigma^{\nu\alpha}}{D\sigma}u_\alpha u^\mu
\ee
Now using the definition of covariant derivative on the spin tensor, moving the $\sigma$ derivatives from $\Sigma$ to $u$ and simplifying using  spin supplementary condition $u_\mu \Sigma^{\mu\nu}=0$\footnote{In the literatures, we find two kinds of spin supplementarity conditions(SSC), which are $u_\mu \Sigma^{\mu\nu}=0$ and $P_\mu \Sigma^{\mu\nu}=0$. Different choices of SSC, will modify  the  $ \mathcal{O}(\Sigma^2)$ terms in the RHS of eq.\eqref{trajectory_final} and eq.\eqref{spin_final}, which are irrelevant for our analysis. }, the above equation takes the following form:
\be 
\f{d\Sigma^{\mu\nu}}{d\sigma}+\Gamma^\mu_{\alpha\beta}\Sigma^{\alpha\nu}u^\beta +\Gamma^{\nu}_{\alpha\beta}\Sigma^{\mu\alpha}u^\beta =\Sigma^{\mu\alpha}\f{du_\alpha}{d\sigma}u^\nu -\Sigma^{\nu\alpha}\f{du_\alpha}{d\sigma}u^\mu -\Gamma^{\alpha}_{\rho\sigma}\Sigma^{\mu\rho}u^\sigma u_\alpha u^\nu +\Gamma^{\alpha}_{\rho\sigma}\Sigma^{\nu\rho}u^\sigma u_\alpha u^\mu  \non\\
\ee
Substituting the trajectory equation \eqref{trajectory_exact} in the RHS of the above equation, we get
\be
\f{d\Sigma^{\mu\nu}}{d\sigma}+\Gamma^\mu_{\alpha\beta}\Sigma^{\alpha\nu}u^\beta +\Gamma^{\nu}_{\alpha\beta}\Sigma^{\mu\alpha}u^\beta = \f{1}{m}\Big(\Sigma^{\mu\alpha}u^{\nu}-\Sigma^{\nu\alpha}u^{\mu}\Big)\Bigg(\f{D^2 \Sigma_{\alpha\beta}}{D\sigma^2}u^{\beta}-\f{1}{2}R_{\alpha\beta\rho\sigma}u^{\beta}\Sigma^{\rho\sigma}\Bigg)\non\\ \label{spin_exact}
\ee
In the above equation the LHS is basically $\f{D\Sigma^{\mu\nu}}{D\sigma}$, which is equal to terms quadratic in spin as written in RHS. This above result also implies that $\f{D^2 \Sigma^{\mu\nu}}{D\sigma^2}$ in eq.\eqref{trajectory_exact} is also quadratic in spin. Hence if we are interested in the trajectory and spin evolution equations which are linear in spin\footnote{Here we want emphasize that we are not making any small spin approximation while ignoring terms quadratic in spin. We are ignoring those terms as those contribute at order $\mathcal{O}(G^{2})$ to the correction of trajectory and spin in our analysis of \S\ref{S:order2waveform} and would not affect our $\mathcal{O}(G^{2})$ gravitational waveform.}, eq.\eqref{trajectory_exact} and eq.\eqref{spin_exact} simplifies to
\be
\f{d^2 x^\mu}{d\sigma^2} +\Gamma^\mu_{\rho\sigma}\f{dx^\rho}{d\sigma}\f{dx^\sigma}{d\sigma}= -\f{1}{2m}R^{\mu}\ _{\nu\rho\sigma}\f{dx^\nu}{d\sigma} \Sigma^{\rho\sigma}+\ \mathcal{O}(\Sigma^2) \label{trajectory_final}
\ee
and,
\be
\f{d\Sigma^{\mu\nu}}{d\sigma}+\Gamma^\mu_{\alpha\beta}\Sigma^{\alpha\nu}\f{dx^\beta}{d\sigma} +\Gamma^{\nu}_{\alpha\beta}\Sigma^{\mu\alpha}\f{dx^\beta}{d\sigma} = 0+\ \mathcal{O}(\Sigma^2) \label{spin_final}
\ee

Now we show that the above equations also follow from the covariant conservation of the canonical version of the matter energy-momentum tensor in eq.\eqref{TXspin}. For a given world-line action $S_X$ of a moving object in gravitational background with metric $g_{\mu\nu}(x)$, the canonical matter energy-momentum tensor is defined by:
\be 
T^{X\alpha\beta}_c(x) &\equiv &\ \f{2}{\sqrt{-g}}\f{\delta S_X}{\delta g_{\alpha\beta}(x)}
\ee
which satisfies the following conservation equation,
\be
\nabla_\alpha T_c^{X\alpha\beta}(x)=0 \label{conservation}
\ee
In the above expressions,  $g\equiv det(g_{\mu\nu})$ and $\nabla_\alpha $ represents the covariant derivative with background metric $g_{\mu\nu}(x)$. Let $\phi_\beta(x)$ be an arbitrary rank-1 tensor with a sufficient fall off at the boundary of spacetime, such that $\big[\sqrt{-g}\ T_c^{X\alpha\beta}(x)\phi_{\beta}(x)\big]_{|x|\rightarrow\infty}=0$ will be satisfied. Now start with the following identity,
\be
&&\int d^4x \sqrt{-g}\nabla_\alpha\Big(T_c^{X\alpha\beta}(x)\phi_{\beta}(x)\Big)=\int d^4x \sqrt{-g}\Big[ \big(\nabla_\alpha T_c^{X\alpha\beta}(x)\big)\phi_{\beta}(x)\big) +T_c^{X\alpha\beta}(x)\nabla_{(\alpha}\phi_{\beta)}(x)\Big]\non\\
&&\Rightarrow  \int d^4x \sqrt{-g}\ T_c^{X\alpha\beta}(x)\nabla_{(\alpha}\phi_{\beta)}(x)\ =\ 0
\ee
Above to get the last line we used the fact that LHS of first line is a boundary term which vanishes and the first term in the RHS of the first line vanishes due to the conservation of matter energy-momentum tensor. In our convention, the relation between matter energy-momentum tensor given in eq.\eqref{TXspin} and the canonical energy-momentum tensor is: $T^{X\alpha\beta}(x)\equiv \sqrt{-g}\ T_c^{X\alpha\beta}(x)$. Now after substituting the matter energy-momentum tensor of eq.\eqref{TXspin} for a single particle in the above relation, we get
\be
&& \int d\sigma\ \int d^4x   \Bigg[ \ m\f{d X^{\alpha}(\sigma)}{d \sigma}\f{d X^{\beta}(\sigma)}{d \sigma}\ \delta^{(4)}\big(x-X(\sigma)\big)\ +\ \f{d X^{(\alpha}(\sigma)}{d \sigma}\ \Sigma^{\beta)\gamma}(\sigma)\non\\
&&\times \ \p_{\gamma}\delta^{(4)}\big(x-X(\sigma)\big)+\Gamma^{(\alpha}_{\gamma\delta}(X)\Sigma^{\beta)\gamma}(\sigma)\f{dX^{\delta}(\sigma)}{d\sigma}\delta^{(4)}\big(x-X(\sigma)\big)\Bigg]\non\\
 &&\ \times  \Bigg\lbrace \p_{(\alpha}\phi_{\beta)}(x)-\Gamma^{\rho}_{\alpha\beta}(x)\phi_{\rho}(x)\Bigg\rbrace\ =\ 0
\ee
In the above expression, we first perform integration by parts to remove the derivative over delta function and then do the spacetime integration using the delta function. Next, after using the identity $\f{dX^\alpha}{d\sigma}\f{\p}{\p X^\alpha}=\f{d}{d\sigma}$ and integration by parts for the derivatives w.r.t $\sigma$ we get,
\be
&& \int d\sigma\ \phi_\rho(X)\Bigg[ -m\f{d^2 X^{\rho}(\sigma)}{d\sigma^2}-m\Gamma^{\rho}_{\alpha\beta}(X)\f{dX^\alpha(\sigma)}{d\sigma}\f{dX^\beta(\sigma)}{d\sigma}\non\\
&& +\p_\gamma \Gamma^\rho_{\alpha\beta}(X)\f{dX^\alpha(\sigma)}{d\sigma}\Sigma^{\beta\gamma}(\sigma) -\Gamma^{\rho}_{\alpha\beta}(X)\Gamma^{\alpha}_{\gamma\delta}(X)\f{dX^\delta(\sigma)}{d\sigma}\Sigma^{\beta\gamma}(\sigma)\Bigg]\non\\
&& +\f{1}{2}\int d\sigma\ \p_\gamma \phi_\rho(X)\Bigg[\f{d\Sigma^{\rho\gamma}(\sigma)}{d\sigma}+ \Gamma^{\rho}_{\alpha\beta}(X)\f{dX^{\beta}(\sigma)}{d\sigma}\Sigma^{\alpha\gamma}(\sigma)-\Gamma^{\gamma}_{\alpha\beta}(X)\f{dX^{\beta}(\sigma)}{d\sigma}\Sigma^{\alpha\rho}(\sigma)\Bigg]\nonumber\\
&&\  =\ 0
\ee
Since $\phi_\rho(x)$ is an arbitrary rank-1 tensor, in the above expression  the coefficients of $\phi_\rho(X)$ and $\p_\gamma\phi_\rho(X)$ will vanish individually. Hence replacing $\rho\rightarrow\mu$ and $\gamma\rightarrow\nu$ and using the antisymmetry property of $\Sigma(\sigma)$ we get the following two equations from the above expression:
\be
&&\f{d^2 X^{\mu}(\sigma)}{d\sigma^2}+\Gamma^{\mu}_{\alpha\beta}(X)\f{dX^\alpha(\sigma)}{d\sigma}\f{dX^\beta(\sigma)}{d\sigma}=-\f{1}{2m}R^{\mu}\ _{\alpha\beta\gamma}(X)\ \f{dX^{\alpha}(\sigma)}{d\sigma}\Sigma^{\beta\gamma}(\sigma)\non\\
&& \f{d\Sigma^{\mu\nu}(\sigma)}{d\sigma}+\Gamma^{\mu}_{\alpha\beta}(X)\f{dX^\alpha(\sigma)}{d\sigma}\Sigma^{\beta\nu}(\sigma)+\Gamma^{\nu}_{\alpha\beta}(X)\f{dX^\alpha(\sigma)}{d\sigma}\Sigma^{\mu\beta}(\sigma)\ =\ 0
\ee
where,
\be
R^{\mu}\ _{\alpha\beta\gamma}(X)=\p_\beta \Gamma^{\mu}_{\alpha\gamma}(X) -\p_\gamma \Gamma^{\mu}_{\alpha\beta}(X)+\Gamma^{\delta}_{\alpha\gamma}(X)\Gamma^{\mu}_{\delta\beta}(X)-\Gamma^{\delta}_{\alpha\beta}(X)\Gamma^{\mu}_{\delta\gamma}(X)
\ee
Now if we compare the above equations with eq.\eqref{trajectory_final} and eq.\eqref{spin_final}, we observe that they are identical. This also proves that the geodesic equation and spin evolution equation are consistent with our matter energy-momentum tensor.

\section{Detail analysis of gravitational energy-momentum tensor given in eq.\eqref{Thomegalnomega}}\label{A}
To evaluate the expression in eq.\eqref{Thomegalnomega} we need to use the results of the following three integrals:
\be
J_{1}^{\alpha\beta}&=&\ \int_{\omega}^{L^{-1}} \f{d^{4}\ell}{(2\pi)^{4}}\ \big[G_{r}(-\ell)\big]^{2}\ \f{1}{p_{a}.\ell+i\epsilon}\f{1}{p_{b}.\ell-i\epsilon}\ \ell^{\alpha}\ell^{\beta}\non\\
&=&\ -\f{1}{2}\Big[p_{a}^{\alpha}\f{\p}{\p p_{a\beta}}+p_{b}^{\alpha}\f{\p}{\p p_{b\beta}}+\eta^{\alpha\beta}\Big]\ \int_{\omega}^{L^{-1}}\f{d^{4}\ell}{(2\pi)^{4}}\ G_{r}(-\ell)\ \f{1}{p_{a}.\ell+i\epsilon}\f{1}{p_{b}.\ell-i\epsilon} \non\\
&=&\ \f{1}{8\pi}\delta_{\eta_a \eta_b ,1}\ln\big\lbrace L(\omega +i\epsilon \eta_a)\big\rbrace \ \f{1}{[(p_a.p_b)^2-p_a^2 p_b^2]^{3/2}}\Big[p_a.p_b\lbrace p_a^\alpha p_b^\beta +p_b^\alpha p_a^\beta\rbrace -p_b^2 p_a^\alpha p_a^\beta\non\\
&&\ -p_a^2 p_b^\alpha p_b^\beta -\eta^{\alpha\beta}\lbrace (p_a.p_b)^2 -p_a^2 p_b^2\rbrace\Big] \label{J1}\\
J_{2}^{\alpha\beta\gamma}&=&\int_{\omega}^{L^{-1}} \f{d^{4}\ell}{(2\pi)^{4}}\ \big[G_{r}(-\ell)\big]^{2}\ \f{1}{p_{b}.\ell-i\epsilon}\f{1}{(p_{a}.\ell+i\epsilon)^{2}}\ \ell^{\alpha}\ell^{\beta}\ell^{\gamma}\non\\
&=&\ \f{1}{2}\Bigg[p_{b}^{\gamma}\f{\p}{\p p_{b\alpha}}\f{\p}{\p p_{a\beta}}+p_{a}^{\gamma}\f{\p}{\p p_{a\alpha}}\f{\p}{\p p_{a\beta}}+\Big(\eta^{\alpha\gamma}\f{\p}{\p p_{a\beta}}+\eta^{\beta\gamma}\f{\p}{\p p_{a\alpha}}\Big)\Bigg]\non\\
&&\ \times \ \int_{\omega}^{L^{-1}}\f{d^{4}\ell}{(2\pi)^{4}}\ G_{r}(-\ell)\ \f{1}{p_{a}.\ell+i\epsilon}\f{1}{p_{b}.\ell-i\epsilon}\non\\
&=& \f{1}{8\pi}\delta_{\eta_a \eta_b ,1}\ln\big\lbrace L(\omega +i\epsilon \eta_a)\big\rbrace \f{1}{[(p_a.p_b)^2 -p_a^2 p_b^2]^{5/2}}\Big[ 3(p_a.p_b)^2 p_a^\alpha p_b^\beta p_b^\gamma -3p_a^2 p_a.p_b p_b^\alpha p_b^\beta p_b^\gamma\non\\
&&\ -3p_b^2 p_a.p_b p_a^\alpha p_a^\beta p_b^\gamma +3p_a^2 p_b^2 p_b^\alpha p_a^\beta p_b^\gamma +3(p_a.p_b)^2 p_b^\alpha p_b^\beta p_a^\gamma -3p_b^2 p_a.p_b p_a^\alpha p_b^\beta p_a^\gamma\non\\
&&\ -3p_b^2 p_a.p_b p_b^\alpha p_a^\beta p_a^\gamma +3(p_b^2)^2 p_a^\alpha p_a^\beta p_a^\gamma \Big]\non\\
&&\ -\f{1}{8\pi} \delta_{\eta_a \eta_b ,1}\ln\big\lbrace L(\omega +i\epsilon \eta_a)\big\rbrace \f{1}{[(p_a.p_b)^2 -p_a^2 p_b^2]^{3/2}}\Big[ p_a^\alpha p_b^\beta p_b^\gamma -2p_b^\alpha p_a^\beta p_b^\gamma +p_a.p_b \eta^{\alpha\beta}p_b^\gamma\non\\
&&\  +p_b^\alpha p_b^\beta p_a^\gamma -p_b^2 \eta^{\alpha\beta}p_a^\gamma +p_a.p_b p_b^\beta \eta^{\alpha\gamma} -p_b^2 p_a^\beta \eta^{\alpha\gamma}+p_a.p_b p_b^\alpha \eta^{\beta\gamma}-p_b^2 p_a^\alpha \eta^{\beta\gamma}\Big]\label{J2}\\
J_{3}^{\alpha}&=&\int_{\omega}^{L^{-1}} \f{d^{4}\ell}{(2\pi)^{4}}\ G_{r}(-\ell)\ \f{1}{p_{b}.\ell-i\epsilon}\f{1}{(p_{a}.\ell+i\epsilon)^{2}}\ \ell^{\alpha}\non\\
&=&\ \f{1}{4\pi}\delta_{\eta_a \eta_b ,1}\ln\big\lbrace L(\omega +i\epsilon \eta_a)\big\rbrace \f{1}{[(p_a.p_b)^2 -p_a^2 p_b^2]^{3/2}}\Big[p_a.p_b p_b^\alpha -p_b^2 p_a^\alpha\Big]\label{J3}
\ee
To evaluate the expression in eq.\eqref{Thomegalnomega}, we divide it into sum over four integrals:
\be
\Delta_{(1)}^{(\omega\ln\omega)}\widehat{T}_{2}^{h\mu\nu}(k)&=&\ \mathcal{L}_1^{\mu\nu}+\mathcal{L}_2^{\mu\nu}+\mathcal{L}_3^{\mu\nu}+\mathcal{L}_4^{\mu\nu}
\ee
The first three lines after the equality in eq.\eqref{Thomegalnomega} takes the following form after removing the terms containing $p_a.\ell$ and $p_b.\ell$ in the numerator\footnote{ We do not keep the terms containing $p_a.\ell$ or $p_b.\ell$ in the numerator as they cancel with the denominator of the integrand and then it can be shown that the integration result for those terms vanishes after the $\ell^0$ contour integration.},
\be
\mathcal{L}_1^{\mu\nu}&=&\ \f{i}{2}(8\pi G)\sum_{a,b=1}^{M+N}\int_{\omega}^{L^{-1}}\f{d^{4}\ell}{(2\pi)^{4}}\ \Big{\lbrace}G_{r}(-\ell)\Big{\rbrace}^{3}\ (-2k.\ell) \f{1}{p_{b}.\ell -i\epsilon}\f{1}{p_{a}.\ell +i\epsilon}\non\\
&&\Big[ 2\ell^{\mu}\ell^{\nu}p_a.p_b p_{a\alpha}J_b^{\alpha\beta}\ell_{\beta}-p_a^2 \ell^\mu \ell^\nu p_{b\alpha}J_b^{\alpha\beta}\ell_\beta +2\ell^2 p_b^\mu p_a^\nu p_{a\alpha}J_b^{\alpha\beta}\ell_\beta +2\ell^2 p_a.p_b p_a^\nu J_b^{\mu\alpha}\ell_\alpha\non\\
&& -\eta^{\mu\nu}\ell^2 p_a.p_b p_{a\alpha}J_b^{\alpha\beta}\ell_\beta +\f{1}{2}p_a^2 \eta^{\mu\nu}\ell^2 p_{b\alpha}J_{b}^{\alpha\beta}\ell_\beta  -2\ell^2 p_a^\mu p_a^\nu p_{b\alpha}J_b^{\alpha\beta}\ell_\beta\Big]\non\\
&&-\f{i}{2}(8\pi G)\sum_{a,b=1}^{M+N}\int_{\omega}^{L^{-1}}\f{d^{4}\ell}{(2\pi)^{4}}\ \Big{\lbrace}G_{r}(-\ell)\Big{\rbrace}^{3}\ (-2k.\ell) \f{1}{p_{b}.\ell -i\epsilon}\f{1}{p_{a}.\ell +i\epsilon}\non\\
&&\ \Big[2\ell^{\mu}\ell^{\nu}p_a.p_b p_{b\rho}J_a^{\rho\sigma}\ell_\sigma  -p_b^2 \ell^\mu \ell^\nu p_{a\rho}J_a^{\rho\sigma}\ell_\sigma +\f{1}{2}p_b^2 \ell^2 \eta^{\mu\nu}p_{a\rho}J_a^{\rho\sigma}\ell_{\sigma} -\ell^2 p_b^2 p_a^\mu J_a^{\nu\sigma}\ell_\sigma\non\\
&& -\ell^2 p_b^2 p_a^\nu J_a^{\mu\sigma}\ell_\sigma +2\ell^2 p_a.p_b p_b^{\mu}J_a^{\nu\sigma}\ell_\sigma +2\ell^2 p_a^\nu p_b^\mu p_{b\rho}J_a^{\rho\sigma}\ell_\sigma -\ell^2 p_a.p_b \eta^{\mu\nu}p_{b\rho}J_a^{\rho\sigma}\ell_\sigma\Big]
\ee
In the above expression the particle indices $a,b$ are dummy, so we can exchange $a\leftrightarrow b$ in the second integral above. In doing so we observe that many terms are same within the square bracket of the first and second integrands, but there is relative sign between the two integrals, hence those terms will cancel each other\footnote{Obviously for the cancelled terms between the two integrals there will be a sign difference in front of $i\epsilon$ in $\lbrace p_a.\ell \pm i\epsilon\rbrace^{-1}$ and $\lbrace p_b.\ell \pm i\epsilon\rbrace^{-1}$ but the contribution turns out to be the same as only the relative signs in front of $i\epsilon$ between the two denominator matter.\label{logic1}}. After cancelling those terms we combine the two integrals into one and get the following expression,
\be 
\mathcal{L}_1^{\mu\nu}&=&\ -\f{i}{2}(8\pi G)\sum_{a,b=1}^{M+N}\int_{\omega}^{L^{-1}}\f{d^{4}\ell}{(2\pi)^{4}}\ \Big{\lbrace}G_{r}(-\ell)\Big{\rbrace}^{2}\ (-2k.\ell) \f{1}{p_{b}.\ell -i\epsilon}\f{1}{p_{a}.\ell +i\epsilon}\non\\
&&\ \Big[ 2 p_b^\mu p_a^\nu p_{a\alpha}J_b^{\alpha\beta}\ell_\beta -2 p_a^\mu p_a^\nu p_{b\alpha}J_b^{\alpha\beta}\ell_\beta + p_a^2 p_b^\mu J_b^{\nu\sigma}\ell_\sigma + p_a^2 p_b^\nu J_b^{\mu\sigma}\ell_\sigma  -2 p_b^\nu p_a^\mu p_{a\rho}J_b^{\rho\sigma}\ell_\sigma \non\\ 
&&\ +2p_a.p_b p_a^\nu J_b^{\mu\alpha}\ell_\alpha -2p_a.p_b p_a^\mu J_b^{\nu\alpha}\ell_\alpha\Big]\label{L1}
\ee
Fourth to sixth lines after the equality in eq.\eqref{Thomegalnomega} takes the following form after removing the terms containing $(p_a.\ell)^2$ and $p_b.\ell$ in the numerator,
\be
\mathcal{L}_2^{\mu\nu}&=&\ \f{i}{2}(8\pi G)\sum_{a,b=1}^{M+N}\int_{\omega}^{L^{-1}}\f{d^{4}\ell}{(2\pi)^{4}}\ \Big{\lbrace}G_{r}(-\ell)\Big{\rbrace}^{2}\ (p_a.k) \f{1}{p_{b}.\ell -i\epsilon}\f{1}{(p_{a}.\ell +i\epsilon)^2}\non\\
&&\Big[ 2\ell^{\mu}\ell^{\nu}p_a.p_b p_{a\alpha}J_b^{\alpha\beta}\ell_{\beta}-p_a^2 \ell^\mu \ell^\nu p_{b\alpha}J_b^{\alpha\beta}\ell_\beta +2\ell^2 p_b^\mu p_a^\nu p_{a\alpha}J_b^{\alpha\beta}\ell_\beta +2\ell^2 p_a.p_b p_a^\nu J_b^{\mu\alpha}\ell_\alpha\non\\
&& -\eta^{\mu\nu}\ell^2 p_a.p_b p_{a\alpha}J_b^{\alpha\beta}\ell_\beta +\f{1}{2}p_a^2 \eta^{\mu\nu}\ell^2 p_{b\alpha}J_{b}^{\alpha\beta}\ell_\beta  -2\ell^2 p_a^\mu p_a^\nu p_{b\alpha}J_b^{\alpha\beta}\ell_\beta +2\ell^\mu p_a^\nu p_a.\ell p_{b\alpha}J_b^{\alpha\beta}\ell_\beta\Big]\non\\
&&-\f{i}{2}(8\pi G)\sum_{a,b=1}^{M+N}\int_{\omega}^{L^{-1}}\f{d^{4}\ell}{(2\pi)^{4}}\ \Big{\lbrace}G_{r}(-\ell)\Big{\rbrace}^{2}\ (p_a.k) \f{1}{p_{b}.\ell -i\epsilon}\f{1}{(p_{a}.\ell +i\epsilon)^2}\non\\
&&\ \Big[2\ell^{\mu}\ell^{\nu}p_a.p_b p_{b\rho}J_a^{\rho\sigma}\ell_\sigma  -p_b^2 \ell^\mu \ell^\nu p_{a\rho}J_a^{\rho\sigma}\ell_\sigma +\f{1}{2}p_b^2 \ell^2 \eta^{\mu\nu}p_{a\rho}J_a^{\rho\sigma}\ell_{\sigma} -\ell^2 p_b^2 p_a^\mu J_a^{\nu\sigma}\ell_\sigma\non\\
&& -\ell^2 p_b^2 p_a^\nu J_a^{\mu\sigma}\ell_\sigma +2\ell^2 p_a.p_b p_b^{\mu}J_a^{\nu\sigma}\ell_\sigma +2\ell^2 p_a^\nu p_b^\mu p_{b\rho}J_a^{\rho\sigma}\ell_\sigma -\ell^2 p_a.p_b \eta^{\mu\nu}p_{b\rho}J_a^{\rho\sigma}\ell_\sigma\non\\
&&\ +p_b^2 \ell^\mu p_a.\ell J_a^{\nu\alpha}\ell_\alpha\Big]\label{L2}
\ee
Seventh to nineth lines after the equality in eq.\eqref{Thomegalnomega} takes the following form after removing the terms containing $p_a.\ell$ and $p_b.\ell$ in the numerator,
\be
\mathcal{L}_3^{\mu\nu}&=&\ \f{i}{2}(8\pi G)\sum_{a,b=1}^{M+N}\int_{\omega}^{L^{-1}}\f{d^{4}\ell}{(2\pi)^{4}}\ \Big{\lbrace}G_{r}(-\ell)\Big{\rbrace}^{2}\  \f{1}{p_{b}.\ell -i\epsilon}\f{1}{p_{a}.\ell +i\epsilon}\non\\
&&\Big[-2\ell^{\mu}k^\nu p_a.p_b p_{a\alpha}J_{b}^{\alpha\beta}\ell_{\beta}+p_a^2 k^\nu \ell^\mu p_{b\alpha}J_b^{\alpha\beta}\ell_\beta -4k^\mu \ell^\nu p_a.p_b p_{a\alpha}J_b^{\alpha\beta}\ell_\beta \non\\
&&\ +2p_a^2 k^\mu \ell^\nu p_{b\alpha}J_b^{\alpha\beta}\ell_\beta +2p_a.p_b p_a^\mu \ell^\nu k_\alpha J_b^{\alpha\beta}\ell_\beta +2p_b.k p_a^\mu \ell^\nu p_{a\alpha}J_b^{\alpha\beta}\ell_\beta\non\\
&&\ -2p_a.k p_a^\mu \ell^\nu p_{b\alpha}J_b^{\alpha\beta}\ell_\beta -p_a^2 p_b^\mu \ell^\nu k_\alpha J_b^{\alpha\beta}\ell_\beta -p_a^2 p_b.k \ell^\nu J_b^{\mu\alpha}\ell_\alpha\non\\
&&\ +2p_a.p_b p_a^\nu \ell^\mu k_\alpha J_b^{\alpha\beta}\ell_\beta +2p_b.k p_a^\nu \ell^\mu p_{a\alpha}J_b^{\alpha\beta}\ell_\beta -2p_a.k p_a^\nu \ell^\mu p_{b\alpha}J_b^{\alpha\beta}\ell_\beta \non\\
&&\ +6k.\ell p_a^\mu p_a^\nu p_{b\alpha}J_b^{\alpha\beta}\ell_\beta -\f{5}{2}k.\ell p_a^2 \eta^{\mu\nu}p_{b\alpha}J_b^{\alpha\beta}\ell_\beta -2k.\ell p_a.p_b p_a^\mu J_b^{\nu\alpha}\ell_\alpha\non\\
&&\ -2k.\ell p_b^\nu p_a^\mu p_{a\alpha}J_b^{\alpha\beta}\ell_\beta +k.\ell p_a^2 p_b^\mu J_b^{\nu\alpha}\ell_\alpha +k.\ell p_a^2 p_b^\nu J_b^{\mu\alpha}\ell_\alpha\non\\
&& -4k.\ell p_b^\mu p_a^\nu p_{a\alpha}J_b^{\alpha\beta}\ell_\beta -4k.\ell p_a.p_b p_a^\nu J_b^{\mu\alpha}\ell_\alpha +5k.\ell \eta^{\mu\nu} p_a.p_b p_{a\alpha}J_b^{\alpha\beta}\ell_\beta \Big]\non\\
&& -\f{i}{2}(8\pi G)\sum_{a,b=1}^{M+N}\int_{\omega}^{L^{-1}}\f{d^{4}\ell}{(2\pi)^{4}}\ \Big{\lbrace}G_{r}(-\ell)\Big{\rbrace}^{2}\  \f{1}{p_{b}.\ell -i\epsilon}\f{1}{p_{a}.\ell +i\epsilon}\non\\
&&\ \Big[-2p_a.p_b \ell^\mu k^\nu p_{b\alpha}J_a^{\alpha\beta}\ell_\beta +p_b^2 \ell^\mu k^\nu p_{a\alpha}J_a^{\alpha\beta}\ell_\beta -4k^\mu \ell^\nu p_a.p_b p_{b\alpha}J_a^{\alpha\beta}\ell_\beta\non\\
&&\ +2p_b^2 k^\mu \ell^\nu p_{a\alpha}J_a^{\alpha\beta}\ell_\beta +2p_b.k p_a.p_b \ell^\nu J_a^{\mu\alpha}\ell_\alpha +2p_b.k p_a^\mu \ell^\nu p_{b\alpha}J_a^{\alpha\beta}\ell_\beta\non\\
&&\ -2p_b.k p_b^\mu \ell^\nu p_{a\alpha}J_a^{\alpha\beta}\ell_\beta -p_b^2 p_a.k \ell^\nu J_a^{\mu\alpha}\ell_\alpha -p_b^2 \ell^\nu p_a^\mu k_\alpha J_a^{\alpha\beta}\ell_\beta \non\\
&&\ +2p_a.p_b p_b.k \ell^\mu J_a^{\nu\alpha}\ell_\alpha +2p_a^\nu \ell^\mu p_b.k p_{b\alpha}J_a^{\alpha\beta}\ell_\beta -p_b^2 p_a.k \ell^\mu J_a^{\nu\alpha}\ell_\alpha \non\\
&&\ -p_b^2 \ell^\mu p_a^\nu k_\alpha J_a^{\alpha\beta}\ell_\beta +2p_b^\mu p_b^\nu k.\ell p_{a\alpha}J_a^{\alpha\beta}\ell_\beta -\f{5}{2}p_b^2 k.\ell \eta^{\mu\nu}p_{a\alpha}J_a^{\alpha\beta}\ell_\beta\non\\
&&\ -4k.\ell p_a.p_b p_b^{\mu}J_a^{\nu\alpha}\ell_\alpha -4k.\ell p_b^\mu p_a^\nu p_{b\alpha}J_a^{\alpha\beta}\ell_\beta -2k.\ell p_b^\nu p_a.p_b J_a^{\mu\alpha}\ell_\alpha\non\\
&&\ -2k.\ell p_b^\nu p_a^\mu p_{b\alpha}J_a^{\alpha\beta}\ell_\beta +3k.\ell p_b^2 p_a^\nu J_a^{\mu\alpha}\ell_\alpha +3k.\ell p_b^2 p_a^\mu J_a^{\nu\alpha}\ell_\alpha\non\\
&&\ +5\eta^{\mu\nu}p_a.p_b k.\ell p_{b\alpha}J_a^{\alpha\beta}\ell_\beta \Big]
\ee
In the above expression the particle indices $a,b$ are dummy, so we can exchange $a\leftrightarrow b$ in the second integral above. In doing so we observe that many terms are same within the square bracket of the first and second integrands, but there is relative sign between the two integrals, hence those terms will cancel each other and we get\footnote{Here also the same logic holds as described in footnote-\eqref{logic1}.},
\be
\mathcal{L}_3^{\mu\nu}&=&\ \f{i}{2}(8\pi G)\sum_{a,b=1}^{M+N}\int_{\omega}^{L^{-1}}\f{d^{4}\ell}{(2\pi)^{4}}\ \Big{\lbrace}G_{r}(-\ell)\Big{\rbrace}^{2}\  \f{1}{p_{b}.\ell -i\epsilon}\f{1}{p_{a}.\ell +i\epsilon}\non\\
&&\Big[ 2p_a.p_b p_a^\mu \ell^\nu k_\alpha J_b^{\alpha\beta}\ell_\beta +2p_b.k p_a^\mu \ell^\nu p_{a\alpha}J_b^{\alpha\beta}\ell_\beta +2p_a.p_b p_a^\nu \ell^\mu k_\alpha J_b^{\alpha\beta}\ell_\beta +2p_b.k p_a^\nu \ell^\mu p_{a\alpha}J_b^{\alpha\beta}\ell_\beta\non\\
&& -2p_a.k p_a^\nu \ell^\mu p_{b\alpha}J_b^{\alpha\beta}\ell_\beta +4k.\ell p_a^\mu p_a^\nu p_{b\alpha}J_b^{\alpha\beta}\ell_\beta +2k.\ell p_a.p_b p_a^\mu J_b^{\nu\alpha}\ell_\alpha  +2k.\ell p_b^\nu p_a^\mu p_{a\alpha}J_b^{\alpha\beta}\ell_\beta\non\\
&& -2k.\ell p_a^2 p_b^\mu J_b^{\nu\alpha}\ell_\alpha -2k.\ell p_a^2 p_b^\nu J_b^{\mu\alpha}\ell_\alpha -2k.\ell p_b^\mu p_a^\nu p_{a\alpha}J_b^{\alpha\beta}\ell_\beta -2k.\ell p_a.p_b p_a^\nu J_b^{\mu\alpha}\ell_\alpha \non \\
&& -2p_a.k p_a.p_b \ell^\nu J_b^{\mu\alpha}\ell_\alpha -2p_a.k p_b^\mu \ell^\nu p_{a\alpha}J_b^{\alpha\beta}\ell_\beta -2p_a.k p_a.p_b \ell^\mu J_b^{\nu\alpha}\ell_\alpha -2p_b^\nu \ell^\mu p_a.k p_{a\alpha}J_b^{\alpha\beta}\ell_\beta\non\\
&& +p_a^2 p_b.k \ell^\mu J_{b}^{\nu\alpha}\ell_\alpha + p_a^2 \ell^\mu p_b^\nu k_\alpha J_b^{\alpha\beta}\ell_\beta \Big]\label{L3}
\ee
The last two lines in eq.\eqref{Thomegalnomega} represents $\mathcal{L}_4^{\mu\nu}$, which takes the following form after contracting the indices of the last line of eq.\eqref{Thomegalnomega} and removing the terms containing $p_a.\ell$ and $p_b.\ell$ in the numerator, as they give vanishing contribution when we perform the $\ell$ integration.
\be
\mathcal{L}_4^{\mu\nu}&=&\ \f{i}{2}(8\pi G)\sum_{a,b=1}^{M+N}\int_{\omega}^{L^{-1}}\f{d^{4}\ell}{(2\pi)^{4}}\ \Big{\lbrace}G_{r}(-\ell)\Big{\rbrace}^{2}\ \f{1}{p_{b}.\ell -i\epsilon}\f{1}{p_{a}.\ell +i\epsilon}\non\\
&&\ \Big[2\ell^{\mu}\ell^{\nu}p_a.p_b p_{b\rho}J_a^{\rho\sigma}k_\sigma + p_b^2 p_a^{\nu}\ell^{\mu}\ell_{\rho}J_a^{\rho\sigma}k_\sigma -p_b^2 \ell^\mu \ell^\nu p_{a\rho}J_a^{\rho\sigma}k_\sigma +\f{1}{2}p_b^2 \ell^2 \eta^{\mu\nu}p_{a\rho}J_a^{\rho\sigma}k_{\sigma}\non\\
&&\ -\ell^2 p_b^2 p_a^\mu J_a^{\nu\sigma}k_\sigma -\ell^2 p_b^2 p_a^\nu J_a^{\mu\sigma}k_\sigma +2\ell^2 p_a.p_b p_b^{\mu}J_a^{\nu\sigma}k_\sigma +2\ell^2 p_a^\nu p_b^\mu p_{b\rho}J_a^{\rho\sigma}k_\sigma \non\\
&&\ -\ell^2 p_a.p_b \eta^{\mu\nu}p_{b\rho}J_a^{\rho\sigma}k_\sigma\Big]\label{L4}
\ee
Let us first evaluate the sum of the contributions in eq.\eqref{L1} and eq.\eqref{L3} ,
\be
&&\mathcal{L}_1^{\mu\nu}+\mathcal{L}_3^{\mu\nu}\non\\
&=&\ \f{i}{2}(8\pi G)\sum_{a,b=1}^{M+N}\int_{\omega}^{L^{-1}}\f{d^{4}\ell}{(2\pi)^{4}}\ \Big{\lbrace}G_{r}(-\ell)\Big{\rbrace}^{2}\  \f{1}{p_{b}.\ell -i\epsilon}\f{1}{p_{a}.\ell +i\epsilon}\non\\
&&\ \Big[ 2k.\ell p_b^\mu p_a^\nu p_{a\alpha}J_b^{\alpha\beta}\ell_\beta  -2k.\ell p_b^\nu p_a^\mu p_{a\rho}J_b^{\rho\sigma}\ell_\sigma + 2p_a.p_b p_a^\mu \ell^\nu k_\alpha J_b^{\alpha\beta}\ell_\beta +2p_b.k p_a^\mu \ell^\nu p_{a\alpha}J_b^{\alpha\beta}\ell_\beta\non\\
&& +2p_a.p_b p_a^\nu \ell^\mu k_\alpha J_b^{\alpha\beta}\ell_\beta +2p_b.k p_a^\nu \ell^\mu p_{a\alpha}J_b^{\alpha\beta}\ell_\beta -2p_a.k p_a^\nu \ell^\mu p_{b\alpha}J_b^{\alpha\beta}\ell_\beta -2k.\ell p_a.p_b p_a^\mu J_b^{\nu\alpha}\ell_\alpha  \non\\
&&  +2k.\ell p_a.p_b p_a^\nu J_b^{\mu\alpha}\ell_\alpha -2p_a.k p_a.p_b \ell^\nu J_b^{\mu\alpha}\ell_\alpha -2p_a.k p_b^\mu \ell^\nu p_{a\alpha}J_b^{\alpha\beta}\ell_\beta -2p_a.k p_a.p_b \ell^\mu J_b^{\nu\alpha}\ell_\alpha\non\\
&& -2p_b^\nu \ell^\mu p_a.k p_{a\alpha}J_b^{\alpha\beta}\ell_\beta +p_a^2 p_b.k \ell^\mu J_{b}^{\nu\alpha}\ell_\alpha + p_a^2 \ell^\mu p_b^\nu k_\alpha J_b^{\alpha\beta}\ell_\beta \Big]
\ee
Now using the result of the integral given in eq.\eqref{J1} we get,
\be
&&\mathcal{L}_1^{\mu\nu}+\mathcal{L}_3^{\mu\nu}\non\\
&=& \f{iG}{2}\sum_{\substack{a,b=1 \\ b\neq a\\ \eta_a \eta_b =1}}^{M+N}\f{1}{\big[(p_a.p_b)^2 -p_a^2 p_b^2\big]^{3/2}}\ \ln\big\lbrace L(\omega +i\epsilon \eta_a)\big\rbrace \non\\
&& \Big[ 2p_a.p_b p_a.k p_b^\mu p_a^\nu p_{a\alpha}J_b^{\alpha\beta}p_{b\beta}-4p_a.p_b p_a.k p_b^\nu p_a^\mu p_{a\alpha}J_b^{\alpha\beta}p_{b\beta}\non\\
&&-4p_a^2 p_b.k p_b^\mu p_a^\nu p_{a\alpha}J_b^{\alpha\beta}p_{b\beta}-4(p_a.p_b)^2 p_b^\mu p_a^\nu p_{a\alpha}J_b^{\alpha\beta}k_\beta +2p_a^2 p_b^2 p_b^\mu p_a^\nu p_{a\alpha}J_b^{\alpha\beta}k_\beta \non\\
&& -p_a^2 p_b^2 p_a^\mu p_b^\nu p_{a\alpha}J_b^{\alpha\beta}k_\beta -4(p_a.p_b)^2 p_a^\mu p_a^\nu p_{b\alpha}J_b^{\alpha\beta}k_\beta +4p_a.p_b p_b^2 p_a^\mu p_a^\nu p_{a\alpha}J_b^{\alpha\beta}k_\beta\non\\
&& +p_a.p_b p_a^2 p_a^\mu p_b^\nu p_{b\alpha}J_b^{\alpha\beta}k_{\beta}+2p_a.p_b p_a^2 p_b^\mu p_a^\nu p_{b\alpha}J_b^{\alpha\beta}k_\beta  +4(p_a.p_b)^2 p_a.k p_a^\mu p_{b\alpha}J_b^{\alpha\nu}\non\\
&& +4p_a.p_b p_b.k p_a^\mu p_a^\nu p_{a\alpha}J_b^{\alpha\beta}p_{b\beta} +3p_b.k p_a^2 p_b^2 p_a^\mu p_{a\alpha}J_b^{\alpha\nu}-4p_b.k (p_a.p_b)^2 p_a^\nu p_{a\alpha}J_b^{\alpha\mu}\non\\
&& +2p_b.k  p_a^2 p_b^2 p_a^\nu p_{a\alpha}J_b^{\alpha\mu}+2p_a.k p_b^2 p_a^\mu p_a^\nu p_{b\alpha}J_b^{\alpha\beta}p_{a\beta}-3p_a^2 p_a.p_b p_b.k p_a^\mu p_{b\alpha}J_b^{\alpha\nu}\non\\
&&  +2p_a^2 p_a.p_b p_b.k p_a^\nu p_{b\alpha}J_b^{\alpha\mu} -4p_a.p_b p_b^2 p_a.k p_a^\mu p_{a\alpha}J_b^{\alpha\nu} +4p_a.p_b \lbrace (p_a.p_b)^2 -p_a^2 p_b^2\rbrace p_a^\mu J_b^{\nu\alpha}k_\alpha\non\\
&& +4p_a.k (p_a.p_b)^2 p_b^\nu p_{a\alpha}J_b^{\alpha\mu}-2p_a.k p_a^2 p_b^2 p_b^\nu p_{a\alpha}J_b^{\alpha\mu} -2p_a.k p_a.p_b p_a^2 p_b^\nu p_{b\alpha}J_b^{\alpha\mu}\non\\
&& +4p_a.k (p_a.p_b)^2 p_b^\mu p_{a\alpha}J_b^{\alpha\nu} -2p_a.k p_a^2 p_b^2 p_b^\mu p_{a\alpha}J_b^{\alpha\nu} -2p_a.k p_a.p_b p_a^2 p_b^\mu p_{b\alpha}J_b^{\alpha\nu}\non\\
&& +4p_a.k p_a^2 p_b^\mu p_b^\nu p_{a\alpha}J_b^{\alpha\beta}p_{b\beta} -p_a^2 p_a.p_b p_b.k p_b^\mu p_{a\alpha}J_b^{\alpha\nu}+(p_a^2)^2 p_b.k p_b^\mu p_{b\alpha}J_b^{\alpha\nu}\non\\
&& +p_a^2 p_b.k \lbrace(p_a.p_b)^2 -p_a^2 p_b^2\rbrace J_b^{\mu\nu}
+p_a^2 \lbrace(p_a.p_b)^2 -p_a^2 p_b^2\rbrace \  p_b^\nu J_b^{\mu\rho}k_\rho \non\\
&&  -p_a^2 p_a.p_b p_b^\mu p_b^\nu p_{a\alpha}J_b^{\alpha\beta}k_\beta +(p_a^2)^2 p_b^\mu p_b^\nu p_{b\alpha}J_b^{\alpha\beta}k_\beta +2p_a.k \lbrace(p_a.p_b)^2 -p_a^2 p_b^2\rbrace  p_a^\nu p_{b\alpha}J_b^{\alpha\mu}\Big]\non\\
\ee
In the above expression we substitute $J_{b}^{\mu\nu}=r_b^\mu p_{b}^\nu -r_b^\nu p_b^\mu +\Sigma_b^{\mu\nu}$ and simplify using SSC, $p_{b\mu}\Sigma_b^{\mu\nu}=0$. Since $a,b$ are dummy indices we interchange them and then the full expression can be written in terms of $r_a$ and $\Sigma_a$. After all these steps finally we get,
\be
&&\mathcal{L}_1^{\mu\nu}+\mathcal{L}_3^{\mu\nu}\non\\
&=& \f{iG}{2}\sum_{\substack{a,b=1 \\ b\neq a\\ \eta_a \eta_b =1}}^{M+N}\f{1}{\big[(p_a.p_b)^2 -p_a^2 p_b^2\big]^{3/2}}\ \ln\big\lbrace L(\omega +i\epsilon \eta_a)\big\rbrace \non\\
&&\Big[ 2p_a.p_b p_b.k p_a^\mu p_b^\nu p_b.r_a p_a^2 +8p_b^2 p_a.k p_a^\mu p_b^\nu p_a.p_b p_a.r_a -8(p_a.p_b)^2 p_a^\mu p_b^\nu p_b.r_a p_a.k \non\\
&& +4(p_a.p_b)^3 p_a^\mu p_b^\nu r_a.k -4p_a^2 p_b^2 p_a^\mu p_b^\nu p_a.p_b r_a.k  \non\\
&& -2p_a^2 p_b^2 p_b.k p_b^\nu p_a^\mu p_a.r_a -8p_a.p_b p_b.k p_a^\nu p_b^\mu p_b.r_a p_a^2 +8(p_a.p_b)^2 p_b.k p_a^\nu p_b^\mu p_a.r_a \non\\
&& +2p_a^2 p_b^2 p_b^\mu p_a^\nu p_b.r_a p_a.k +4p_a^2 p_b^2 p_b^\mu p_a^\nu p_a.p_b r_a.k -2p_a.p_b p_b^2 p_b^\mu p_a^\nu p_a.r_a p_a.k\non\\
&& -4(p_a.p_b)^3 p_b^\mu p_a^\nu r_a.k -8(p_a.p_b)^2 p_b^\mu p_b^\nu p_a.r_a p_a.k +8p_a.p_b p_a^2 p_b^\mu p_b^\nu p_b.r_a p_a.k \non\\
&& +2p_b.k p_a^2 p_b^\mu p_b^\nu p_a.r_a p_a.p_b -2p_b.k (p_a^2)^2 p_b^\mu p_b^\nu p_b.r_a +4p_a.p_b\lbrace (p_a.p_b)^2 -p_a^2 p_b^2\rbrace p_a.k p_b^\mu r_a^\nu \non\\
&& +4p_a.k (p_a.p_b)^3 p_b^\nu r_a^\mu -4p_a.k p_a^2 p_b^2 p_a.p_b p_b^\nu r_a^\mu -2p_b.k \lbrace (p_a.p_b)^2 -p_a^2 p_b^2 \rbrace p_a^2 p_b^\nu r_a^\mu \non\\
&& +8p_b.k (p_a.p_b)^2 p_a^\nu p_b.r_a p_a^\mu -8p_b.k p_a.p_b p_b^2 p_a.r_a p_a^\mu p_a^\nu -2p_b^2 p_a.p_b p_a.k p_a^\mu p_a^\nu p_b.r_a \non\\
&& +2(p_b^2)^2 p_a.k p_a.r_a p_a^\mu p_a^\nu -4p_b.k (p_a.p_b)^3 p_a^\nu r_a^\mu +4p_b.k p_a^2 p_b^2 p_a.p_b p_a^\nu r_a^\mu \non\\
&& +p_b^2 \lbrace (p_a.p_b)^2 -p_a^2 p_b^2 \rbrace p_a.k p_a^\nu r_a^\mu -4p_b.k (p_a.p_b)^3 p_a^\mu r_a^\nu +4p_b.k p_a^2 p_b^2 p_a.p_b p_a^\mu r_a^\nu \non\\
&& +p_b^2 \lbrace (p_a.p_b)^2 -p_a^2 p_b^2 \rbrace p_a.k p_a^\mu r_a^\nu -4(p_a.p_b)^2 p_a^\mu p_b^\nu p_{b\alpha}\Sigma_a^{\alpha\beta}k_\beta +2p_a^2 p_b^2 p_a^\mu p_b^\nu p_{b\alpha}\Sigma_a^{\alpha\beta}k_\beta\non\\
&& -p_a^2 p_b^2 p_b^\mu p_a^\nu p_{b\alpha}\Sigma_a^{\alpha\beta}k_\beta +4p_a.p_b p_a^2 p_b^\mu p_b^\nu p_{b\alpha}\Sigma_a^{\alpha\beta}k_\beta +3p_a.k p_a^2 p_b^2 p_b^\mu p_{b\alpha}\Sigma_a^{\alpha\nu}\non\\
&& -4p_a.k (p_a.p_b)^2 p_b^\nu p_{b\alpha}\Sigma_a^{\alpha\mu} +2p_a.k p_a^2 p_b^2 p_b^\nu p_{b\alpha}\Sigma_a^{\alpha\mu} -4p_a.p_b p_a^2 p_b.k p_b^\mu p_{b\alpha}\Sigma_a^{\alpha\nu}\non\\
&& +4p_a.p_b\lbrace (p_a.p_b)^2 -p_a^2 p_b^2 \rbrace p_b^\mu \Sigma_a^{\nu\alpha}k_\alpha +4p_b.k (p_a.p_b)^2 p_a^\nu p_{b\alpha}\Sigma_a^{\alpha\mu} -2p_b.k p_a^2 p_b^2 p_a^\nu p_{b\alpha}\Sigma_a^{\alpha\mu}\non\\
&& +4p_b.k (p_a.p_b)^2 p_a^\mu p_{b\alpha}\Sigma_a^{\alpha\nu} -2p_b.k p_a^2 p_b^2 p_a^\mu p_{b\alpha}\Sigma_a^{\alpha\nu} -p_b^2 p_a.p_b p_a.k p_a^\mu p_{b\alpha}\Sigma_a^{\alpha\nu} \non\\
&& +p_b^2 p_a.k \lbrace (p_a.p_b)^2 -p_a^2 p_b^2 \rbrace J_a^{\mu\nu} +p_b^2 \lbrace (p_a.p_b)^2 -p_a^2 p_b^2 \rbrace p_a^\nu \Sigma_a^{\mu\rho}k_\rho -p_b^2 p_a.p_b p_a^\mu p_a^\nu p_{b\alpha}\Sigma_{a}^{\alpha\beta}k_\beta \Big]\label{L1L3final}\non\\
\ee
From eq.\eqref{L4} using the result of the integration \eqref{J1}, we get
\be
\mathcal{L}_4^{\mu\nu}&=&\ \f{iG}{2}\sum_{\substack{a,b=1 \\ b\neq a\\ \eta_a \eta_b =1}}^{M+N}\f{1}{\big[(p_a.p_b)^2 -p_a^2 p_b^2\big]^{3/2}}\ln\big\lbrace L(\omega +i\epsilon \eta_a)\big\rbrace \Big[ 2(p_a.p_b)^2 p_{b\rho}J_a^{\rho\sigma}k_\sigma (p_a^\mu p_b^\nu +p_a^\nu p_b^\mu)\non\\
&&\ -2p_a.p_b p_a^2 p_{b\rho}J_a^{\rho\sigma}k_\sigma p_b^\mu p_b^\nu -p_a.p_b p_b^2 p_{b\rho}J_a^{\rho\sigma}k_\sigma p_a^\mu p_a^\nu -p_a^2 p_b^2 p_{b\rho}J_a^{\rho\sigma}k_\sigma p_a^\nu p_b^\mu \non\\
&&\ -p_a.p_b p_b^2 p_{a\rho}J_a^{\rho\sigma}k_\sigma p_a^\mu p_b^\nu +p_a^2 p_b^2 p_{a\rho}J_a^{\rho\sigma}k_\sigma p_b^\mu p_b^\nu \Big]\non\\
&& -(iG)\ \sum_{\substack{a,b=1 \\ b\neq a\\ \eta_a \eta_b =1}}^{M+N}\f{1}{\big[(p_a.p_b)^2 -p_a^2 p_b^2\big]^{1/2}}\ \ln\big\lbrace L(\omega +i\epsilon \eta_a)\big\rbrace  \Big[-p_b^2 p_a^\mu J_a^{\nu\alpha} k_\alpha -\f{1}{2}p_b^2 p_a^\nu J_a^{\mu\alpha}k_\alpha\non\\
&&\ +2p_a.p_b p_b^\mu J_a^{\nu\alpha} k_\alpha +2p_a^\nu p_b^\mu p_{b\rho}J_a^{\rho\sigma}k_\sigma \Big]
\ee
Now substituting $J_a^{\mu\nu}=r_a^\mu p_a^\nu -r_a^\nu p_a^\mu +\Sigma_a^{\mu\nu}$ in the above expression and using SSC : $p_{a\mu}\Sigma_a^{\mu\nu}=0$, we get
\be
\mathcal{L}_4^{\mu\nu}&=&\ \f{iG}{2}\sum_{\substack{a,b=1 \\ b\neq a\\  \eta_a \eta_b =1}}^{M+N}\f{1}{\big[(p_a.p_b)^2 -p_a^2 p_b^2\big]^{3/2}}\ln\big\lbrace L(\omega +i\epsilon \eta_a)\big\rbrace \Big[ 2(p_a.p_b)^2 p_b.r_a p_a.k (p_a^\mu p_b^\nu +p_a^\nu p_b^\mu)\non\\
&& -2(p_a.p_b)^3 r_a.k (p_a^\mu p_b^\nu +p_a^\nu p_b^\mu) -2p_a.p_b p_a^2 p_b.r_a p_a.k p_b^\mu p_b^\nu +2(p_a.p_b)^2 p_a^2 r_a.k p_b^\mu p_b^\nu \non\\
&& -p_a.p_b p_b^2 p_b.r_a p_a.k p_a^\mu p_a^\nu +(p_a.p_b)^2 p_b^2 r_a.k p_a^\mu p_a^\nu -p_a^2 p_b^2 p_b.r_a p_a.k p_a^\nu p_b^\mu +p_a^2 p_b^2 p_a.p_b r_a.k p_a^\nu p_b^\mu \non\\
&& -p_a.p_b p_b^2 p_a.r_a p_a.k p_a^\mu p_b^\nu +p_a.p_b p_a^2 p_b^2 r_a.k p_a^\mu p_b^\nu +p_a^2 p_b^2 p_a.r_a p_a.k p_b^\mu p_b^\nu -(p_a^2)^2 p_b^2 r_a.k p_b^\mu p_b^\nu \non\\
&& +2(p_a.p_b)^2 p_{b\rho}\Sigma_a^{\rho\sigma}k_\sigma (p_a^\mu p_b^\nu +p_a^\nu p_b^\mu) -2p_a.p_b p_a^2 p_{b\rho}\Sigma_a^{\rho\sigma}k_\sigma p_b^\mu p_b^\nu -p_a.p_b p_b^2 p_{b\rho}\Sigma_a^{\rho\sigma}k_\sigma p_a^\mu p_a^\nu \non\\
&&-p_a^2 p_b^2 p_{b\rho}\Sigma_a^{\rho\sigma}k_\sigma p_a^\nu p_b^\mu \Big]\non\\
&&-(iG)\ \sum_{\substack{a,b=1 \\ b\neq a\\ \eta_a \eta_b =1}}^{M+N}\f{1}{\big[(p_a.p_b)^2 -p_a^2 p_b^2\big]^{1/2}}\ \ln\big\lbrace L(\omega +i\epsilon \eta_a)\big\rbrace  \Big[-p_b^2 p_a^\mu r_a^\nu p_a.k +\f{3}{2}p_b^2 p_a^\mu p_a^\nu r_a.k\non\\
&&\ -\f{1}{2}p_b^2 p_a^\nu r_a^\mu p_a.k  +2p_a.p_b p_b^\mu r_a^\nu p_a.k -4p_a.p_b p_b^\mu p_a^\nu r_a.k +2p_a^\nu p_b^\mu p_b.r_a p_a.k\non\\
&&\   -p_b^2 p_a^\mu \Sigma_a^{\nu\alpha} k_\alpha -\f{1}{2}p_b^2 p_a^\nu \Sigma_a^{\mu\alpha}k_\alpha +2p_a.p_b p_b^\mu \Sigma_a^{\nu\alpha} k_\alpha +2p_a^\nu p_b^\mu p_{b\rho}\Sigma_a^{\rho\sigma}k_\sigma \Big]\label{L4final}
\ee
Contribution from eq.\eqref{L2} after using the results of integrations \eqref{J1},\eqref{J2} and \eqref{J3} takes the following form,
\be
\mathcal{L}_2^{\mu\nu}&=&\ \f{iG}{2}\sum_{\substack{a,b=1 \\ b\neq a\\ \eta_a \eta_b =1}}^{M+N}\f{p_a.k}{\big[(p_a.p_b)^2 -p_a^2 p_b^2\big]^{5/2}}\ln\big\lbrace L(\omega +i\epsilon \eta_a)\big\rbrace \non\\
&& \Big[ 6(p_a.p_b)^3 p_a^\mu p_b^\nu p_{a\alpha}J_b^{\alpha\beta}p_{b\beta}-3(p_a.p_b)^2 p_a^2 p_b^\mu p_b^\nu p_{a\alpha}J_b^{\alpha\beta}p_{b\beta}-6(p_a.p_b)^2 p_b^2 p_a^\mu p_a^\nu p_{a\alpha}J_b^{\alpha\beta}p_{b\beta}\non\\
&&+3p_a.p_b p_a^2 p_b^2 p_b^\mu p_a^\nu p_{a\alpha}J_b^{\alpha\beta}p_{b\beta}-3p_a.p_b p_a^2 p_b^2 p_a^\mu p_b^\nu p_{a\alpha}J_b^{\alpha\beta}p_{b\beta}+3p_a^2 (p_b^2)^2 p_a^\mu p_a^\nu p_{a\alpha}J_b^{\alpha\beta}p_{b\beta} \Big]\non\\
&& -\f{iG}{2}\sum_{\substack{a,b=1 \\ b\neq a\\ \eta_a \eta_b =1}}^{M+N}\f{p_a.k}{\big[(p_a.p_b)^2 -p_a^2 p_b^2\big]^{5/2}}\ln\big\lbrace L(\omega +i\epsilon \eta_a)\big\rbrace \non\\
&&\Big[ 3p_b^2 (p_a.p_b)^2 p_a^\mu p_b^\nu p_{a\alpha}J_a^{\alpha\beta}p_{b\beta}+3p_a^2 p_b^2 p_a.p_b p_b^\mu p_b^\nu p_{a\alpha}J_a^{\alpha\beta}p_{b\beta}-3(p_b^2)^2 p_a.p_b p_a^\mu p_a^\nu p_{a\alpha}J_a^{\alpha\beta}p_{b\beta}\non\\
&& -3(p_b^2)^2 p_a^2 p_b^\mu p_a^\nu p_{a\alpha}J_a^{\alpha\beta}p_{b\beta}+6(p_a.p_b)^3 p_b^\mu p_b^\nu p_{b\alpha}J_a^{\alpha\beta}p_{a\beta}-6(p_a.p_b)^2 p_b^2 p_b^\mu p_a^\nu p_{b\alpha}J_a^{\alpha\beta}p_{a\beta}\Big]\non\\
&& - \f{iG}{2}\sum_{\substack{a,b=1 \\ b\neq a\\ \eta_a \eta_b =1}}^{M+N}\f{p_a.k}{\big[(p_a.p_b)^2 -p_a^2 p_b^2\big]^{3/2}}\ln\big\lbrace L(\omega +i\epsilon \eta_a)\big\rbrace\ \Big[ 2p_a.p_b p_a^\mu p_b^\nu p_{a\alpha}J_b^{\alpha\beta}p_{b\beta}\non\\
&& +p_a^2 p_b^\mu p_b^\nu p_{a\alpha}J_b^{\alpha\beta}p_{b\beta}+2(p_a.p_b)^2 p_b^{\nu}p_{a\alpha}J_b^{\alpha\mu}-p_a^2 p_a.p_b p_b^\nu p_{b\alpha}J_b^{\alpha\mu}+2p_a.p_b p_b^2 p_a^\nu p_{a\alpha}J_b^{\alpha\mu}\non\\
&& -p_a^2 p_b^2 p_a^\nu p_{b\alpha}J_b^{\alpha\mu} +2(p_a.p_b)^2 p_b^\mu p_{a\alpha}J_b^{\alpha\nu} -p_a^2 p_a.p_b p_b^{\mu}p_{b\alpha}J_b^{\alpha\nu}-2p_a.p_b p_b^2 p_a^\mu p_{a\alpha}J_b^{\alpha\nu}\non\\
&& +p_a^2 p_b^2 p_a^\mu p_{b\alpha}J_b^{\alpha\nu}-2(p_a.p_b)^2 p_a^\nu p_{b\alpha}J_b^{\alpha\mu} -6p_b^2 p_a^\mu p_a^\nu p_{a\alpha}J_b^{\alpha\beta}p_{b\beta}+2p_a.p_b p_b^\mu p_a^\nu p_{a\alpha}J_b^{\alpha\beta}p_{b\beta} \Big]\non\\
&& + \f{iG}{2}\sum_{\substack{a,b=1 \\ b\neq a\\  \eta_a \eta_b =1}}^{M+N}\f{p_a.k}{\big[(p_a.p_b)^2 -p_a^2 p_b^2\big]^{3/2}}\ln\big\lbrace L(\omega +i\epsilon \eta_a)\big\rbrace\ \Big[ -p_b^2 p_a^\mu p_b^\nu p_{a\alpha}J_a^{\alpha\beta}p_{b\beta}\non\\
&& +6p_b^2 p_b^\mu p_a^\nu p_{a\alpha}J_a^{\alpha\beta}p_{b\beta}+2p_a.p_b p_b^\mu p_b^\nu p_{b\alpha}J_a^{\alpha\beta}p_{a\beta}+2(p_a.p_b)^2 p_b^\nu p_{b\alpha}J_a^{\alpha\mu}-p_b^2 p_a.p_b p_b^\nu p_{a\alpha}J_a^{\alpha\mu}\non\\
&& -(p_b^2)^2 p_a^\nu p_{a\alpha}J_a^{\alpha\mu}-2(p_a.p_b)^2 p_b^\mu p_{b\alpha}J_a^{\alpha\nu}+4p_b^2 p_a.p_b p_b^\mu p_{a\alpha}J_a^{\alpha\nu}-2(p_b^2)^2 p_a^\mu p_{a\alpha}J_a^{\alpha\nu}\non\\
&& +p_b^2 p_a.p_b p_a^\mu  p_{b\alpha}J_a^{\alpha\nu} -p_a^2 p_b^2 p_b^\mu p_{b\alpha}J_a^{\alpha\nu} -p_b^2\lbrace (p_a.p_b)^2 -p_a^2 p_b^2 \rbrace J_a^{\mu\nu}\Big]
\ee
Now in the above expression we substitute total angular momenta in terms of orbital and spin  and interchange $a\leftrightarrow b$ in appropriate places  to make the result $r_a$ and $\Sigma_a$ dependent. Then the above expression finally reduces to,
\be
\mathcal{L}_2^{\mu\nu}&=&\ \f{iG}{2}\sum_{\substack{a,b=1 \\ b\neq a\\ \eta_a \eta_b =1}}^{M+N}\f{1}{\big[(p_a.p_b)^2 -p_a^2 p_b^2\big]^{5/2}}\ln\big\lbrace L(\omega +i\epsilon \eta_a)\big\rbrace \non\\
&&\Big[ p_b.k \Big\lbrace 6(p_a.p_b)^3 p_b^\mu p_a^\nu p_b.r_a p_a^2 -6(p_a.p_b)^4 p_b^\mu p_a^\nu p_a.r_a -3(p_a.p_b)^2 p_a^2 p_b^2 p_a^\mu p_a^\nu p_b.r_a \non\\
&& +3(p_a.p_b)^3 p_b^2 p_a^\mu p_a^\nu p_a.r_a -6(p_a.p_b)^2 (p_a^2)^2 p_b^\mu p_b^\nu p_b.r_a +6(p_a.p_b)^3 p_a^2 p_b^\mu p_b^\nu p_a.r_a\non\\
&& +3p_a.p_b (p_a^2)^2 p_b^2 p_a^\mu p_b^\nu p_b.r_a -3(p_a.p_b)^2 p_a^2 p_b^2 p_a^\mu p_b^\nu p_a.r_a -3p_a.p_b (p_a^2)^2 p_b^2 p_b^\mu p_a^\nu p_b.r_a\non\\
&& +3(p_a.p_b)^2 p_a^2 p_b^2 p_b^\mu p_a^\nu p_a.r_a +3p_b^2 (p_a^2)^3 p_b^\mu p_b^\nu p_b.r_a -3p_b^2 (p_a^2)^2 p_a.p_b p_b^\mu p_b^\nu p_a.r_a\Big\rbrace\non\\
&& -p_a.k \Big\lbrace 3p_b^2 (p_a.p_b)^3 p_a^\mu p_b^\nu p_a.r_a -3p_b^2 (p_a.p_b)^2 p_a^\mu p_b^\nu p_a^2 p_b.r_a +3p_a^2 p_b^2 (p_a.p_b)^2 p_b^\mu p_b^\nu p_a.r_a \non\\
&& -3(p_a^2)^2 p_b^2 p_a.p_b p_b^\mu p_b^\nu p_b.r_a -3(p_b^2)^2 (p_a.p_b)^2 p_a^\mu p_a^\nu p_a.r_a +3(p_b^2)^2 p_a^2 p_a.p_b p_a^\mu p_a^\nu p_b.r_a \non\\
&& -3(p_b^2)^2 p_a^2 p_a.p_b p_b^\mu p_a^\nu p_a.r_a +3(p_a^2)^2 (p_b^2)^2 p_b^\mu p_a^\nu p_b.r_a +6(p_a.p_b)^3 p_b^\mu p_b^\nu p_a^2 p_b.r_a \non\\
&& -6(p_a.p_b)^4 p_b^\mu p_b^\nu p_a.r_a -6(p_a.p_b)^2 p_a^2 p_b^2 p_b^\mu p_a^\nu p_b.r_a +6(p_a.p_b)^3 p_b^2 p_b^\mu p_a^\nu p_a.r_a\Big\rbrace\Big]\non\\
&& - \f{iG}{2}\sum_{\substack{a,b=1 \\ b\neq a\\ \eta_a \eta_b =1}}^{M+N}\f{1}{\big[(p_a.p_b)^2 -p_a^2 p_b^2\big]^{3/2}}\ln\big\lbrace L(\omega +i\epsilon \eta_a)\big\rbrace\ \non\\
&& \Big[ p_b.k \Big\lbrace -2(p_a.p_b)^2 p_b^\mu p_a^\nu p_a.r_a +p_a^2 p_b^2 p_a^\mu p_a^\nu p_b.r_a -3p_b^2 p_a.p_b p_a^\mu p_a^\nu p_a.r_a \non\\
&& +4(p_a.p_b)^2 p_a^\nu p_b.r_a p_a^\mu -2(p_a.p_b)^3 p_a^\nu r_a^\mu +p_a^2 p_b^2 p_a.p_b p_a^\nu r_a^\mu +4p_a.p_b p_a^2 p_b^\nu p_b.r_a p_a^\mu \non\\
&& -p_a^2 p_b^2 p_b^\nu p_a.r_a p_a^\mu +(p_a^2)^2 p_b^2 p_b^\nu r_a^\mu -2(p_a.p_b)^3 p_a^\mu r_a^\nu +p_a^2 p_b^2 p_a.p_b p_a^\mu r_a^\nu \non\\
&& +2(p_a.p_b)^2 p_a^2 p_b^\mu r_a^\nu +p_a^2 p_b^2 p_b^\mu p_a.r_a p_a^\nu -(p_a^2)^2 p_b^2 p_b^\mu r_a^\nu -4(p_a.p_b)^2 p_b^\nu p_a.r_a p_a^\mu \non\\
&& -6(p_a^2)^2 p_b^\mu p_b^\nu p_b.r_a +6p_a^2 p_b^\mu p_b^\nu p_a.p_b p_a.r_a +2(p_a.p_b)^2 p_a^\nu p_{b\alpha}\Sigma_a^{\alpha\mu}\non\\
&& +2p_a.p_b p_a^2 p_b^\nu p_{b\alpha}\Sigma_a^{\alpha\mu}+2(p_a.p_b)^2 p_a^\mu p_{b\alpha}\Sigma_a^{\alpha\nu}-2p_a.p_b p_a^2 p_b^\mu p_{b\alpha}\Sigma_a^{\alpha\nu}\Big\rbrace \non\\
&& -p_a.k \Big\lbrace -2p_b^2 p_a^\mu p_b^\nu p_a.r_a p_a.p_b +p_a^2 p_b^2 p_a^\mu p_b^\nu p_b.r_a +10 p_b^2 p_b^\mu p_a^\nu p_a.r_a p_a.p_b \non\\
&& -7p_a^2 p_b^2 p_b^\mu p_a^\nu p_b.r_a +2p_a.p_b p_b^\mu p_b^\nu p_b.r_a p_a^2 -2(p_a.p_b)^2 p_b^\mu p_b^\nu p_a.r_a +2(p_a.p_b)^2 p_b^\nu p_a^\mu p_b.r_a \non\\
&& -2(p_a.p_b)^3 p_b^\nu r_a^\mu +p_a^2 p_b^2 p_a.p_b p_b^\nu r_a^\mu -3(p_b^2)^2 p_a.r_a p_a^\mu p_a^\nu +(p_b^2)^2 p_a^2 p_a^\nu r_a^\mu \non\\
&& -2(p_a.p_b)^2 p_b.r_a p_b^\mu p_a^\nu +2(p_a.p_b)^3 p_b^\mu r_a^\nu -3p_a^2 p_b^2 p_a.p_b p_b^\mu r_a^\nu +2p_a^2 (p_b^2)^2 p_a^\mu r_a^\nu \non\\
&& +p_b^2 p_a.p_b p_a^\mu p_b.r_a p_a^\nu -p_b^2 (p_a.p_b)^2 p_a^\mu r_a^\nu -p_b^2 \lbrace (p_a.p_b)^2 -p_a^2 p_b^2 \rbrace J_a^{\mu\nu}+2(p_a.p_b)^2 p_b^\nu p_{b\alpha}\Sigma_a^{\alpha\mu}\non\\
&& -2(p_a.p_b)^2 p_b^\mu p_{b\alpha}\Sigma_a^{\alpha\nu} +p_b^2 p_a.p_b p_a^\mu p_{b\alpha}\Sigma_a^{\alpha\nu}-p_a^2 p_b^2 p_b^\mu p_{b\alpha}\Sigma_a^{\alpha\nu}\Big\rbrace\Big]\label{L2final}
\ee
\section{Soft radiation from high frequency gravitational wave emission}\label{radiation}
Let us first write down the expression of eq.\eqref{Thathspin} with the replacement of $G_{r}(k-\ell)G_{r}(\ell)$ by  $-2\pi i\delta(\ell^2)\big[H(\ell^0)-H(-\ell^0)\big]G_{r}(k-\ell)$,
\be
\widehat{T}_{extra}^{h\mu\nu}(k) &\equiv & (8\pi G) (2\pi i)\sum_{a,b=1}^{M+N}\int \f{d^{4}\ell}{(2\pi)^{4}}\ \delta(\ell^2)\big[H(\ell^0)-H(-\ell^0)\big] G_{r}(k-\ell) \f{1}{p_{b}.\ell-i\epsilon}\ \f{1}{p_{a}.(k-\ell)-i\epsilon}\non\\
&&\times \Big{\lbrace} p_{b\alpha}p_{b\beta}-\f{1}{2}p_{b}^{2}\eta_{\alpha\beta}+ip_{b(\alpha}J_{b,\beta)\gamma}\ell^{\gamma}-\f{i}{2}\eta_{\alpha\beta}p_{b}^{\delta}J_{b,\delta\gamma}\ell^{\gamma}\Big{\rbrace}\ \mathcal{F}^{\mu\nu, \alpha\beta , \rho\sigma}(k,\ell)\non\\
&&\times \Big{\lbrace}p_{a\rho}p_{a\sigma}-\f{1}{2}p_{a}^{2}\eta_{\rho\sigma}+ip_{a(\rho}J_{a,\sigma)\delta}(k-\ell)^{\delta}-\f{i}{2}\eta_{\rho\sigma}p_{a}^{\kappa}J_{a,\kappa\tau}(k-\ell)^{\tau}\Big{\rbrace}
\ee
We want to analyze the above expression in the integration region $\omega <<|\ell^\mu|<<L^{-1}$. Due to the presence of $\big[H(\ell^0)-H(-\ell^0)\big]$ inside the integrand, the part of the integrand containing even number of $\ell$ vanishes. On the other hand to produce a $\ln\omega$ factor we need an integrand with four power of $\ell$ in the denominator, which will vanish due to the presence of $\big[H(\ell^0)-H(-\ell^0)\big]$. Hence just from this argument it is clear that we can not receive any non-vanishing $\ln\omega$ or $\omega\ln\omega$ contribution from the above expression. Still for completeness, let us analysze the non-vanishing contribution of the above expression up to order $\mathcal{O}(\omega^0)$ in the integration range $\omega <<|\ell^\mu|<<L^{-1}$:
\be
&&\Delta \widehat{T}_{extra}^{h\mu\nu}(k) \non\\
&=& -(8\pi G) (2\pi i)\sum_{a,b=1}^{M+N}\int_\omega^{L^{-1}} \f{d^{4}\ell}{(2\pi)^{4}}\ \delta(\ell^2)\big[H(\ell^0)-H(-\ell^0)\big] \f{1}{2k.\ell -i\epsilon} \f{1}{p_{b}.\ell-i\epsilon}\ \f{1}{p_{a}.\ell+i\epsilon}\non\\
&& \Bigg[\Big{\lbrace} p_{b\alpha}p_{b\beta}-\f{1}{2}p_{b}^{2}\eta_{\alpha\beta}\Big{\rbrace}\ \Delta_{(\ell\ell)}\mathcal{F}^{\mu\nu, \alpha\beta , \rho\sigma}(k,\ell) \Big{\lbrace}p_{a\rho}p_{a\sigma}-\f{1}{2}p_{a}^{2}\eta_{\rho\sigma}+ip_{a(\rho}J_{a,\sigma)\delta}k^{\delta}-\f{i}{2}\eta_{\rho\sigma}p_{a}^{\kappa}J_{a,\kappa\tau}k^{\tau}\Big{\rbrace}\non\\
&& -\Big{\lbrace} p_{b\alpha}p_{b\beta}-\f{1}{2}p_{b}^{2}\eta_{\alpha\beta}\Big{\rbrace}\ \Delta_{(k\ell)}\mathcal{F}^{\mu\nu, \alpha\beta , \rho\sigma}(k,\ell) \Big{\lbrace}ip_{a(\rho}J_{a,\sigma)\delta}\ell^{\delta}-\f{i}{2}\eta_{\rho\sigma}p_{a}^{\kappa}J_{a,\kappa\tau}\ell^{\tau}\Big{\rbrace}\non\\
&& +\Big{\lbrace} ip_{b(\alpha}J_{b,\beta)\gamma}\ell^\gamma -\f{i}{2}\eta_{\alpha\beta}p_{b}^\delta J_{b,\delta\gamma}\ell^\gamma\Big{\rbrace}\ \Delta_{(k\ell)}\mathcal{F}^{\mu\nu, \alpha\beta , \rho\sigma}(k,\ell) \Big{\lbrace}p_{a\rho}p_{a\sigma}-\f{1}{2}p_{a}^{2}\eta_{\rho\sigma}\Big{\rbrace}\Bigg]\non\\
&& -(8\pi G) (2\pi i)\sum_{a,b=1}^{M+N}\int_\omega^{L^{-1}} \f{d^{4}\ell}{(2\pi)^{4}}\ \delta(\ell^2)\big[H(\ell^0)-H(-\ell^0)\big] \f{1}{2k.\ell -i\epsilon} \f{1}{p_{b}.\ell-i\epsilon}\ \f{p_a.k}{(p_{a}.\ell+i\epsilon)^2}\non\\
&& \Bigg[-\Big{\lbrace} p_{b\alpha}p_{b\beta}-\f{1}{2}p_{b}^{2}\eta_{\alpha\beta}\Big{\rbrace}\ \Delta_{(\ell\ell)}\mathcal{F}^{\mu\nu, \alpha\beta , \rho\sigma}(k,\ell) \Big{\lbrace}ip_{a(\rho}J_{a,\sigma)\delta}\ell^{\delta}-\f{i}{2}\eta_{\rho\sigma}p_{a}^{\kappa}J_{a,\kappa\tau}\ell^{\tau}\Big{\rbrace}\non\\
&& +\Big{\lbrace} ip_{b(\alpha}J_{b,\beta)\gamma}\ell^\gamma -\f{i}{2}\eta_{\alpha\beta}p_{b}^\delta J_{b,\delta\gamma}\ell^\gamma\Big{\rbrace}\ \Delta_{(\ell\ell)}\mathcal{F}^{\mu\nu, \alpha\beta , \rho\sigma}(k,\ell) \Big{\lbrace}p_{a\rho}p_{a\sigma}-\f{1}{2}p_{a}^{2}\eta_{\rho\sigma}\Big{\rbrace}\Bigg]\label{Textra}
\ee
Now in the above expression after contraction of various terms within the square bracket, we find some terms containing $\ell^2$, which will vanish due to the presence of $\delta(\ell^2)$. On the other hand if we get $p_b.\ell$ (or $p_a.\ell$) then it cancels with the denominator $\lbrace p_b.\ell -i\epsilon\rbrace ^{-1}$ (or$\lbrace p_a.\ell +i\epsilon\rbrace ^{-1}$ ) and if only one such denominator present then after cancellation of it the rest of the coefficient vanishes after using $\sum\limits_b p_{b}^\alpha =0$ or $\sum\limits_b J_b^{\alpha\beta}=0$ (or $\sum\limits_a p_{a}^\alpha =0$ or $\sum\limits_a J_a^{\alpha\beta}=0$). Hence after eliminating those terms and  interchanging $a\leftrightarrow b$ in some places\footnote{Though the signs of $i\epsilon$ are different for denominators between $p_a.\ell$ and $p_b.\ell$, still $a\leftrightarrow b$ interchange makes sense as the integrals have to evaluate with  $\delta (\ell^2)$. Hence we can set $\epsilon =0$ from the beginning.}, the above expression simplifies to:
\be
&&\Delta \widehat{T}_{extra}^{h\mu\nu}(k) \non\\
&=& -(8\pi G) (2\pi i)\sum_{a,b=1}^{M+N}\int_\omega^{L^{-1}} \f{d^{4}\ell}{(2\pi)^{4}}\ \delta(\ell^2)H(\ell^0) \f{\ell^\mu \ell^\nu}{k.\ell -i\epsilon} \f{1}{p_{b}.\ell-i\epsilon}\ \f{1}{p_{a}.\ell+i\epsilon}\Big\lbrace (p_a.p_b)^2 -\f{1}{2}p_a^2 p_b^2 \Big\rbrace \non\\
&& +(8\pi^2 G) \sum_{a,b=1}^{M+N}\int_\omega^{L^{-1}} \f{d^{4}\ell}{(2\pi)^{4}}\ \delta(\ell^2)H(\ell^0) \f{1}{k.\ell -i\epsilon} \f{1}{p_{b}.\ell-i\epsilon}\ \f{1}{p_{a}.\ell+i\epsilon}\ \Big\lbrace 2\ell^\mu \ell^\nu p_a.p_b p_{a\rho}J_b^{\rho\sigma} k_\sigma \non\\
&& -p_a^2 \ell^\mu \ell^\nu p_{b\rho}J_b^{\rho\sigma}k_\sigma  + 2p_a.p_b p_a^\mu \ell^\nu k_\alpha J_b^{\alpha\beta}\ell_\beta +2p_b.k p_a^\mu \ell^\nu p_{a\alpha}J_b^{\alpha\beta}\ell_\beta +2p_a.p_b p_a^\nu \ell^\mu k_\alpha J_b^{\alpha\beta}\ell_\beta\non\\
&& +2p_b.k p_a^\nu \ell^\mu p_{a\alpha}J_b^{\alpha\beta}\ell_\beta -2p_a.k p_a.p_b \ell^\nu J_b^{\mu\alpha}\ell_\alpha -2p_a.k p_b^\mu \ell^\nu p_{a\alpha}J_b^{\alpha\beta}\ell_\beta -2p_a.k p_a.p_b \ell^\mu J_b^{\nu\alpha}\ell_\alpha\non\\
&& -2p_b^\nu \ell^\mu p_a.k p_{a\alpha}J_b^{\alpha\beta}\ell_\beta  \Big\rbrace\non\\
&& +(8\pi^2 G) \sum_{a,b=1}^{M+N}\int_\omega^{L^{-1}} \f{d^{4}\ell}{(2\pi)^{4}}\ \delta(\ell^2)H(\ell^0) \f{1}{k.\ell -i\epsilon} \f{1}{p_{b}.\ell-i\epsilon}\ \f{p_a.k}{(p_{a}.\ell+i\epsilon)^2}\non\\
&& \Big[  2\ell^{\mu}\ell^{\nu}p_a.p_b p_{a\alpha}J_b^{\alpha\beta}\ell_{\beta}-p_a^2 \ell^\mu \ell^\nu p_{b\alpha}J_b^{\alpha\beta}\ell_\beta  - 2\ell^{\mu}\ell^{\nu}p_a.p_b p_{b\rho}J_a^{\rho\sigma}\ell_\sigma  +p_b^2 \ell^\mu \ell^\nu p_{a\rho}J_a^{\rho\sigma}\ell_\sigma \Big]\non\\
&& +(8\pi^2 G) \sum_{a,b=1}^{M+N}\int_\omega^{L^{-1}} \f{d^{4}\ell}{(2\pi)^{4}}\ \delta(\ell^2)H(\ell^0)  \f{1}{p_{b}.\ell-i\epsilon}\ \f{1}{p_{a}.\ell+i\epsilon}\ \Big\lbrace  4 p_a^\mu p_a^\nu p_{b\alpha}J_b^{\alpha\beta}\ell_\beta +2 p_a.p_b p_a^\mu J_b^{\nu\alpha}\ell_\alpha\non\\
&& +2 p_b^\nu p_a^\mu p_{a\alpha}J_b^{\alpha\beta}\ell_\beta -2 p_a^2 p_b^\mu J_b^{\nu\alpha}\ell_\alpha -2 p_a^2 p_b^\nu J_b^{\mu\alpha}\ell_\alpha -2 p_b^\mu p_a^\nu p_{a\alpha}J_b^{\alpha\beta}\ell_\beta -2 p_a.p_b p_a^\nu J_b^{\mu\alpha}\ell_\alpha\Big\rbrace \label{Textra_final}
\ee
The first line in the above expression has been identified with leading order soft radiation from real hard gravitational radiation in appendix-B of \cite{1912.06413}. So here generalizing appendix-B of \cite{1912.06413} to the next order, we try to understand whether the rest of the terms above can be understood as  subleading order soft radiation from real hard gravitational radiation. Taking care of gravitational flux up to subleading order, the energy-momentum tensor for soft gravitational radiation with momentum $k$ becomes,
\be
&& \widehat{T}_{R}^{\mu\nu}(k)\non\\
&=& \f{G}{\pi^2}\sum_{a,b=1}^{M+N} \int d^4 \ell \ \delta(\ell^2)H(\ell^0) \Bigg[ \f{p_a^\rho p_a^\sigma -ip_a^{(\rho}(k-\ell)_\alpha J_a^{\alpha\sigma)}}{p_a.(\ell -k)+i\epsilon}\Bigg]\ \Bigg[ \f{p_b^\kappa p_b^\tau -ip_b^{(\kappa}\ell_\beta J_b^{\beta\tau)}}{p_b.\ell -i\epsilon}\Bigg]\non\\
&& \ \times \f{\ell^\mu \ell^\nu \sum\limits_{r}\varepsilon^{r}_{\rho\sigma}\varepsilon_{\kappa\tau}^{r*}-i\ell^{(\mu}k_{\gamma}\big[\Sigma_g^{\gamma\nu)}\big]_{\rho\sigma ,\kappa\tau}}{i(\ell.k -i\epsilon)}\label{TR}
\ee
where polarisation sum and spin tensor of soft gravitational field are given by,
\be
\sum\limits_{r}\varepsilon^{r}_{\rho\sigma}\varepsilon_{\kappa\tau}^{r*} &=&\f{1}{2}\big(\eta_{\rho\kappa}\eta_{\sigma\tau}+\eta_{\rho\tau}\eta_{\sigma\kappa}-\eta_{\rho\sigma}\eta_{\kappa\tau}\big)\\
\big[\Sigma_g^{\gamma\nu)}\big]_{\rho\sigma ,\kappa\tau}&=& -\f{i}{2}\Big[ \eta_{\rho\kappa}(\delta^\gamma_\sigma \delta^\nu_\tau -\delta^\nu_\sigma \delta^\gamma_\tau)+ \eta_{\rho\tau}(\delta^\gamma_\sigma\delta^\nu_\kappa - \delta^\nu_\sigma\delta^\gamma_\kappa)+ \eta_{\sigma\kappa}(\delta^\gamma_\rho \delta^\nu_\tau -\delta^\nu_\rho \delta^\gamma_\tau)+ \eta_{\sigma\tau}(\delta^\gamma_\rho\delta^\nu_\kappa - \delta^\nu_\rho\delta^\gamma_\kappa)\Big]\non\\
\ee
Substituting the above results in eq.\eqref{TR} and analyzing in the integration region $\omega<<|\ell^\mu|<<L^{-1}$, we get the following non-vanishing contribution up to order $\omega^0$:
\be
&& \widehat{T}_{R}^{\mu\nu}(k)\non\\
&=& -\f{iG}{\pi^2}\sum_{a,b=1}^{M+N}\int_\omega^{L^{-1}} d^{4}\ell\ \delta(\ell^2)H(\ell^0) \f{\ell^\mu \ell^\nu}{k.\ell -i\epsilon} \f{1}{p_{b}.\ell-i\epsilon}\ \f{1}{p_{a}.\ell+i\epsilon}\Big\lbrace (p_a.p_b)^2 -\f{1}{2}p_a^2 p_b^2 \Big\rbrace \non\\
&& + \f{G}{2\pi^2} \sum_{a,b=1}^{M+N}\int_\omega^{L^{-1}}d^{4}\ell\ \delta(\ell^2)H(\ell^0) \f{1}{k.\ell -i\epsilon} \f{1}{p_{b}.\ell-i\epsilon}\ \f{1}{p_{a}.\ell+i\epsilon}\ \Big\lbrace 2\ell^\mu \ell^\nu p_a.p_b p_{a\rho}J_b^{\rho\sigma} k_\sigma \non\\
&& -p_a^2 \ell^\mu \ell^\nu p_{b\rho}J_b^{\rho\sigma}k_\sigma  + 2p_a.p_b p_a^\mu \ell^\nu k_\alpha J_b^{\alpha\beta}\ell_\beta +2p_b.k p_a^\mu \ell^\nu p_{a\alpha}J_b^{\alpha\beta}\ell_\beta +2p_a.p_b p_a^\nu \ell^\mu k_\alpha J_b^{\alpha\beta}\ell_\beta\non\\
&& +2p_b.k p_a^\nu \ell^\mu p_{a\alpha}J_b^{\alpha\beta}\ell_\beta -2p_a.k p_a.p_b \ell^\nu J_b^{\mu\alpha}\ell_\alpha -2p_a.k p_b^\mu \ell^\nu p_{a\alpha}J_b^{\alpha\beta}\ell_\beta -2p_a.k p_a.p_b \ell^\mu J_b^{\nu\alpha}\ell_\alpha\non\\
&& -2p_b^\nu \ell^\mu p_a.k p_{a\alpha}J_b^{\alpha\beta}\ell_\beta \Big\rbrace\non\\
&& + \f{G}{2\pi^2} \sum_{a,b=1}^{M+N}\int_\omega^{L^{-1}}d^{4}\ell\ \delta(\ell^2)H(\ell^0) \f{\ell^\mu \ell^\nu}{k.\ell -i\epsilon} \f{1}{p_{b}.\ell-i\epsilon}\ \f{p_a.k}{(p_{a}.\ell+i\epsilon)^2}\non\\
&& \Big[ \Big\lbrace 2p_a.p_b p_{a\alpha}J_b^{\alpha\beta}\ell_{\beta}-p_a^2 p_{b\alpha}J_b^{\alpha\beta}\ell_\beta \Big\rbrace -\Big\lbrace 2p_a.p_b p_{b\rho}J_a^{\rho\sigma}\ell_\sigma  -p_b^2  p_{a\rho}J_a^{\rho\sigma}\ell_\sigma  \Big\rbrace\Big]
\ee
Now if we compare the above expression with eq.\eqref{Textra_final}, we observe that leaving the last two lines of eq.\eqref{Textra_final}, rest of the terms matches. But we do not need to worry about these extra terms as they don't contribute at order $\omega\ln\omega$.

\bibliography{classicalref.bib}
\bibliographystyle{utphys}

\end{document}